\documentclass[runningheads]{llncs}
\usepackage{amsmath}
\usepackage{amsfonts}
\usepackage{graphicx}
\usepackage{framed}
\usepackage[colorlinks, anchorcolor=blue, linkcolor=blue, citecolor=blue, urlcolor=cyan]{hyperref}
\usepackage{tikz}
\usepackage{lscape}
\usepackage{hyperref}
\usepackage{breakcites}
\usepackage{colortbl}
\usepackage{amssymb}

\hypersetup{
    colorlinks=true,    
    urlcolor=blue       
}

\usepackage{pifont}
\usepackage{lineno}
\usepackage{multirow}
\usepackage{enumitem}

\newcommand{\Cuckoo}{\mathsf{Cuckoo}}
\newcommand{\Simple}{\mathsf{Simple}}
\newcommand{\PRF}{\mathsf{PRF}}
\newcommand{\HW}{\mathsf{HW}}
\newcommand{\Zero}{\mathsf{zero}}
\newcommand{\Func}{\mathcal{F}}
\newcommand{\FuncROT}{\mathcal{F}_\mathsf{ROT}}

\newcommand{\FuncbOPPRF}{\mathcal{F}_\mathsf{bOPPRF}}
\newcommand{\FuncbssPMT}{\mathcal{F}_\mathsf{bssPMT}}

\newcommand{\FuncMS}{\mathcal{F}_\mathsf{shuffle}}

\newcommand{\FuncMZS}{\mathcal{F}_\mathsf{MZS}}
\newcommand{\FuncbMZS}{\mathcal{F}_\mathsf{bMZS}}
\newcommand{\FuncbpMZS}{\mathcal{F}_\mathsf{bpMZS}}
\newcommand{\FuncbpMZSp}{\mathcal{F}_\mathsf{bpMZSp}}
\newcommand{\FuncbpNMZS}{\mathcal{F}_\mathsf{bpNMZS}}
\newcommand{\FuncPZS}{\mathcal{F}_\mathsf{PZS}}

\newcommand{\FuncrPZS}{\mathcal{F}_\mathsf{rPZS}}

\newcommand{\ProtobMZS}{\Pi_\mathsf{bMZS}}
\newcommand{\ProtoPZS}{\Pi_\mathsf{PZS}}
\newcommand{\ProtobpMZS}{\Pi_\mathsf{bpMZS}}
\newcommand{\ProtobpMZSp}{\Pi_\mathsf{bpMZSp}}
\newcommand{\ProtobpNMZS}{\Pi_\mathsf{bpNMZS}}

\newcommand{\Sd}{\mathcal{S}}
\newcommand{\Rcv}{\mathcal{R}}
\newcommand{\Adv}{\mathcal{A}}

\begin{document}


%
%
%
%
%

\title{\Large \bf Multi-Party Private Set Operations from Predicative Zero-Sharing}

\titlerunning{Multi-Party Private Set Operations from Predicative Zero-Sharing}

\author{Minglang Dong\inst{1} \and
Yu Chen\inst{1} \and
Cong Zhang\inst{2} \and 
Yujie Bai\inst{1} \and
Yang Cao\inst{1}
}

\authorrunning{M. Dong et al.}

\institute{School of Cyber Science and Technology, Shandong University, Qingdao 266237, China
\and
Institute for Advanced Study, BNRist, Tsinghua University, Beijing, China \\ 
\email{\{minglang\_dong,baiyujie\}@mail.sdu.edu.cn, yuchen@sdu.edu.cn, zhangcong@mail.tsinghua.edu.cn}
}

\maketitle

\begin{abstract}
Typical protocols in the multi-party private set operations (MPSO) setting enable $m > 2$ parties to perform certain secure computation on the intersection or union of their private sets, realizing a very limited range of MPSO functionalities. Most works in this field focus on just one or two specific functionalities, resulting in a large variety of isolated schemes and a lack of a unified framework in MPSO research. 
In this work, we present an MPSO framework, which allows $m$ parties, each holding a set, to securely compute any set formulas (arbitrary compositions of a finite number of binary set operations, including intersection, union and difference) on their private sets. Our framework is highly versatile and can be instantiated to accommodate a broad spectrum of MPSO functionalities. To the best of our knowledge, this is the first framework to achieve such a level of flexibility and generality in MPSO, without relying on generic secure multi-party computation (MPC) techniques.

Our framework exhibits favorable theoretical and practical performance. The computation and communication complexity scale linearly with the set size $n$, and it achieves optimal complexity that is on par with the naive solution for widely used functionalities, such as multi-party private set intersection (MPSI), MPSI with cardinality output (MPSI-card), and MPSI with cardinality and sum (MPSI-card-sum), in the standard semi-honest model. Furthermore, the instantiations of our framework mainly from symmetric-key techniques yield efficient protocols for MPSI, MPSI-card, MPSI-card-sum, and multi-party private set union (MPSU), with online performance surpassing or matching the state of the art.

At the technical core of our framework is a newly introduced primitive called predicative zero-sharing. This primitive captures the universality of a number of MPC protocols and is composable. We believe it may be of independent interest.
\end{abstract}


\section{Introduction}

In the setting of multi-party private set operations (MPSO), a set of $m$ $(m > 2)$ parties, each holding a private set of items, wish to perform secure computation on their private sets without revealing any additional information. In the end, only one of the parties (denoted as the leader) learns the resulting set and other parties (denoted as clients) learn nothing.
MPSO is an expansive research field with a variety of rich functionalities.
The typical functionalities that have been studied in the MPSO literature can be
divided into two categories:

\begin{itemize}
    \item Multi-party private set intersection (MPSI)~\cite{FreedmanNP04,KS-CRYPTO-2005,SangSTX06,LiW07,SangS09,BA-ASIACCS-2012,HazayV17,KMPRT-CCS-2017,InbarOP18,GhoshN19,GPRTY-CRYPTO-2021,NTY-CCS-2021,ChandranD0OSS21,GordonHL22,BNOP-AsiaCCS-2022,VCE22,BayEHSV22,Zhang23c,WuYC24}, which is to compute intersection, and its variants --- MPSI with cardinality output (MPSI-card)~\cite{KS-CRYPTO-2005,BA-ASIACCS-2012,ChenDGB22,GaoTY24,GiorgiLOV24}, which is to compute intersection cardinality, MPSI with cardinality and sum (MPSI-card-sum)~\cite{ChenDGB22,GiorgiLOV24}, which is to compute intersection cardinality and sum (of the associated payloads), and circuit-MPSI~\cite{KS-CRYPTO-2005,BA-ASIACCS-2012,ChandranD0OSS21,Zhang23c,SC-abs-2309}, which allows parties to learn secret shares of the indicator vector for intersection with respect to leader's elements, that can be further fed into generic MPC (with leader's elements) to compute arbitrary function on intersection; 
    \item Multi-party private set union (MPSU)~\cite{KS-CRYPTO-2005,Frikken-ACNS-2007,BA-ASIACCS-2012,SCK12,VCE22,LG-ASIACRYPT-2023,GaoNT24,DongCZB24}, which is to compute union, and its variants --- MPSU with cardinality output (MPSU-card)~\cite{BA-ASIACCS-2012}, which is to compute union cardinality, and circuit-MPSU~\cite{BA-ASIACCS-2012}, which allows parties to learn secret shares of elements in union, that can be further fed into generic MPC to compute arbitrary function on union. 
\end{itemize}

There are several major problems in the field of MPSO: 
\begin{itemize}
    \item \textbf{Unrealistic security assumptions.}
    Despite the vast body of existing works in the MPSO literature, many rely on assumptions of unconditional trust, which is fraught with security risks. For example, some works~\cite{LG-ASIACRYPT-2023,Zhang23c,GaoTY24,SC-abs-2309} assume non-collusion among particular parties, which is unlikely to hold in practice. Therefore, an important theme in the MPSO research is to achieve security against arbitrary collusion.

    \item \textbf{Unsatisfied application needs.} Many real-world applications want to compute more than intersection and union (or partial/aggregate information on them), however, the existing protocols cannot meet these demands, both in terms of functionality and efficiency. For instance, a social services organization intends to determine the list of people on welfare with cancer~\cite{KS-CRYPTO-2005}. To fulfill this task, all hospitals should collectively calculate the union of lists of cancer patients, meanwhile keeping the union privacy, then an intersection operation between the unrevealed union of cancer patients and the welfare rolls is performed. This problem is apparently beyond the above two categories and still lacks practical solutions to date. 
    
    \item \textbf{Fragmented landscape of protocols.} Most existing works focus on only one or two specific functionalities, resulting in a large variety of isolated schemes and a lack of a unified framework in MPSO.
\end{itemize}

Given the numerous possible compositions of a finite number of binary set operations (including intersection, union and difference) on $m$ sets, ideally, MPSO should enable $m$ parties to securely compute arbitrary set formulas on their private sets. All the aforementioned functionalities are special cases of this generic functionality (hereafter, we use MPSO to refer particularly to this generic functionality). The seminal work of Kissner and Song~\cite{KS-CRYPTO-2005} has explored the MPSO functionality. Unfortunately, they failed to fully realize it. The set formulas being computed in their protocol only allow to include union and
intersection set operations, excluding the difference operation. Namely, their protocol only realizes a restricted MPSO functionality. For instance, computing $X_1 \setminus (X_2 \cap X_3)$ is not feasible in their protocol.
Furthermore, their protocol relies heavily on additively homomorphic encryption (AHE) and high-degree polynomial calculations, leading to prohibitively large computational costs, hence is totally impractical.

A follow-up work by Blanton and Aguiar~\cite{BA-ASIACCS-2012} redesigns the circuits for computing intersection, union and difference as oblivious sorting and adjacent comparisons in a sorted set, and implements these circuits using generic MPC protocols. Thanks to the composability of generic MPC, their protocols are also composable to compute arbitrary compositions of binary set operations, thereby fully realizing the MPSO functionality. However, the protocol’s heavy reliance on generic MPC incurs substantial computational costs, and even the simplest cases --- MPSI and MPSU --- exhibit poor practical efficiency. For instances, their experiments report a runtime of $24.8$ seconds for both MPSI and MPSU with $3$ parties, each holding $2^{11}$ items of $32$ bits, which is the largest experiment in terms of the number of parties and set size reported in their paper. Moreover, their protocols are only secure in the honest majority setting.

Motivated by the above, we raise the following question:

\begin{center}
\emph{Can we fully realize the MPSO functionality with security against arbitrary collusion and acceptable performance in the semi-honest model?}
\end{center}


\subsection{Our Contributions}
In this work, we answer the above question affirmatively. Our technical route is: First, defining a predicate formula representation for any set formulas; Second, presenting a composable primitive --- predicative zero-sharing --- and its composition technique; Then, instantiating predicative zero-sharing as a primitive tailored for MPSO --- membership zero-sharing --- with lightweight building blocks; Finally, constructing a framework based on oblivious transfer (OT) and symmetric-key operations in the standard semi-honest model, which fully realizes not only the MPSO functionality, but also the extended MPSO-card and circuit-MPSO functionalities.
Our contributions can be detailed as follows:
\begin{trivlist}
\item \textbf{Predicate Formula Representation.} The first challenge in realizing MPSO is to identify a suitable representation for any set formulas, which determines the generality and practicality of the resulting framework. The prior work~\cite{KS-CRYPTO-2005} represents set formulas using intersection, union, and element reduction operations, whose arbitrary compositions can only express a limited subset of set formulas, thereby restricting its generality. The follow-up work~\cite{BA-ASIACCS-2012} adopts the naive representation based on intersection, union, and difference operations to achieve full generality. However, to support composability, it relies heavily on generic MPC, which significantly hinders practicality. In this work, we introduce a new representation called canonical predicate formula (CPF), which is designed with a particular structure to enable an MPSO framework achieving  the best of both worlds: generality and practicality. Specifically, this representation is a subset of first-order set predicate formulas (which are first-order predicate formulas where each atomic proposition is a set membership predicate $x \in X_i$, connected by $\mathsf{AND}$, $\mathsf{OR}$ and $\mathsf{NOT}$ operators), defined as a disjunction of several subformulas that are in a certain form, representing a partition of the desired set. We prove that any set formulas can be transformed into CPF representations, and the number of subformulas in CPF dominates the performance of protocols.

\item \textbf{Predicative Zero-Sharing and Relaxation.} The second challenge is to devise a composable primitive based on our predicate formula representation. We introduce a novel primitive called predicative zero-sharing, which is a family of protocols, each associated with a first-order predicate formula and encoding the truth-value of the formula on the parties’ inputs into a secret-sharing over a finite field among the parties. Specifically, if the formula is true, the parties hold a secret-sharing of 0, otherwise a secret sharing of a random value. We put forward a simpler simulation-based security definition for predicative zero-sharing protocols, which is composed of three requirements: correctness, privacy and independence, and give a rigorous proof of its equivalence to the standard security definition for a broader class of MPC protocols (predicative zero-sharing is its subset). This simpler security definition simplifies the security proof of our predicative zero-sharing protocols. Moreover,
under this simpler security definition, we can relax the security of predicative zero-sharing by removing the independence requirement. This relaxed version of predicative zero-sharing admits the abstraction of much more MPC protocols, such as random oblivious transfer (ROT), equality-conditional randomness generation (ECRG)~\cite{Jia0ZG24},\footnote{We found that the ECRG functionality satisfies the definition of predicative zero-sharing while the construction in~\cite{Jia0ZG24} only achieves the security of relaxed predicative zero-sharing. This is because ECRG is a probabilistic functionality whereas~\cite{Jia0ZG24} proved its security using the definition for deterministic functionalities.} and so on. We present a composition technique to compose several relaxed predicative zero-sharing protocols into a single relaxed predicative zero-sharing protocol based on $\mathsf{AND}$ and $\mathsf{OR}$ operators. We also present a transformation technique to transform any relaxed predicative zero-sharing protocol into a standard version. Combining these two techniques, we can construct predicative zero-sharing for any first-order predicate formulas, from relaxed predicative zero-sharing associated
with all literals (atomic propositions or their negations) within the formula.

\item \textbf{Membership Zero-Sharing.} To enable the instantiation of predicative zero-sharing, we introduce membership zero-sharing, a particular class of predicative zero-sharing tailored for MPSO, by specifying the associated predicate formula as a first-order set predicate formula $Q$. In this setting, one party (denoted as $P_\mathsf{pivot}$) inputs an element and the other parties input sets. The output secret-sharing among the parties encodes whether $P_\mathsf{pivot}$'s input element, together with all input sets, satisfy $Q$. For example, consider $3$ parties where $P_1$ inputs an element $x$, $P_2$ inputs a set $X_2$, and $P_3$ inputs a set $X_3$. Suppose $Q$ is in the form of $x \in X_2 \land x \notin X_3$, if $x \in X_2 \setminus X_3$, $P_1,P_2,P_3$ hold a secret-sharing of 0, otherwise they hold a secret-sharing of a random value.
Given that any first-order set predicate formula $Q$ is only composed of two types of literals --- set membership predicates $x \in Y$ and the negations $x \notin Y$, by instantiating relaxed membership zero-sharing associated with $x \in Y$ and $x \notin Y$ respectively, we can build membership zero-sharing protocols for any first-order set predicate formulas, by following the recipe for predicative zero-sharing. Our instantiations are both built on lightweight components, including oblivious programmable pseudorandom function (OPPRF), batch secret-shared private membership test (batch ssPMT), and ROT. This contributes to the good efficiency of our framework.

\item \textbf{MPSO, MPSO-card and Circuit-MPSO.} In analogy with MPSI (resp. MPSU) functionality to MPSI-card and circuit-MPSI (resp. MPSU-card and circuit-MPSU), we extend MPSO functionality into two new functionalities --- MPSO-card and circuit-MPSO, where MPSO-card computes the resulting set's cardinality and circuit-MPSO reveals secret shares of the resulting set, which can be further fed into generic MPC to compute arbitrary function on the resulting set.
Based on the CPF representation for any set formulas and membership zero-sharing for any
first-order set predicate formulas, we put forth a framework fully realizing MPSO, MPSO-card and circuit-MPSO functionalities. A high level of our framework proceeds as follows.
We begin with the simplest case, where the desired set is a subset of the input set of the leader.\footnote{MPSI is a typical example of this case, as the intersection is a subset of any input sets.}
In this case, the leader acts as $P_\mathsf{pivot}$, and for each element in its input set, the leader invokes the membership zero-sharing associated with the CPF representation of the desired set, with the other parties inputting their sets. As a result, for each elements in the leader's set that belongs to the desired set, the parties hold a secret-sharing of 0. Since all these elements exactly compose the desired set, the partiescan simply reconstruct all secret-sharings to the leader, who computes the resulting set by identifying all elements with corresponding secrets as 0. This construction can be optimized using the hashing to bins technique (see Figure). We extend this simplest case of our framework to achieve full MPSO functionality, by leveraging the structural properties of our CPF representation, which guarantee that the set represented by each subformula $Q_i$ in the CPF is a subset of some party $P_j$’s input set, and all these sets form a partition of the desired set. For each subformula $Q_i$, the parties invoke membership zero-sharing with $P_j$ acting as $P_\mathsf{pivot}$, and the invocation is similar to the simplest case.
After the membership zero-sharing invocations for all subformulas, the union of all output secret-sharings encode a partition of the desired set. However, a straightforward reconstruction in this setting may reveal information through the order of secret-sharings, therefore, the parties have to invoke a multi-party secret-shared shuffle protocol to randomly permute and re-share all secret-sharings. Finally, the shuffled secret-sharings are reconstructed to the leader. Since the resulting set remains being secret-shared before the last reconstruction step, this MPSO protocol is easy to be extended to MPSO-card and circuit-MPSO protocols.
\end{trivlist} 

In addition to the above contributions, perhaps surprisingly, we also make independent contributions in the following sub-fields, 
by instantiating our framework to yield the aforementioned typical protocols:

\begin{trivlist}
\item \textbf{MPSI.} 
The MPSI protocol from our framework has the best computation and communication complexity among all MPSI protocols based on OT and symmetric-key operations in the standard semi-honest model. Particularly, this is the first MPSI construction to achieve the optimal complexity that is on par with the naive solution (the leader's computation and communication complexity are both $O(m n)$ and each client's computation and communication complexity are both $O(n)$, where $n$ is the set size and $m$ is the number of parties) without extensive use of public-key operations, in standard semi-honest model. The previous MPSI protocol~\cite{KMPRT-CCS-2017} with this optimal complexity is only secure in the weaker augmented semi-honest model. In this work, we close this gap. Our MPSI protocol is also highly efficient in the online phase, which 
requires only $8.9$ seconds and $738$ MB of communications for $10$ parties with sets of $2^{20}$ items each, regardless of the item length, while the state-of-the-art MPSI protocol~\cite{WuYC24} requires $32.9$ seconds and $1921$ MB of communications for their full protocol.

\item \textbf{MPSI-card, MPSI-card-sum and Circuit-MPSI.} 
The MPSI-card and MPSI-card-sum protocols from our framework are the first MPSI-card and MPSI-card-sum constructions with the optimal computation and communication complexity in the standard semi-honest model. Our MPSI-card has $14.0 - 20.3 \times$ lower communication than the state-of-the-art MPSI-card protocol~\cite{ChenDGB22}. We provide the first MPSI-card-sum implementation and it only doubles the computation and communication costs of our MPSI while realizing a richer functionality. Concretely, our MPSI-card requires only $9.2$ seconds while our MPSI-card-sum requires $16.7$ seconds in online phase for $10$ parties with sets of $2^{20}$ items each, regardless of the item length. Additionally, the circuit-MPSI protocol from our framework is the first circuit-MPSI construction in dishonest majority setting.

\item \textbf{MPSU.} 
The MPSU protocol from our framework has the best computation and communication complexity among all MPSU protocols based on OT and symmetric-key operations in the standard semi-honest model. It could be seen as an instance of the secret-sharing based MPSU paradigm, which abstracts all existing MPSU protocols relying only
on symmetric-key primitives~\cite{LG-ASIACRYPT-2023, DongCZB24}. Our protocol achieves the optimal complexity of this paradigm for the first time (with $O(m^2 n)$ computation and communication complexity of leader and $O(m n)$ computation and communication complexity of clients). Our MPSU protocol is comparable with the state-of-the-art MPSU protocol~\cite{DongCZB24}. Concretely, it requires only $13.2$ seconds in online phase for $5$ parties, each holding $2^{20}$ items of $64$ bits.

\item \textbf{MPSU-card and circuit-MPSU.} 
The MPSU-card and circuit-MPSU protocols from our framework are the only efficient constructions for MPSU-card and circuit-MPSU, with performance that is nearly the same as our MPSU protocol.
\end{trivlist} 

\subsection{Related Work}
Despite the immense amount of existing works on the typical functionalities in this field, many are insecure against arbitrary collusion~\cite{LiW07,BA-ASIACCS-2012,SCK12,ChandranD0OSS21,LG-ASIACRYPT-2023,Zhang23c,GaoTY24,SC-abs-2309}, or have non-negligible false positives~\cite{BayEHSV22,VCE22}.
We only focus on works achieving semi-honest security against arbitrary collusion without non-negligible false positives. Distribution of research attention among these works is extremely imbalanced: MPSI has been extensively studied~\cite{FreedmanNP04,KS-CRYPTO-2005,SangSTX06,SangS09,HazayV17,KMPRT-CCS-2017,InbarOP18,GhoshN19,GPRTY-CRYPTO-2021,NTY-CCS-2021,GordonHL22,BNOP-AsiaCCS-2022,WuYC24}, while MPSU only receives relatively little attention~\cite{KS-CRYPTO-2005,Frikken-ACNS-2007,GaoNT24,DongCZB24}. MPSI-card~\cite{KS-CRYPTO-2005,ChenDGB22,GiorgiLOV24} and MPSI-card-sum~\cite{ChenDGB22,GiorgiLOV24} are extremely understudied sub-fields, with only a couple of secure MPSI-card protocol~\cite{KS-CRYPTO-2005,ChenDGB22,GiorgiLOV24} and MPSI-card-sum protocol~\cite{ChenDGB22,GiorgiLOV24} against arbitrary collusion. Even worse, no prior work has realized circuit-MPSI, MPSU-card, circuit-MPSU in the dishonest majority setting. We provide more details on the classic and state-of-the-art protocols below. A comprehensive theoretical comparison between related protocols and ours is provided in Appendix~\ref{appdix:compare}.


\begin{trivlist}
\item \textbf{MPSI.} Freedman et al.~\cite{FreedmanNP04} introduced the first MPSI protocol based on oblivious polynomial evaluation (OPE), which is implemented using AHE. Kisser and Song~\cite{KS-CRYPTO-2005} proposed an MPSI protocol using the OPE technique along with the polynomial representations. These two protocols both require quadratic computation complexity with respect to the set size $n$ for each party, resulting in complete impracticality.

Kolesnikov et al.~\cite{KMPRT-CCS-2017} proposed two MPSI protocols in the augmented semi-honest model and standard semi-honest model, respectively. 
The former achieves the optimal complexity of MPSI, 
while it is only secure in the augmented semi-honest model. The latter fails to achieve optimal complexity as it requires the clients' complexity to depend on the corruption threshold $t$. Garimella et al.~\cite{GPRTY-CRYPTO-2021} improved these protocols using oblivious key-value store (OKVS)~\cite{PRTY20,GPRTY-CRYPTO-2021,RR-CCS-2022,BPSY-USENIX-2023} and showed that the augmented semi-honest protocol actually enjoys malicious security. 
Following these works, Nevo et al.~\cite{NTY-CCS-2021} proposed an efficient MPSI protocol in the malicious model, where the client's communication complexity depends only on $n$ (while the computation complexity still depends on $t$). 

Inbar et al.~\cite{InbarOP18} proposed two MPSI protocols in the augmented semi-honest and standard semi-honest model, based on OT and garbled Bloom filters. In these two protocols, each party's computation complexity is $O(m n)$. The Ben-Efraim et al.~\cite{BNOP-AsiaCCS-2022} extended the former to the malicious model.

Recently, Wu et al.~\cite{WuYC24} proposed two semi-honest MPSI protocols based on OPRF and OKVS. In these two protocols, the client's complexity depends on $t$.


\item \textbf{MPSI-card and MPSI-card-sum.} Chen et al.~\cite{ChenDGB22} proposed the first MPSI-card and MPSI-card-sum protocols based on OT and symmetric-key operations, which are also the only practical MPSI-card and MPSI-card-sum protocols in the standard semi-honest model. In their protocols, the leader's complexity is $O(m n + t n \log n)$ and the client's complexity is $O(t n)$.

\item \textbf{MPSU.} Kisser and Song~\cite{KS-CRYPTO-2005} introduced the first MPSU protocol, based on polynomial representations and AHE. The substantial number of AHE operations and high-degree polynomial calculations incur unacceptable efficiency.

Recently, Gao et al.~\cite{GaoNT24} proposed an MPSU protocol in the standard semi-honest model. This protocol relies on public-key operations and has super-linear computation and communication complexity for each party in term of $n$.

Recently, Dong et al.~\cite{DongCZB24} proposed two MPSU protocols in the standard semi-honest model. The first protocol, based on OT and symmetric-key operations, eliminates the non-collusion assumption in~\cite{LG-ASIACRYPT-2023}, at the cost of increasing client's complexity to quadratic in terms of $m$.
The second protocol achieves linear complexity. However, it relies on public-key operations with a lower efficiency. 


\end{trivlist} 

\section{Preliminaries}\label{sec:preliminaries}

\subsection{Notation}
Let $m$ denote the number of parties. 
We use $P_i$ ($1 \le i \le m$) to denote the parties, $X_i$ to represent the sets they hold, where each set has $n$ $l$-bit elements. $[x] = (x_1,\cdots,x_m)$ denotes an additive secret-sharing among $m$ parties, i.e., each $P_i$ holds a share $x_i$ such that $x_1 + \cdots x_m = x$. $x \Vert y$ denotes the concatenation of two strings. We use $\lambda, \sigma$ as the computational and statistical security parameters respectively, and use $\overset{s}{\approx}$ (resp.$\overset{c}{\approx}$) to denote that two distributions are statistically (resp. computationally) indistinguishable. For a vector $\textbf{a}$, $a_i$ denotes the $i$-th component, $\HW(\textbf{a})$ denotes the hamming weight of \textbf{a}, $\Zero(\textbf{a})$ denotes the number of 0 in \textbf{a}, and $\pi(\textbf{a}) = (a_{\pi(1)}, \cdots, a_{\pi(n)})$, where $\pi$ is a permutation over $n$ items. 
 The notation $\textbf{a} \oplus \textbf{b}$ denotes a component-wise XOR, i.e., $(a_1 \oplus b_1, \cdots, a_n \oplus b_n)$.

\subsection{Security Model}

In this work, we consider semi-honest and static adversaries $\Adv$ with the capability to corrupt an arbitrary subset of parties. To capture the security of a protocol in the simulation-based model~\cite{Goldreich-FoC2,Canetti01}, we use the following notations:
\begin{itemize}
    \item Let $f = (f_1, \cdots, f_m)$ be a probabilistic polynomial-time $m$-ary functionality and let $\Pi$ be a $m$-party protocol for computing $f$. 
    \item The view of $P_i$ ($1 \le i \le m$) during an execution of $\Pi$ on all parties' inputs $\textbf{x} = (x_1, \cdots, x_m)$ is denoted by $\mathsf{View}_i^\Pi(\textbf{x})$, including the $i$-th party's input $x_i$, its internal random tape and all messages that it received.
    \item The output of $P_i$ during an execution of $\Pi$ on $\textbf{x}$ is denoted by $\mathsf{Output}_i^\Pi(\textbf{x})$. The joint output of parties is $\mathsf{Output}^\Pi(\textbf{x}) = (\mathsf{Output}_1^\Pi(\textbf{x}), \cdots, \mathsf{Output}_m^\Pi(\textbf{x})).$
\end{itemize}
\begin{definition}\label{def:prob-secure}
We say that $\Pi$ securely computes $f$ in the presence of $\mathcal{A}$ if there exists a PPT algorithm $\mathsf{Sim}$ s.t. for every $\textbf{P}_{\mathcal{A}} = \{P_{i_1},\cdots,P_{i_t}\} \subset \{P_1,\cdots,P_m\}$, 
\begin{gather*}
    \{\mathsf{Sim}(\textbf{P}_{\mathcal{A}},\textbf{x}_{\mathcal{A}},f_{\mathcal{A}}(\textbf{x})),f(\textbf{x})\}_{\textbf{x}} \overset{c}{\approx} \{\mathsf{View}_{\mathcal{A}}^\Pi(\textbf{x}),\mathsf{Output}^\Pi(\textbf{x})\}_{\textbf{x}},
\end{gather*}
\end{definition}
where $\textbf{x}_{\mathcal{A}} = (x_{i_1}, \cdots, x_{i_t}), f_{\mathcal{A}} = (f_{i_1}, \cdots, f_{i_t}), \mathsf{View}_{\mathcal{A}}^\Pi(\textbf{x}) = (\mathsf{View}_{i_1}^\Pi(\textbf{x}),\cdots, \mathsf{View}_{i_t}^\Pi(\textbf{x}))$. 

\subsection{Multi-party Private Set Operations}

\begin{figure}[!hbtp]
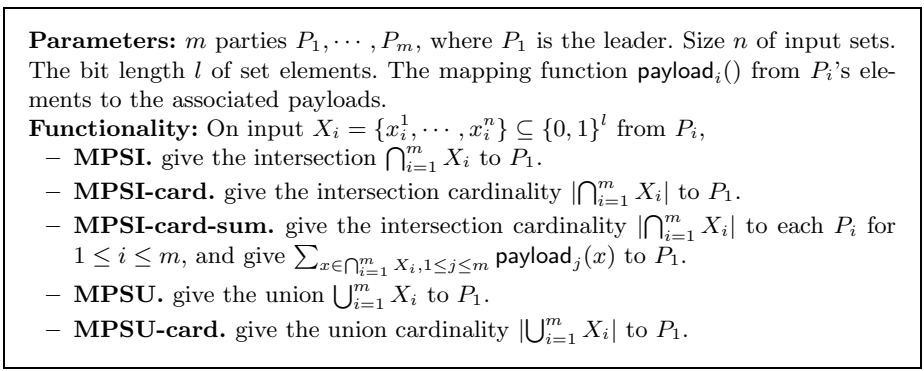

\begin{framed}
\begin{minipage}[center]{\textwidth}
\begin{trivlist}
\item \textbf{Parameters:} $m$ parties $P_1, \cdots, P_m$, where $P_1$ is the leader. Size $n$ of input sets. The bit length $l$ of set elements. The mapping function $\mathsf{payload}_i()$ from $P_i$'s elements to the associated payloads.
\item \textbf{Functionality:} On input $X_i = \{x_i^1,\cdots, x_i^n\} \subseteq \{0,1\}^l$ from $P_i$, 
\begin{itemize}[itemsep=2pt,topsep=0pt,parsep=0pt]
    \item \textbf{MPSI.} give the intersection $\bigcap_{i=1}^m X_i$ to $P_1$.
    \item \textbf{MPSI-card.} give the intersection cardinality $\lvert \bigcap_{i=1}^m X_i \rvert$ to $P_1$.
    \item \textbf{MPSI-card-sum.} give the intersection cardinality $\lvert \bigcap_{i=1}^m X_i \rvert$ to each $P_i$ for $1 \le i \le m$, and give $\sum_{x \in \bigcap_{i=1}^m X_i, 1 \le j \le m} \mathsf{payload}_j(x)$ to $P_1$.
    \item \textbf{MPSU.} give the union $\bigcup_{i=1}^m X_i$ to $P_1$.
    \item \textbf{MPSU-card.} give the union cardinality $\lvert \bigcup_{i=1}^m X_i \rvert$ to $P_1$.
\end{itemize}
\end{trivlist}
\end{minipage}
\end{framed}
\caption{Typical Functionalities in MPSO}\label{fig:func-mpso}
\end{figure}

MPSO is a special case of secure multi-party computation (MPC). Figure~\ref{fig:func-mpso} formally defines the typical ideal functionalities computing the intersection, intersection cardinality, intersection sum with cardinality, union, and union cardinality over the parties' private sets.

\subsection{Random Oblivious Transfer}
Oblivious transfer (OT)~\cite{Rabin05} is a foundational primitive in MPC, the functionality of 1-out-of-2 random OT (ROT) is given in Figure~\ref{fig:func-rot}.

\begin{figure}[!hbtp]
\begin{framed}
\begin{minipage}[center]{\textwidth}
\begin{trivlist}
\item \textbf{Parameters.} Sender $\Sd$, Receiver $\Rcv$. A field $\mathbb{F}$.
\item \textbf{Functionality.} On input $e \in \{0,1\}$ from $\Rcv$, sample $r_0, r_1 \gets \mathbb{F}$. Give $(r_0, r_1)$ to $\Sd$ and give $r_e$ to $\Rcv$.
\end{trivlist}
\end{minipage}
\end{framed}
\caption{1-out-of-2 Random OT Functionality $\FuncROT$}\label{fig:func-rot}
\end{figure}

\subsection{Batch Oblivious Programmable Pseudorandom Function}\label{sec:bopprf}


Oblivious pseudorandom function (OPRF)~\cite{FIPR-TCC-2005,CM-CRYPTO-2020,RS-EUROCRYPT-2021} is a central primitive in the area of PSO. Kolesnikov et al.~\cite{KKRT-CCS-2016} introduced batched OPRF, which provides a batch of OPRF instances. In the $i$-th instance, the sender $\Sd$ learns a PRF key $k_i$, while the receiver $\Rcv$ inputs $x_i$ and learns $\PRF(k_i, x_i)$. 

Oblivious programmable pseudorandom function (OPPRF)~\cite{KMPRT-CCS-2017,PSTY-EUROCRYPT-2019,CGS22,RS-EUROCRYPT-2021,RR-CCS-2022} is an extension of OPRF, which lets $\Sd$ program a PRF $F$ so that it has specific uniform outputs for some specific inputs and pseudorandom outputs for all other inputs. This kind of PRF that outputs programmed values on a certain programmed set of inputs is called programmable PRF (PPRF)~\cite{PSTY-EUROCRYPT-2019}. $\Rcv$ evaluates OPPRF with no knowledge of whether it learns a programmed output of $F$ or just a pseudorandom value. The batch OPPRF functionality is given in Figure~\ref{fig:func-bopprf}.

\begin{figure}[!hbtp]
\begin{framed}
\begin{minipage}[center]{\textwidth}
\begin{trivlist}
\item \textbf{Parameters.} Sender $\Sd$. Receiver $\Rcv$. Batch size $B$. The bit length $l$ of keys. The bit length $\gamma$ of values.
\item \textbf{Sender's inputs.} $\Sd$ inputs $B$ sets of key-value pairs including:
\begin{itemize}[itemsep=2pt,topsep=0pt,parsep=0pt]
\item Disjoint key sets $K_1,\cdots,K_B$.
\item The value sets $V_1,\cdots,V_B$, where $\lvert K_i \rvert = \lvert V_i \rvert$, $i \in [B]$.
\end{itemize}
\item \textbf{Receiver's inputs.} $\Rcv$ inputs $B$ queries $\textbf{x} \subseteq (\{0,1\}^l)^B$.
\item \textbf{Functionality:} On input $(K_1, \cdots, K_B)$ and $(V_1,\cdots,V_B)$ from $\Sd$ and $\textbf{x} \subseteq (\{0,1\}^l)^B$ from $\Rcv$,
\begin{itemize}[itemsep=2pt,topsep=0pt,parsep=0pt]
\item Generate a uniform PPRF key $k_i$ and an auxiliary information $\mathsf{hint}_i$ for $i \in [B]$;
\item Give vector $\textbf{k} = (k_1, \cdots, k_B)$ and $(\mathsf{hint}_1, \cdots, \mathsf{hint}_B)$ to $\Sd$.
\item Sample a PPRF $F: \{0,1\}^* \times \{0,1\}^l \to \{0,1\}^{\gamma}$ such that $F(k_i,K_i(j)) = V_i(j)$ for $i \in [B], 1 \le j \le \lvert K_i \rvert$;
\item Define $f_i = F(k_i,x_i)$, for $i \in [B]$;
\item Give vector $\textbf{f} = (f_1, \cdots, f_B)$ to $\Rcv$.
\end{itemize}
\end{trivlist}
\end{minipage}
\end{framed}
\caption{Batch OPPRF Functionality $\FuncbOPPRF$}\label{fig:func-bopprf}
\end{figure}

\subsection{Batch Secret-Shared Private Membership Test}

Batch secret-shared private membership test (batch ssPMT)~\cite{DongCZB24} is a two-party protocol that implements multiple instances of ssPMT\cite{CGS22,LG-ASIACRYPT-2023} between a sender $\Sd$ and a receiver $\Rcv$. Given a batch size of $B$, $\Sd$ inputs $B$ sets $X_1,\cdots,X_B$, while $\Rcv$ inputs $B$ elements $x_1,\cdots,x_B$. As a result, $\Sd$ and $\Rcv$ receive secret shares of a bit vector of size $B$, where the $i$-th bit is 1 if $x_i \in X_i$, 0 otherwise. The batch ssPMT functionality is given in Figure~\ref{fig:func-bssPMT}.
Dong et al.~\cite{DongCZB24} proposed an efficient construction with linear complexities, based on batch OPPRF and secret-shared private equality test (ssPEQT)~\cite{PSTY-EUROCRYPT-2019,CGS22}.

\begin{figure}[!hbtp]
\begin{framed}
\begin{minipage}[center]{\textwidth}
\begin{trivlist}
\item \textbf{Parameters.} Sender $\Sd$. Receiver $\Rcv$. Batch size $B$. The bit length $l$ of set elements.
\item \textbf{Inputs.} $\Sd$ inputs $B$ disjoint sets $X_1, \cdots, X_B$ and $\Rcv$ inputs $\textbf{x} \subseteq (\{0,1\}^l)^B$.
\item \textbf{Functionality.} On inputs $X_1, \cdots, X_B$ from $\Sd$ and input $\textbf{x}$ from $\Rcv$, for $1 \le i \le B$, sample two random bits $e_S^i, e_R^i$ under the constraint that if $x_i \in X_i,e_S^i \oplus e_R^i = 1$, otherwise $e_S^i \oplus e_R^i = 0$. Give $\textbf{e}_S = (e_S^1, \cdots, e_S^B)$ to $\Sd$ and $\textbf{e}_R = (e_R^1, \cdots, e_R^B)$ to $\Rcv$.
\end{trivlist}
\end{minipage}
\end{framed}
\caption{Batch ssPMT Functionality $\FuncbssPMT$}\label{fig:func-bssPMT}
\end{figure}

\subsection{Multi-Party Secret-Shared Shuffle}\label{sec:ms}

Multi-party secret-shared shuffle functionality works by randomly permuting the share vectors of all parties and then refreshing all shares, ensuring that the permutation remains unknown to any coalition of $m-1$ parties. The formal functionality is given in Figure \ref{fig:func-ms}. 
Eskandarian et al.~\cite{EB22} proposed an online-efficient protocol where the parties generate share correlations in the offline phase, so that the leader's online complexity scales linearly with $n$ and $m$, while the clients' online complexity scales linearly with $n$ and is independent of $m$.

\begin{figure}[!hbtp]
\begin{framed}
\begin{minipage}[center]{\textwidth}
\begin{trivlist}
\item \textbf{Parameters.} $m$ parties $P_1, \cdots P_m$. The dimension of vector $n$. The item length $l$.
\item \textbf{Functionality.} On input $\textbf{x}_i = {(x_i^1, \cdots, x_i^n)}$ from each $P_i$, sample a random permutation $\pi : [n] \to [n]$. For $1 \le i \le m$, sample $\textbf{x}_{i}' \gets (\{0,1\}^l)^n$ satisfying $\bigoplus_{i=1}^m \textbf{x}_{i}' = \pi(\bigoplus_{i=1}^m \textbf{x}_i)$. Give $\textbf{x}_{i}'$ to $P_i$.
\end{trivlist}
\end{minipage}
\end{framed}
\caption{Multi-Party Secret-Shared Shuffle Functionality $\FuncMS$}\label{fig:func-ms}
\end{figure}

\subsection{Hashing to Bins}

The hashing to bins technique was introduced by Pinkas et al.~\cite{PSZ-USENIX-2014,PSZ-USENIX-2015} to construct two-party PSI. At a high level, the receiver $\Rcv$ uses hash functions $h_1, h_2, h_3$ to assign its items to $B$ bins via Cuckoo hashing~\cite{PR04}, so that each bin has at most one item.\footnote{The Cuckoo hashing process uses eviction and the choice of bins for each item depends on the entire set.} On the other hand, the sender $\Sd$ assigns each of its items $x$ to all bins $h_1(x), h_2(x), h_3(x)$ via simple hashing. This guarantees that for each item $x$ of $\Rcv$, if $x$ is mapped into the $b$-th bin of Cuckoo hash table ($b \in \{h_1(x), h_2(x), h_3(x)\}$), and $x$ is in $\Sd$'s set, then the $b$-th of simple hash table certainly contains $x$.

We denote simple hashing with the following notation:
\begin{gather*}
\mathcal{T}^1, \cdots, \mathcal{T}^B \gets \Simple_{h_1,h_2,h_3}^B(X)
\end{gather*}
This expression represents hashing the items of $X$ into $B$ bins using simple hashing with hash functions $h_1, h_2, h_3: \{0, 1\}^* \to [B]$. The output is a hash table denoted by $\mathcal{T}^1, \cdots, \mathcal{T}^B$, where for each $x \in X$, $\mathcal{T}^{h_i(x)} \supseteq \{x \Vert i \vert i= 1, 2, 3\}$.\footnote{Appending the index of the hash function is helpful for dealing with edge cases like $h_1(x) = h_2(x) = i$, which happen with non-negligible probability.}

We denote Cuckoo hashing with the following notation:
\begin{gather*}
\mathcal{C}^1, \cdots, \mathcal{C}^B \gets \Cuckoo_{h_1,h_2,h_3}^B(X)
\end{gather*}
This expression represents hashing the items of $X$ into $B$ bins using Cuckoo hashing with hash functions $h_1, h_2, h_3: \{0, 1\}^* \to [B]$. The output is a Cuckoo hash table denoted by $\mathcal{C}^1, \cdots, \mathcal{C}^B$, where for each $x \in X$ there is some $i \in \{1, 2, 3\}$ such that $\mathcal{C}^{h_i(x)} = \{x \Vert i\}$. Some Cuckoo hash positions are irrelevant, corresponding to empty bins. We use these symbols throughout subsequent sections. 

\section{Predicate Formula Representation of Set Formulas}

In this section, we formally introduce our predicate formula representation for any set formulas. We define several notions to facilitate the subsequent discussion of our work and present several theorems.

\subsection{Constructible Set}

We formalize the notion of the resulting sets that can be derived from any set formulas being computed over the parties' private sets in the context of MPSO. We refer to these resulting sets as constructible sets.

\begin{definition} Let $X_1,\cdots,X_m$ be $m$ sets. A set $Y$ is called a constructible set (over $X_1,\cdots,X_m$) if it can be derived from $X_1,\dots,X_m$ through a finite number of set operations, including intersection, union, and difference.
\end{definition}\label{constructible}

In particular, if a constructible set $Y$ satisfies $Y \subseteq X_i$ for some $1 \le i \le m$, we call it an $X_i$-constructible set (over $X_1,\cdots,X_m$).

\begin{definition} Let $\varphi(x,X_1,\cdots,X_m)$ be a first-order predicate formula. If $\varphi$ is composed of atomic propositions of the form $M(x,X_i): x \in X_i$, we call it a (first-order) set predicate formula.
\end{definition}\label{constructible}

Any constructible set can be represented by a set predicate formula. This corresponding relationship is formalized in the following theorem.

\begin{theorem}\label{theorem:set1}
Let $X_1,\cdots,X_m$ be $m$ sets and $Y$ is a constructible set. There exists a set predicate formula $\varphi(x,X_1,\cdots,X_m)$, s.t. for any urelement $x$,
$$x \in Y \iff \varphi(x,X_1,\cdots,X_m) = 1.$$
\end{theorem}

We prove this theorem in Appendix~\ref{proof:set1}.

\subsection{Canonical Predicate Formula Representation}

\begin{definition}\label{def:cpf}
A set predicate formula $\varphi(x,X_1,\cdots,X_m)$ is called set-separable with respect to $X_i$ for some $1 \le i \le m$ if it can be written in the form: $$\varphi(x,X_1,\cdots,X_m) = (x \in X_i) \land \psi(x, X_1, \cdots, X_{i-1},X_{i+1},\cdots,X_m),$$ where $\psi(x, X_1, \cdots, X_{i-1},X_{i+1},\cdots,X_m)$ is a set predicate formula not involving $X_i$, which we call the separation formula of $\varphi(x,X_1,\cdots,X_m)$ with respect to $X_i$.
\end{definition}

\begin{corollary}
If a constructible set $Y$ corresponds to a set predicate formula which is set-separable with respect to $X_i$, then $Y$ is an $X_i$-constructible set.
\end{corollary}

\begin{definition} Let set predicate formula $\psi(x,X_1,\cdots,X_m)$ is a disjunction of one or more subformulas,\footnote{A disjunction of one subformulas is itself.} denoted as $\psi = Q_1 \lor \cdots \lor Q_s (s \ge 1)$. Let $Y_i$ be the corresponding set represented by $Q_i$, then if each subformula $Q_i$ is set-separable with respect to some $X_j$ ($1 \leq j \leq m$), and the set of $\{Y_1, \cdots, Y_s\}$ forms a partition of $Y$, we call $\psi(x,X_1,\cdots,X_m)$ a canonical predicate formula (CPF) representation (over $X_1,\cdots,X_m$).
\end{definition}\label{def:set}

\begin{theorem}\label{theorem:set2}
Let $X_1,\cdots,X_m$ be $m$ sets and $Y$ is a constructible set. There exists a CPF representation $\psi(x,X_1,\cdots,X_m)$ s.t. for any urelement $x$,
$$x \in Y \iff \psi(x,X_1,\cdots,X_m) = 1$$
\end{theorem}

We prove this theorem by showing how to construct $\psi$ in Appendix~\ref{proof:set2}.

In order to illustrate Theorem~\ref{theorem:set2}, consider three constructible sets in the three-party setting: the intersection $Y = X_1 \cap X_2 \cap X_3$, the union $Y = X_1 \cup X_2 \cup X_3$ and a complex set formula $Y = ((X_1 \cap X_2) \cup (X_1 \cap X_3)) \setminus (X_1 \cap X_2 \cap X_3)$. We provide the respective CPF representation $\psi(x,X_1,X_2,X_3)$ for each case below.
\begin{trivlist}
    \item \textbf{Intersection.} $\psi(x,X_1,X_2,X_3) = (x \in X_1) \land (x \in X_2) \land (x \in X_3)$. 
    In this case, $\psi$ is a disjunction of one subformula $Q_1 = \psi$, corresponding to the set $Y_1 = Y$. $Q_1$ is set-separable with respect to $X_1$, $X_2$ and $X_3$. $Y_1$ itself is a partition of $Y$.
    \item \textbf{Union.} $\psi(x,X_1,X_2,X_3) = (x \in X_1) \lor ((x \notin X_1) \land (x \in X_2)) \lor ((x \notin X_1) \land (x \notin X_2) \land (x \in X_3))$. $\psi$ is a disjunction of three subformulas $Q_1, Q_2, Q_3$, where each $Q_i = (x \notin X_1) \land \cdots \land (x \notin X_{i-1}) \land (x \in X_i)$ represents $Y_i = X_i \setminus (X_1 \cup \cdots \cup X_{i-1})$. $Q_i$ is set-separable with respect to $X_i$. $\{Y_1,Y_2,Y_3\}$ is a partition of $Y$.
    \item \textbf{Complex set formula.} There are two CPF representations for this case: 
    \begin{itemize}
      \item $\psi(x,X_1,X_2,X_3) = ((x \in X_1) \land (x \in X_2) \land (x \notin X_3)) \lor ((x \in X_1) \land (x \in X_3) \land (x \notin X_2))$. $\psi$ is a disjunction of two subformulas $Q_1, Q_2$ with the corresponding sets $Y_1 = X_1 \cap X_2 \setminus X_3$ and $Y_2 = X_1 \cap X_3 \setminus X_2$. $Q_1$ is set-separable with respect to $X_1$ and $X_2$, while $Q_2$ is set-separable with respect to $X_1$ and $X_3$. $\{Y_1,Y_2\}$ is a partition of $Y$.
      \item $\psi(x,X_1,X_2,X_3) = (x \in X_1) \land [((x \in X_2) \land (x \notin X_3)) \lor ((x \in X_3) \land (x \notin X_2))]$. $\psi$ is set-separable with respect to $X_1$, so it is a disjunction of one subformula $Q_1 = \psi$, which obviously satisfies the definition of CPF representation.
    \end{itemize}
\end{trivlist}
The third example demonstrates that the CPF representation for a given constructible set is not unique. Different CPF representations can impact our protocols' efficiency. A key principle is to minimize the number of subformulas in the CPF representation to optimize performance.

\section{Predicative Zero-Sharing}
In this section, we introduce a new notion called predicative zero-sharing. 
By zero-sharing, we refer to a ``redundant'' secret-sharing that distributes one bit into secret shares over a finite field, where this bit is 0 only if some condition holds (e.g. the truth-value of a first-order predicate formula is true). Predicative zero-sharing is a family of protocols, each associated with a first-order predicate formula, encoding the truth-value of the formula on the parties' inputs into a zero-sharing among the parties. This class of protocols can be composed based on $\mathsf{AND}$ and $\mathsf{OR}$ operators.

\subsection{Definitions}\label{def:pzs}
A predicative zero-sharing protocol allows a set of $m$ $(m \ge 2)$ parties with private inputs to receive secret shares of 0, on condition that the truth-value of the associated first-order predicate formula $Q$ in terms of their inputs is true, otherwise receive secret shares of a uniformly random value. The formal definition of predicative zero-sharing functionality is given in Figure~\ref{fig:func-pzs}. 

\begin{figure}[!hbtp]
\begin{framed}
\begin{minipage}[center]{\textwidth}
\begin{trivlist}
\item \textbf{Parameters:} $m$ parties $P_1, \cdots P_m$ with inputs $\textbf{x} = (x_1, \cdots, x_m)$. A field $\mathbb{F}$. A first-order predicate formula $Q$.
\item \textbf{Functionality:} On input $x_i$ from each $P_i$, sample $s_i \gets \mathbb{F}$ s.t. if $Q(\textbf{x}) = 1$, $s_1 + \cdots + s_m = 0$. Give $s_i$ to $P_i$ for $1 \le i \le m$.
\end{trivlist}
\end{minipage}
\end{framed}
\caption{Ideal functionality for predicative zero-sharing $\FuncPZS^Q$}\label{fig:func-pzs}
\end{figure}

\subsection{Security}
Given the probabilistic functionality, a protocol must meet Definition \ref{def:prob-secure} to securely compute predicative zero-sharing.
However, we observe that for predicative zero-sharing, a simpler security definition with three requirements, including correctness, privacy and independence, is equivalent. 
We demonstrate this equivalence through the following theorem. Note that we will use this simpler security definition to prove security of all predicative zero-sharing protocols in this work.

Consider a probabilistic $m$-ary functionality $\Func^f$, which takes the parties’ inputs $\textbf{x} = (x_1, \cdots, x_m)$ and outputs secret shares of $f(\textbf{x})$ to the parties. Let $\Pi$ be a $m$-party protocol for computing $\Func^f$, and $s_i$ and $s^{\Pi}_i$ denote the output of $P_i$ from $\Func^f$, and that during the execution of $\Pi$ on \textbf{x}, respectively ($1 \le i \le m$).

\begin{theorem}\label{security}
If $f$ is a probabilistic functionality in terms of \textbf{x}, and $\Pi$ satisfies:
\begin{itemize}
    \item \textbf{Correctness}. The outputs of $\Pi$ are secret shares of $f(\textbf{x})$, namely, $$\{s_1, \cdots, s_m\}_{\textbf{x}} \overset{s}{\approx} \{s^{\Pi}_1, \cdots, s^{\Pi}_m\}_{\textbf{x}}$$
    \item \textbf{Privacy}. There exists a PPT algorithm $\mathsf{Sim}$ s.t. for every $\textbf{P}_{\mathcal{A}} = \{P_{i_1},\cdots,P_{i_t}\}$,
    \begin{gather*}
    \{\mathsf{Sim}(\textbf{P}_{\mathcal{A}},\textbf{x}_{\mathcal{A}},\textbf{s}_{\mathcal{A}})\}_{\textbf{x}} \overset{c}{\approx} \{\mathsf{View}_{\mathcal{A}}^\Pi(\textbf{x})\}_{\textbf{x}}
    \end{gather*}
    \item \textbf{Independence}. The randomness in $f(\textbf{x})$ is independent of $\mathsf{View}_{\mathcal{A}}^\Pi(\textbf{x})$ for every $\textbf{P}_{\mathcal{A}} = \{P_{i_1},\cdots,P_{i_t}\}$ during an execution of $\Pi$.
\end{itemize}
Then, there exists a PPT algorithm $\mathsf{Sim}$ s.t. for every $\textbf{P}_{\mathcal{A}} = \{P_{i_1},\cdots,P_{i_t}\}$, 
\begin{gather*}
    \{\mathsf{Sim}(\textbf{P}_{\mathcal{A}},\textbf{x}_{\mathcal{A}},\textbf{s}_{\mathcal{A}}),s_1, \cdots, s_m\}_{\textbf{x}} \overset{c}{\approx} \{\mathsf{View}_{\mathcal{A}}^\Pi(\textbf{x}),s^{\Pi}_1, \cdots, s^{\Pi}_m\}_{\textbf{x}}
\end{gather*}
\end{theorem}

We prove this theorem in Appendix~\ref{appdix:security}. 
Note that predicative zero-sharing functionality $\FuncPZS^Q$ is a special case of $\Func^f$, where
\[
f(\textbf{x}) =
\begin{cases}
  0 & \text{if } Q(\textbf{x}) = 1 \\
  s & \text{if } Q(\textbf{x}) = 0
\end{cases}
\]
and $s$ is a uniform value (the randomness in $f$, denoted as $s^\Pi$ in the real execution). The independence requirement in this case is instantiated as: if $Q(\textbf{x}) = 0$, the distribution of the secret $s^{\Pi} = s^{\Pi}_1 + \cdots + s^{\Pi}_m$ during an execution of $\Pi$ is independent of $\mathsf{View}_{\mathcal{A}}^\Pi(\textbf{x})$, and the correctness and independence requirements ensure that if $Q(\textbf{x}) = 0$, $s^{\Pi}$ is uniform and independent of the joint view of any $t \le m-1$ parties in real execution. 

\subsection{Relaxed Predicative Zero-Sharing}\label{relaxed-pzs}
Predicative zero-sharing serves as an abstraction of many existing MPC protocols. Some protocols, like multi-party secret-shared ROT (mss-ROT)~\cite{DongCZB24}, rigidly conform to Theorem~\ref{security}. In contrast, others realize functionality without meeting the independence requirement. We refer to this relaxed predicative zero-sharing functionality associated with $Q$ as $\FuncrPZS^Q$. 



Relaxed predicative zero-sharing accommodates a broader range of existing protocols, such as ROT and equality-conditional randomness generation (ECRG)~\cite{Jia0ZG24}. We demonstrate that ROT implies a relaxed predicative zero-sharing associated with a simple predicate to test whether the choice bit $e = 1$. Let $\Sd$ set its share $s_1 = -r_0$, where $r_0$ is the first message from ROT, and let $\Rcv$ set its share $s_2 = r_e$, the received message. Given that ROT functionality can be written as $-r_0 + r_e = e \cdot (-r_0 + r_1)$, if $e = 0$, $s_1 + s_2 = 0$; else, $s_1 + s_2 = -r_0 + r_1$, which is uniform but dependent on the output messages from ROT in $\Sd$'s view.

Using the standard simulation-based definition, it is hard to depict this security relaxation by defining merely a single functionality that considers all possible coalitions of $t \le m-1$ parties in the multi-party setting ($m>2$).
However, with our new security definition tailored for predicative zero-sharing, the relaxation is precisely formalized by removing the independence requirement.

\subsection{From Relaxed to Standard Predicative Zero-Sharing}\label{transform-pzs}

We give an efficient method for transforming relaxed predicative zero-sharing into standard predicative zero-sharing below.

Assuming that all parties obtain a secret-sharing $[r] = (r_1, \cdots, r_m)$ from a relaxed predicative zero-sharing protocol, with the goal to generate a new secret-sharing $[s] = (s_1, \cdots, s_m)$ meeting the standard predicative zero-sharing definition. All they need to do is to prepare a random secret-sharing $[b]$ in the offline phase (by each $P_i$ sampling a uniform share $b_i$), and perform a secure multiplication $[s] = [r] \cdot [b]$ in the online phase. We optimize the online phase of this secure multiplication through Beaver triples~\cite{Beaver-CRYPTO-1991} in Appendix~\ref{beaver triples}.
\begin{trivlist}
\item \textbf{Correctness and independence.} We set the field size $\lvert \mathbb{F} \rvert \ge 2^\sigma$. 
\begin{itemize}
\item If $Q(\textbf{x}) = 1$, $r = 0$, then $s = 0$.
\item If $Q(\textbf{x}) = 0$, $r$ is uniform. Let $E$ be the event that $s$ is uniform and and independent of the joint view of any $t \le m-1$ parties. Let $E_0$ be the event $r = 0$ and $E_1$ be the event $r \ne 0$.
Since $b$ is uniform and independent, we have $Pr[E] = Pr[E \vert E_0] \cdot Pr[E_0] + Pr[E \vert E_1] \cdot Pr[E_1] = 0 \cdot Pr[E_0] + 1 \cdot Pr[E_1] = Pr[E_1] = 1 - Pr[E_0] = 1 - \frac{1}{\lvert \mathbb{F} \rvert} \ge 1- 2^{-\sigma}$.
\end{itemize}
\item \textbf{Privacy.} The privacy follows immediately from the privacy of the relaxed predicative zero-sharing protocol and the secure multiplication.
\end{trivlist}

\subsection{From Simple to Compound Predicative Zero-Sharing}\label{compound-pzs}

According to the type of the associated first-order predicate formula $Q$, we divide predicative zero-sharing into two categories: If $Q$ is a simple predicate, we call it a simple predicative zero-sharing; If $Q$ is a compound predicate, we call it compound predicative zero-sharing. A compound predicate $Q$ is formed from $q$ literals ($q>1$) and logical connectives $\land$ and $\lor$, where each literal $Q_i$ corresponds to a simple predicate or its negation ($1 \le i \le q$).
We show that as long as we have a relaxed simple predicative zero-sharing protocol for each $Q_i$, we can build a compound predicative zero-sharing protocol for any compound predicate $Q$. 

At a high level, a compound predicative zero-sharing protocol for $Q$ proceeds in three phases: First, the parties execute the relaxed simple predicative zero-sharing protocol for each literal. For literals involving only a subset of the parties, the uninvolved parties set their missing secret share to 0; Second, they collectively manipulate the output secret-sharings by emulating the evaluation of $Q$, composing them into one output secret-sharing that meets the definition of relaxed compound predicative zero-sharing for $Q$, step by step. At the end of each step, they obtain a secret-sharing associated with the currently evaluated formula; Finally, the parties transform the relaxed compound predicative zero-sharing into the standard. The complete construction is described in Figure~\ref{fig:proto-cpzs}.

\begin{theorem}\label{theorem:cpzs}
Protocol $\ProtoPZS^Q$ securely realizes $\FuncPZS^Q$ against any semi-honest adversary corrupting $t < m$ parties in the $(\FuncrPZS^{Q_1},\cdots,\FuncrPZS^{Q_q})$-hybrid model.    
\end{theorem}
\begin{trivlist}
\item \textbf{Correctness and independence.} In each step of the formula emulation stage, 
\begin{itemize}
    \item If $Q'(\textbf{x}) = Q'_i(\textbf{x}) \land Q'_j(\textbf{x})$, the parties compute $[r_i + r_j] = [r_i] + [r_j]$. If $Q'(\textbf{x}) = 1$, namely, $Q'_i(\textbf{x}) = 1 \land Q'_j(\textbf{x}) = 1$, by the functionalities of $\FuncrPZS^{Q'_i}$ and $\FuncrPZS^{Q'_j}$, $r_i = 0 \land r_j = 0$, hence we have $r_i + r_j = 0$; otherwise, $Q'_i(\textbf{x}) = 0 \lor Q'_j(\textbf{x}) = 0$, which results that one of $r_i$ and $r_j$ is random, so $r_i + r_j$ is random.
    \item If $Q'(\textbf{x}) = Q'_i(\textbf{x}) \lor Q'_j(\textbf{x})$, the parties compute $[r_i \cdot r_j] = [r_i] \cdot [r_j]$. If $Q'(\textbf{x}) = 1$, namely, $Q'_i(\textbf{x}) = 1 \lor Q'_j(\textbf{x}) = 1$, we have $r_i = 0 \lor r_j = 0$, hence $r_i \cdot r_j = 0$; otherwise, $Q'_i(\textbf{x}) = 0 \land Q'_j(\textbf{x}) = 0$, then both of $r_i$ and $r_j$ are random. Let $E^{i,j}$ be the event that $r_i \cdot r_j$ is random. Let $E_0^{i}$ be the event $r_i = 0$ and $E_1^{i}$ be the event $r_i \ne 0$. We have $Pr[E^{i,j}] = Pr[E^{i,j} \vert E_0^{i}] \cdot Pr[E_0^{i}] + Pr[E \vert E_1^{i}] \cdot Pr[E_1^{i}] = 0 \cdot Pr[E_0^{i}] + 1 \cdot Pr[E_1^{i}] = Pr[E_1^{i}] = 1- Pr[E_0^{i}]$. To bound the correctness error by $2^{-\sigma}$, we require that the probability of any $E_0^{i}$ occurring is negligible. By union bound, $Pr[\bigvee_i E_0^{i}] \le \sum_{i} Pr[E_0^{i}] = \frac{\lvert \mathsf{OR} \rvert}{\lvert \mathbb{F} \rvert}$. Therefore, we set the field size $\lvert \mathbb{F} \rvert \ge \lvert \mathsf{OR} \rvert \cdot 2^\sigma$, where $\lvert \mathsf{OR} \rvert$ is the number of $\mathsf{OR}$ operators in $Q$.
\end{itemize}
The above correctness of implementing $\mathsf{AND}$ and $\mathsf{OR}$ operators in each step ensures the correctness of generating a relaxed predicative compound zero-sharing for $Q$. Then following the proof of correctness and independence in Section~\ref{transform-pzs}, the protocol satisfies the correctness and independence requirements of the standard predicative compound zero-sharing for $Q$.
\item \textbf{Privacy.} 
The privacy of predicative zero-sharing is straightforward to verify: All interactions happen within the invocations of blocking blocks --- all relaxed simple predicative zero-sharing protocols, the secure multiplication and transformation. Therefore, given the outputs from the ideal functionality, the simulator only needs to invoke the sub-simulators for these blocking blocks in a backward-chaining manner. As long as the privacy of all relaxed simple predicative zero-sharing protocols, the secure multiplication and transformation holds, the adversary's view is indistinguishable in the ideal and real executions.
\end{trivlist}

\begin{figure}[!hbtp]
\begin{framed}
\begin{minipage}[center]{\textwidth}
\begin{trivlist}
\item \textbf{Parameters:}  $m$ parties $P_1, \cdots P_m$ with inputs $\textbf{x} = (x_1, \cdots, x_m)$. A field $\mathbb{F}$. A simple/compound predicate $Q$ composed of $q$ literals $Q_1,\cdots,Q_q$ ($q \ge 1$) and logical connectives.
A Beaver triple $([a], [b], [c])$ generated in the offline phrase.
\item \textbf{Protocol:}
\begin{enumerate}[itemsep=2pt,topsep=0pt,parsep=0pt]
\item \textbf{The simple predicative sharing stage.} In this stage, the parties invoke $\FuncrPZS^{Q_i}$ for each literal $Q_i$, $1 \le i \le q$. If $Q_i$ does not involve all the parties, then the uninvolved parties set their secret shares to 0. As a result, each $Q_i$ has a corresponding secret-sharing among the parties.
\item \textbf{The formula emulation stage.} If $q>1$, the parties collectively emulate the computation of $Q$ in the order of operator precedence, step by step. In each step, the parties generate a secret-sharing associated with a binary clause connected by a given operator, based on the secret-sharings associated with the two contained literals $Q'_i$ and $Q'_j$, which they obtain from previous steps. The actions of parties depend on the type of operator being computed:
    \begin{itemize}
        \item $\mathsf{AND}$ operator: Suppose the parties hold two secret-sharings $[r_i]$ and $[r_j]$ associated with $Q'_i$ and $Q'_j$ respectively.     
        They want to compute a relaxed predicative zero-sharing for $Q'$, where $Q'(\textbf{x}) = Q'_i(\textbf{x}) \land Q'_j(\textbf{x})$. All they need to do is to locally add two shares to obtain the secret-sharing $[r_i + r_j]$. 
        \item $\mathsf{OR}$ operator: Suppose the parties hold two secret-sharings $[r_i]$ and $[r_j]$ associated with $Q'_i$ and $Q'_j$ respectively. They want to compute a relaxed predicative zero-sharing for $Q'$, where $Q'(\textbf{x}) = Q'_i(\textbf{x}) \lor Q'_j(\textbf{x})$. Then they perform a secure multiplication $[r_i \cdot r_j] = [r_i] \cdot [r_j]$.
    \end{itemize}
After obtaining the secret-sharing associated with $Q'$, the parties regard $Q'$ as a new literal, and repeat the above process until there is only one literal in $Q$. The secret-sharing $[r]$ associated with the ultimate literal held by the parties is the relaxed compound predicative zero-sharing for $Q$.
\item \textbf{Transformation from relaxed to standard.} All parties compute $[s]$ by performing a secure multiplication $[s] = [r] \cdot [b]$, which requires one reconstruction in the online phase using Beaver triple technique (c.f. Appendix~\ref{beaver triples}).
\end{enumerate}
\end{trivlist}
\end{minipage}
\end{framed}
\caption{Predicative Zero-Sharing $\ProtoPZS^Q$}\label{fig:proto-cpzs}
\end{figure}



\section{Membership Zero-Sharing}
Predicative zero-sharing is the abstraction of a class of MPC protocols. With the associated first-order predicate formulas determined, predicative zero-sharing can be instantiated. To instantiate predicative zero-sharing in the context of MPSO, we introduce membership zero-sharing, each associated with a set predicate formula, which serves as the technical core of our framework.

\begin{figure}[!hbtp]
\centering
\resizebox{\columnwidth}{!}{
\begin{tikzpicture}
\node (opprf) at (-3,2.5+0.5) {batch OPPRF};
\node (sspmt) at (3,2.5+0.5) {batch ssPMT};
\node (rot) at (6,2.5+0.5) {ROT};

\draw[dashed]  (-4.5,2.8+0.5) rectangle (-1.5,2.25+0.5);
\draw[dashed]  (1.5,2.8+0.5) rectangle (4.5,2.25+0.5);
\draw[dashed]  (5,2.8+0.5) rectangle (7,2.25+0.5);

\node [text width=5cm, align=center] (mem) at (-3,1.5+0.2) {Relaxed Membership Zero-Sharing for $x \in Y$};
\node [text width=5cm, align=center] (non) at (4.5,1.5+0.2) {Relaxed Membership Zero-Sharing for $x \notin Y$};

\draw  (-5,1.78+0.4) rectangle (-1,1.22);
\draw  (2.5,1.78+0.4) rectangle (6.5,1.22);

\node [text width=7cm, align=center] (v6) at (0.5,0) {Relaxed Membership Zero-Sharing for any set predicate formula $Q$};
\node [text width=7cm, align=center]  (v7) at (0.5,-1.75) {Membership Zero-Sharing for any set predicate formula $Q$};

\draw  (-2.5,0.48) rectangle (3.5,-0.48);
\draw  (-2.5,-1.25) rectangle (3.5,-2.25);

\path[->,bend right=15] (mem) edge (-0.5,0.48);
\path[->,bend left=15] (non) edge (1.7,0.48);

\draw[->]  (opprf) edge (mem);
\path[->,bend right=15]  (sspmt) edge (non);
\path[->,bend left=15]  (rot) edge (non);

\draw[->]  (v6) edge (v7);

\node at (0.6,0.85) {composition technique};
\node at (2.5,-0.85) {transformation technique};
\end{tikzpicture}
}
\caption{Technical route of building membership zero-sharing protocols for any
first-order set predicate formulas. The newly introduced primitives are marked with solid boxes. The existing primitives are marked with dashed boxes.}\label{fig:route}
\end{figure}
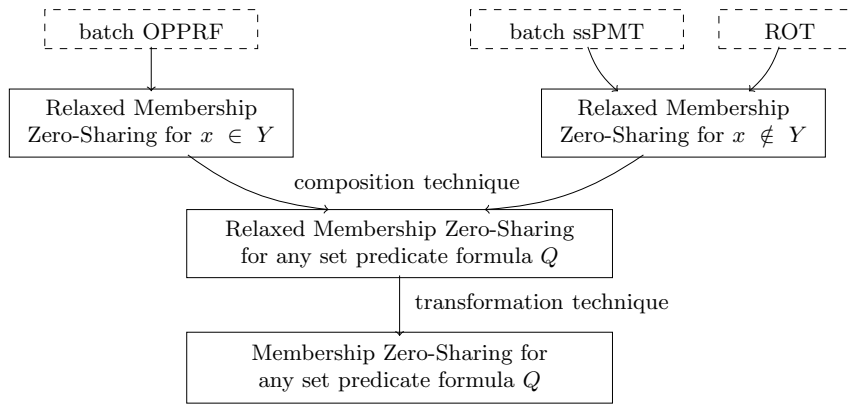

Our goal in this section is to build membership zero-sharing protocols for any first-order set predicate formulas. At a very high level, our construction follows the recipe for predicative zero-sharing in Figure~\ref{fig:proto-cpzs}, with the relaxed predicative zero-sharing components awaiting instantiations. Given that any set predicate formula is only composed of two types of literals --- set membership predicates $x \in Y$ and the negations $x \notin Y$, the task reduces to constructing two relaxed membership zero-sharing protocols, associated with $x \in Y$ and $x \notin Y$ respectively, in the two-party setting. The technical route is outlined in Figure~\ref{fig:route}. 

\subsection{Membership Zero-Sharing}\label{compound-mzs}
A membership zero-sharing protocol allows $m$ parties $(m \ge 2)$, where one party (denoted as $P_\mathsf{pivot}$) holds an element $x$ while each of the others $P_j$ holds a set $X_j$ ($j \in \{1,\cdots,m\} \setminus \{\mathsf{pivot}\}$) as input. If the associated set predicate formula $Q(x,X_1,\cdots,X_{\mathsf{pivot}-1},X_{\mathsf{pivot}+1},\cdots,X_m) = 1$, they receive secret shares of 0, otherwise they receive secret shares of a random value. The formal definition of membership zero-sharing functionality is given in Figure~\ref{fig:func-mzs}.

\begin{figure}[!hbtp]
\begin{framed}
\begin{minipage}[center]{\textwidth}
\begin{trivlist}
\item \textbf{Parameters:} $m$ parties $P_1, \cdots P_m$, where $P_\mathsf{pivot}$ is the only one holding an element instead of a set. A set predicate formula $Q$. A field $\mathbb{F}$.
\item \textbf{Functionality:} On input $x$ from $P_\mathsf{pivot}$, $X_j$ from each $P_j$ ($j \in \{1,\cdots m\} \setminus \{\mathsf{pivot}\}$), sample $s_i \gets \mathbb{F}$ for $1 \le i \le m$ s.t. if $Q(x,X_1,\cdots,X_{\mathsf{pivot}-1},X_{\mathsf{pivot}+1},\cdots,X_m)=1$, $\sum_{1 \le i \le m} s_i = 0$. Give $s_i$ to $P_i$.
\end{trivlist}
\end{minipage}
\end{framed}
\caption{Membership Zero-Sharing Functionality $\FuncMZS^Q$}\label{fig:func-mzs}
\end{figure}

A batched version of membership zero-sharing is defined in Figure~\ref{fig:func-bmzs}, where $P_\mathsf{pivot}$ holds a vector $\textbf{x} = (x_1, \cdots, x_n)$ and each $P_j$ holds $n$ sets as inputs. The parties obtain $n$ secret-sharings, where the $i$-th secret-sharing indicates the truth-value of the same formula $Q$ evaluated on their $i$-th inputs. 
In particular, if $Q$ is a conjunction of $m-1$ set membership predicates (i.e., $\bigwedge_{j \in \{1,\cdots,m\} \setminus \{\mathsf{pivot}\}} x \in X_j$), we refer to it as batch pure membership zero-sharing; if $Q$ is a conjunction of $m-1$ set non-membership predicates (i.e., $\bigwedge_{j \in \{1,\cdots,m\} \setminus \{\mathsf{pivot}\}} x \notin X_j$), we refer to it as batch pure non-membership zero-sharing. We use $\FuncbpMZS$ and $\FuncbpNMZS$ to denote these two functionalities, respectively. The details of batch pure membership zero-sharing and batch pure non-membership zero-sharing are provided in Appendix~\ref{MZS-appendix}. We also introduce a variant of pure membership zero-sharing called pure membership zero-sharing with payloads, where $P_\mathsf{pivot}$ holds an element $x$ while each of the others holds a set of elements and a set of associated payloads. In the end, the parties hold two secret sharings. If the conjunction of set membership predicates holds true (i.e., $x$ belongs to all element sets), the parties receive secret shares of 0 and secret shares of the sum of all payloads associated with $x$; otherwise they receive secret shares of two random values. The ideal functionality of batch pure membership zero-sharing with payloads $\FuncbpMZSp$ is given in Figure~\ref{fig:func-mzsp}, with further details also found in Appendix~\ref{MZS-appendix}.

\begin{figure}[!hbtp]
\begin{framed}
\begin{minipage}[center]{\textwidth}
\begin{trivlist}
\item \textbf{Parameters:} $m$ parties $P_1, \cdots P_m$, where $P_\mathsf{pivot}$ is the only one holding $n$ elements instead of $n$ sets. A set membership predicate formula $Q$. Batch size $n$. A field $\mathbb{F}$.
\item \textbf{Functionality:} On input $\textbf{x} = (x_1, \cdots, x_n)$ from $P_\mathsf{pivot}$ and $\textbf{X}_j = (X_{j,1}, \cdots, X_{j,n})$ from each $P_j$ ($j \in \{1,\cdots,m\} \setminus \{\mathsf{pivot}\}$), sample $\textbf{s}_i = (s_{i,1},\cdots,s_{i,n}) \gets \mathbb{F}^n$ for $1 \le i \le m$, s.t. for $1 \le d \le n$, if $Q(x_d,X_{1,d},X_{\mathsf{pivot}-1,d},X_{\mathsf{pivot}+1,d},\cdots,X_{m,d})=1$, $\sum_{1 \le i \le m, 1 \le d \le n} s_{i,d} = 0$. Give $\textbf{s}_i$ to $P_i$.
\end{trivlist}
\end{minipage}
\end{framed}
\caption{Batch Membership Zero-Sharing Functionality $\FuncbMZS^Q$}\label{fig:func-bmzs}
\end{figure}

\begin{figure}[!hbtp]
\begin{framed}
\begin{minipage}[center]{\textwidth}
\begin{trivlist}
\item \textbf{Parameters:} $m$ parties $P_1, \cdots P_m$, where $P_\mathsf{pivot}$ is the only one holding $n$ elements instead of $2n$ sets. Batch size $n$. A field $\mathbb{F}$ and payload field $\mathbb{F'}$. The mapping function $\mathsf{payload}_j()$ from element sets $\textbf{X}_j$ to the associated payload sets $\textbf{V}_j$.
\item \textbf{Functionality:} On input $\textbf{x} = (x_1, \cdots, x_n)$ from $P_\mathsf{pivot}$, $\textbf{X}_j = (X_{j,1}, \cdots, X_{j,n})$ and $\textbf{V}_j = (V_{j,1}, \cdots, V_{j,n})$ from each $P_j$ ($j \in \{1,\cdots m\} \setminus \{\mathsf{pivot}\}$), sample $\textbf{s}_i = (s_{i,1},\cdots,s_{i,n}) \gets \mathbb{F}^n, \textbf{w}_i = (s_{w,1},\cdots,s_{w,n}) \gets \mathbb{F'}^n$ for $1 \le i \le m$, s.t. for $1 \le d \le n$, if $\bigwedge_{j \in \{1,\cdots,m\} \setminus \{\mathsf{pivot}\}} (x_d \in X_{j,d}) = 1$, $\sum_{1 \le i \le m} s_{i,d} = 0$ and $\sum_{1 \le i \le m} w_{i,d} = \sum_{j \in \{1,\cdots,m\} \setminus \{\mathsf{pivot}\}} v_{j,d}$, where $v_{j,d} = \mathsf{payload}_j(x_d) \in V_{j,d}$. Give $(\textbf{s}_i,\textbf{w}_i)$ to $P_i$.
\end{trivlist}
\end{minipage}
\end{framed}
\caption{Batch Pure Membership Zero-Sharing with Payloads Functionality $\FuncbpMZSp$}\label{fig:func-mzsp}
\end{figure}

\subsection{Relaxed Membership Zero-Sharing for Set Membership Predicate}\label{zs-PMT}

A class of relaxed membership zero-sharing for set membership predicate $x \in Y$ can be defined as a two-party functionality as follows: There are two parties, the sender $\Sd$ with a set $Y$ and the receiver $\Rcv$ with an element $x$. The functionality samples $s,r \gets \mathbb{F}$ and if $x \in Y$, sets $u = -r$, otherwise $u = s - r$. It also generates an auxiliary information $\mathsf{hint}$ based on $s$ and outputs $r, \mathsf{hint}$ to $\Sd$ and $u$ to $\Rcv$. The $\mathsf{hint}$ is part of the syntax that allows for some leakage of the secret $s$ to $\Sd$ when $x \notin Y$, capturing the security relaxation in relaxed predicative zero-sharing.

We construct this protocol using OPPRF: $\Sd$ samples a uniform $r$, and sets $Y$ as the key set and $n$ repeated values $-r$ as the value set. Then $\Sd$ and $\Rcv$ invoke OPPRF, where $\Rcv$ inputs $x$ and receives $u$. In the end, $\Sd$ and $\Rcv$ outputs $r$ and $u$ respectively. By the OPPRF functionality, if $x \in Y$, $u = -r$, otherwise $u$ is pseudorandom. The $\mathsf{hint}$ outputted to $\Sd$ is the PRF key from OPPRF. This protocol can be naturally extended to a batched version by using batch OPPRF.


\subsection{Relaxed Membership Zero-Sharing for Set Non-Membership Predicate}\label{zs-PNMT}

A class of relaxed membership zero-sharing protocol for set non-membership predicate $x \notin Y$ can be defined as a two-party functionality as follows: The sender $\Sd$ inputs a set $Y$ while the receiver $\Rcv$ inputs $x$. The functionality samples $s,r \gets \mathbb{F}$ and if $x \notin Y$, sets $u = -r$, otherwise $u = s - r$. It also generates an auxiliary information $\mathsf{hint}$ based on $s$ and outputs $r, \mathsf{hint}$ to $\Sd$ and $u$ to $\Rcv$.

Intuitively, this functionality shares similarities with the ssPMT --- both yield secret shares of 0 when $x \notin Y$. The key difference lies in that it outputs a zero-sharing over a field $\mathbb{F}$, where the opposite of a secret-sharing of 0 is a secret-sharing of a random value in $\mathbb{F}$, while ssPMT outputs a bit secret-sharing over $\mathbb{F}_2$, where the opposite of a secret-sharing of 0 is a secret-sharing of 1. Given the efficient construction for batch ssPMT in~\cite{DongCZB24}, our goal is to efficiently transform bit secret-sharings into zero-sharings (The batched version proceeds by first having the parties invoke batch ssPMT then execute $n$ transformations).

Recall that in Section~\ref{def:pzs}, ROT is considered as a relaxed simple predicative zero-sharing associated with the predicate $e = 0$. A variant of ROT, involving two choice bits $e_0, e_1$ held by $\Sd$ and $\Rcv$ respectively~\cite{LG-ASIACRYPT-2023,Jia0ZG24,DongCZB24}, is a relaxed simple predicative zero-sharing with the associated predicate $e_0 \oplus e_1 = 0$. After executing the protocol, $\Sd$ receives $r_0, r_1 \in \mathbb{F}$ while $\Rcv$ receives $r_{e_0 \oplus e_1} \in \mathbb{F}$.\footnote{This two-choice-bit ROT is identical to the standard 1-out-of-2 ROT, where $e_0$ is sampled by $\Sd$, indicating whether to swap the order of $r_0$ and $r_1$, as in Figure~\ref{fig:proto-cmzs}.} This two-choice-bit ROT can be used to transform bit secret-sharing into zero-sharing as follows: Let $\Sd$ set $r = -r_0$ and $\Rcv$ set $u = r_{e_0 \oplus e_1}$, then if $e_0 \oplus e_1 = 0$, $r + u = 0$; otherwise $r + u = r_1 - r_0$ is uniform. The $\mathsf{hint}$ outputted to $\Sd$ is $r_1$. 


\subsection{Membership Zero-Sharing for Any Set Predicate Formulas}
Using the above instantiations for the two-party relaxed membership zero-sharing protocols, we present the complete protocol of batch membership zero-sharing for any set predicate formula $Q$ in Figure~\ref{fig:proto-cmzs}. 

\begin{theorem}\label{theorem:bmzs}
Protocol $\ProtobMZS^Q$ securely realizes $\FuncbMZS^Q$ against any semi-honest adversary corrupting $t < m$ parties in the $(\FuncbOPPRF,\FuncbssPMT, \FuncROT)$-hybrid model.    
\end{theorem}

The correctness and independence of membership zero-sharing are inherited from predicative zero-sharing, with a parameter adjustment for correctness: $\lvert \mathbb{F} \rvert \ge \lvert \mathsf{OR} \rvert \cdot n \cdot 2^\sigma$, where $\lvert \mathsf{OR} \rvert$ is the number of $\mathsf{OR}$ operators in $Q$. 

The privacy of membership zero-sharing is straightforward to verify: All interactions happen within the invocations of two relaxed batch membership zero-sharing protocols, which can be further decomposed into three blocking blocks --- batch OPPRF, batch ssPMT and ROT. Therefore, given the outputs from the ideal functionality, the simulator only needs to invoke the sub-simulators for these blocking blocks in a backward-chaining manner. As long as the batch OPPRF, batch ssPMT and ROT protocols are secure, the adversary's view is indistinguishable in ideal and real executions, thus meeting privacy definition.


\begin{figure}[!hbtp]
\begin{framed}
\begin{minipage}[center]{\textwidth}
\begin{trivlist}
\item \textbf{Parameters:} $m$ parties $P_1, \cdots P_m$, where $P_\mathsf{pivot}$ holds $n$ elements instead of $n$ sets. A set predicate formula $Q$ composed of $q \ge 1$ literals $Q_1,\cdots,Q_q$ where each $Q_i$ ($1 \le i \le q$) is in the form $x \in X_j$ or $x \notin X_j$ for some $j \in \{1,\cdots m\} \setminus \{\mathsf{pivot}\}$. $n$ Beaver triples $([\textbf{a}], [\textbf{b}], [\textbf{c}])$, where $[\textbf{a}] = ([a_1], \cdots, [a_n])$, $[\textbf{b}] = ([b_1], \cdots, [b_n])$, $[\textbf{c}] = ([c_1], \cdots, [c_n])$ and $c_i = a_i \cdot b_i$ for $1 \le i \le n$. Batch size $n$. A field $\mathbb{F}$. 
\item \textbf{Protocol:}
\begin{enumerate}[itemsep=2pt,topsep=0pt,parsep=0pt]
\item \textbf{The simple predicative sharing stage.} In this stage, for each $Q_i$, $P_\mathsf{pivot}$ and $P_j$ invoke the relaxed batch membership zero-sharing for $x \in X_j$ or $x \notin X_j$ (according to the form of $Q_i$) of size $n$, and the remaining parties set their secret shares to 0. As a result, the parties hold a vector of $n$ secret-sharings associated with $Q_i$. To be specific, if $Q_i$ is in the form of
    \begin{itemize}
        \item $x \in X_j$: For the $k$-th instance ($1 \le k \le n$), $P_j$ samples $r_{i,k}$ and sets $K_{i,k} = X_{j,k}$ and $V_{i,k} = \{-r_{i,k}, \cdots, -r_{i,k}\}$, where $\lvert K_{i,k} \rvert = \lvert V_{i,k} \rvert$. Then $P_\mathsf{pivot}$ and $P_j$ invoke $\FuncbOPPRF$ where $P_j$ acts as $\Sd$ with inputs $(K_{i,1},\cdots,K_{i,n})$ and $(V_{i,1},\cdots,V_{i,n})$, and $P_\mathsf{pivot}$ acts as $\Rcv$ with input $\textbf{x}$ and receives $\textbf{u}_{i}$. $P_\mathsf{pivot}$ sets its shares $\textbf{r}_{i,\mathsf{pivot}} = \textbf{u}_{i}$. $P_j$ sets its shares $\textbf{r}_{i,j} = (r_{i,1}, \cdots, r_{i,n})$. For each $d \in \{1,\cdots m\} \setminus \{\mathsf{pivot}, j\}$, $P_{d}$ sets its shares $\textbf{r}_{i,d} = \textbf{0}$. 
        \item $x \notin X_j$: $P_\mathsf{pivot}$ and $P_j$ invoke $\FuncbssPMT$, where in the $k$-th instance ($1 \le k \le n$), $P_j$ inputs $X_{j,k}$ and receives $e_{i,k}^0$, while $P_\mathsf{pivot}$ inputs $x_k$ and receives $e_{i,k}^1$. Then they invoke $n$ instances of ROT, where in the $k$-th instance ($1 \le k \le n$), $P_j$ acts as $\Sd$ and receives $r_{i,k}^0,r_{i,k}^1$, while $P_\mathsf{pivot}$ acts as $\Rcv$ with input $e_{i,k}^1$ and receives $r_{i,k}^{e_{i,k}^1}$. $P_\mathsf{pivot}$ sets its shares $\textbf{r}_{i,\mathsf{pivot}} = (r_{i,1}^{e_{i,1}^1},\cdots,r_{i,n}^{e_{i,n}^1})$. $P_j$ sets its shares $\textbf{r}_{i,j} = (-r_{i,1}^{e_{i,1}^0}, \cdots, -r_{i,n}^{e_{i,n}^0})$. For each $d \in \{1,\cdots m\} \setminus \{\mathsf{pivot}, j\}$, $P_{d}$ sets its shares $\textbf{r}_{i,d} = \textbf{0}$. 
    \end{itemize}
The vector of $n$ secret-sharings for $Q_i$ denotes as $[\textbf{r}_i] = (\textbf{r}_{i,1},\cdots,\textbf{r}_{i,m})$.
\item \textbf{The formula emulation stage.} If $q>1$, the parties collectively emulate the computation of $Q$ in the order of operator precedence, step by step. In each step, the parties generate a vector of $n$ secret-sharings associated with a binary clause connected by a given operator, based on the two vectors associated with the contained literals $Q'_i$ and $Q'_j$, which they obtain from previous steps. The actions of the parties depend on the type of operator being computed:
    \begin{itemize}
        \item $\mathsf{AND}$ operator: Suppose the parties hold two vectors of $n$ secret-sharings $[\textbf{r}_i]$ and $[\textbf{r}_j]$ associated with $Q'_i$ and $Q'_j$ respectively. They want to compute $n$ relaxed membership zero-sharings of $Q' = Q'_i \land Q'_j$. Then they locally add the corresponding components of two vectors to obtain $[\textbf{r}_i + \textbf{r}_j]$. 
        \item $\mathsf{OR}$ operator: Suppose the parties hold two vectors of $n$ secret-sharings $[\textbf{r}_i]$ and $[\textbf{r}_j]$ associated with $Q'_i$ and $Q'_j$ respectively. They want to compute $n$ relaxed membership zero-sharings of $Q'= Q'_i \lor Q'_j$. Then the parties perform $n$ secure multiplications between the corresponding components of two vectors, i.e., $[\textbf{r}_i \cdot \textbf{r}_j] = [\textbf{r}_i] \cdot [\textbf{r}_j]$.
    \end{itemize}
After obtaining the vector of secret-sharings associated with $Q'$, the parties regard $Q'$ as a new literal, and repeat the above steps until there is only one literal in $Q$. The vector $[\textbf{r}]$ associated with the ultimate literal held by the parties is the vector of $n$ relaxed membership zero-sharings for $Q$.
\item \textbf{Transformation from relaxed to standard.} All parties compute $[\textbf{s}]$ by performing $[\textbf{s}] = [\textbf{r}] \cdot [\textbf{b}]$, using $n$ Beaver triples $([\textbf{a}], [\textbf{b}], [\textbf{c}])$ (c.f. Appendix~\ref{beaver triples}).
\end{enumerate}
\end{trivlist}
\end{minipage}
\end{framed}
\caption{Batch Membership Zero-Sharing $\ProtobMZS^Q$}\label{fig:proto-cmzs}
\end{figure}

\section{Our MPSO Framework}

\subsection{Overview}\label{sec:overview}
Our framework is based on the CPF representation $\psi(x, X_1, \cdots, X_m)$ of any constructible set $Y$. Recall that $\psi = Q_1 \lor \cdots \lor Q_s$. Assuming that $Q_i$ ($1 \le i \le s$) only contains atomic propositions relevant to $X_{i_1},\cdots,X_{i_q}$ ($1 \le q \le m$), $Q_i$ is set-separable with respect to $X_j$ for some $j \in \{{i_1},\cdots,{i_q}\}$. Namely, $$Q_i(x,X_{i_1},\cdots,X_{i_q}) = (x \in X_j) \land Q'_i(x,X_{i_1},\cdots,X_{j-1},X_{j+1},\cdots,X_{i_q}).$$ where $Q'_i$ is the separation formula of $Q_i$. We use $Y_i$ to denote the set represented by $Q_i$. Our high-level idea is that for each $Q_i$, we designate $P_j$ as $P_{\mathsf{pivot}}$ and let $P_{i_1},\cdots,P_{i_q}$ engage in the batch membership zero-sharing protocol for $Q'_i$, so that for each $x \in X_j$, if $Q'_i = 1$, the parties receive secret shares of 0, otherwise secret shares of a random value. $P_j$ adds $x$ to its associated shares. As a result, for each $x \in X_j$, if $Q_i = 1$, i.e., $Q'_i = 1$, the parties receive secret shares of $x$, otherwise secret shares of a random value. The process is elaborated below.

For each $Q_i$, $P_j$ uses hash functions $h_1, h_2, h_3$ to assign elements to $B = O(n)$ bins via Cuckoo hashing, so that each bin $\mathcal{C}_j^b$ ($1 \le b \le B$) has at most one item. Meanwhile, each $P_{j'}$ ($j' \in \{{i_1},\cdots,{i_q}\}\setminus \{j\}$) assigns each element $y \in X_{j'}$ to bins $\mathcal{T}_{j'}^{h_1(y)}, \mathcal{T}_{j'}^{h_2(y)}, \mathcal{T}_{j'}^{h_3(y)}$. Note that if $P_j$ maps the item $x \in X_j$ into $\mathcal{C}_j^b$, then $b \in \{h_1(x),h_2(x),h_3(x)\}$. If $P_{j'}$ also holds $x$, it must map $x$ into $\mathcal{T}_{j'}^b$. This enables the remaining parties to align their input sets with respect to $X_j$, such that for each $x$ in $\mathcal{C}_j^b$, $x \in X_j'$ if and only if $x $ is in $\mathcal{T}_{j'}^b$. Thereby, we derive that $$Q'_i(x,X_{i_1},\cdots,X_{j-1},X_{j+1},\cdots,X_{i_q}) = Q'_i(x,\mathcal{T}_{i_1}^b,\cdots,\mathcal{T}_{j-1}^b,\mathcal{T}_{j+1}^b,\cdots,\mathcal{T}_{i_q}^b).$$
$P_{i_1},\cdots,P_{i_q}$ engage in the batch membership zero-sharing for $Q'_i$, where $P_j$ acts as $P_{\mathsf{pivot}}$ with inputs $\mathcal{C}_j^1, \cdots, \mathcal{C}_j^B$ and each $P_{j'}$ inputs $\mathcal{T}_{j'}^1, \cdots, \mathcal{T}_{j'}^B$. In the end, they receive $B$ secret-sharings, so that if the element $x$ in each bin $C_j^b$ satisfies $Q_i(x,X_{i_1},\cdots,X_{i_q}) = 1$, i.e. $x \in Y_i$, the parties receive secret shares of 0, otherwise secret shares of a random value. $P_j$ adds $x$ appended with an all-zero string (for the distinction between elements and random values) to the $b$-th secret share (This last element addition step actually depends on the functionality, see below), so that if $x \in Y_i$, the parties hold a secret-sharing of $x$.

Given that $\{Y_1, \cdots, Y_s\}$ form a partition of Y, if the parties execute the above process with the last element addition step for all $Q_1, \cdots, Q_s$, they will hold secret-sharings of all elements in $Y$ (the secret-sharing of each element appears once, interspersed by random secret-sharings for elements not in $Y$ and duplicate elements), arranged in the primary order of $Y_1, \cdots, Y_s$. In each $Y_i$, secret-sharings are arranged in the secondary order of $P_j$'s Cuckoo hash positions, which depends on the whole set $X_j$. On the contrary, if the parties execute the above process without the last step, they hold secret-sharings of 0 instead of elements in the same positions. The decision to execute the last step and the subsequent process are determined by the target functionality, which is divided into three categories:

\begin{enumerate}

    \item \textbf{MPSO.} In this functionality, the parties must reconstruct the elements in $Y$ to $P_1$, thus they have to execute the last element addition step to secret-share elements. However, a straightforward reconstruction of secret-sharings leads to two types of information leakage: 1) The primary order of the reconstructed elements reveals the subset $Y_i$ which each element belongs to. 2) The secondary order of the reconstructed elements reveals the information of $X_j$. The solution is to let all parties invoke the multi-party secret-shared shuffle to randomly permute and re-share secret-sharings before reconstruction.

    \item \textbf{MPSO-card.} In this functionality, the parties must reconstruct the secrets without revealing the actual elements but only the cardinality. To achieve this, the parties skip the last element addition step, so that for each element in $Y$, the parties hold secret-sharings of 0, and for elements not in $Y$ or repeated elements, the parties hold random secret-sharings. These secret-sharings are arranged in a specific sequence, and straightforward reconstruction would cause similar leakage as previous, thus the parties need to invoke the multi-party secret-shared shuffle as well. Afterwards, they reconstruct secrets to the leader, who counts the number of 0s as the cardinality of $Y$.

    \item \textbf{Circuit-MPSO.} There are two ways to realize this functionality.

    \begin{itemize}
      \item \textbf{Approach 1:} The parties skip the last element addition step for all subformulas. They feed secret-sharings along with the elements in indicated Cuckoo hashing bins into generic MPC in order, which implements a circuit identifying 0s from random values, collecting elements in the corresponding positions as $Y$, and computing arbitrary function on $Y$.
      \item \textbf{Approach 2:} The parties execute the last element addition step for all  subformulas. They feed secret-sharings into generic MPC, which implements a circuit first distinguishing elements from random values (by the appended 0s) to identify $Y$, then computing arbitrary function on $Y$.
    \end{itemize}


\end{enumerate}

In the following sections, we progressively introduce our framework in detail. Specifically, we start by constructing the simplest cases --- MPSI/MPSI-card/circuit-MPSI, which are on behalf of a special case where $\psi(x, X_1, \cdots, X_m)$ is a disjunction of one subformula that is set-separable with respect to $X_1$, in Section~\ref{sec:mpsi}. The protocols in this setting can bypass the invocation of multi-party secret-shared shuffle. In addition, we propose an MPSI-card-sum protocol as a variant.
Next, we discuss another special case where $Y$ is represented as the disjunction of several subformulas. We construct MPSU/MPSU-card/circuit-MPSU protocols as illustrations in Section~\ref{sec:mpsu}. Finally, in Section~\ref{sec:MPSO}, we present the complete MPSO/MPSO-card/circuit-MPSO protocols.

\subsection{MPSI, MPSI-card and Circuit MPSI}\label{sec:mpsi}

Consider a constructible set $Y$, which can be represented as a set-separable formula $Q(x,X_1,\cdots,X_m)$ with respect to $X_1$, such as $X_1 \cap X_2 \cap X_3$ can be represented as $(x \in X_1) \land (x \in X_2) \land (x \in X_3)$ and $X_1 \setminus (X_2 \cap X_3)$ can be represented as $(x \in X_1) \land \neg ((x \in X_2) \land (x \in X_3)) = (x \in X_1) \land ((x \notin X_2) \lor (x \in X_3))$. Let $Q'(x,X_2,\cdots,X_m)$ be the separation formula of $Q$ with respect to $X_1$. 

In this case, all elements in $Y$ belong to $P_1$, and the order of secret-sharings is totally determined by $X_1$, so the two types of ``information leakage'' associated with the specific sequence of secret-sharings no longer constitute actual information leakage for $P_1$. Therefore, after $P_1$ invokes batch membership zero-sharing with the other parties, acting as $P_{\mathsf{pivot}}$, to realize MPSO, the parties straightforwardly reconstruct $B$ secret-sharings to $P_1$. For each $1 \le b \le B$, $P_1$ checks whether the $b$-th secret is 0. If so, the element in the $b$-th bin is in $Y$. While in MPSO-card, the parties still need to invoke a multi-party secret-shared shuffle protocol before reconstruction to shuffle the correspondences between elements and secret-sharings, preventing $P_1$ from learning the exact elements in $Y$. 

Another and the most important benefit in this setting is that the costs of the protocols do not scale with the input length of set elements, as long as the parties pre-hash their elements into shorter strings. For correctness, we must ensure that the hashing introduces no collisions among $P_1$ and the other parties' input elements, so the  hash function's output length is at least $\sigma + \log_2 (m-1) + 2 \log_2 n$.

\begin{figure}[!hbtp]
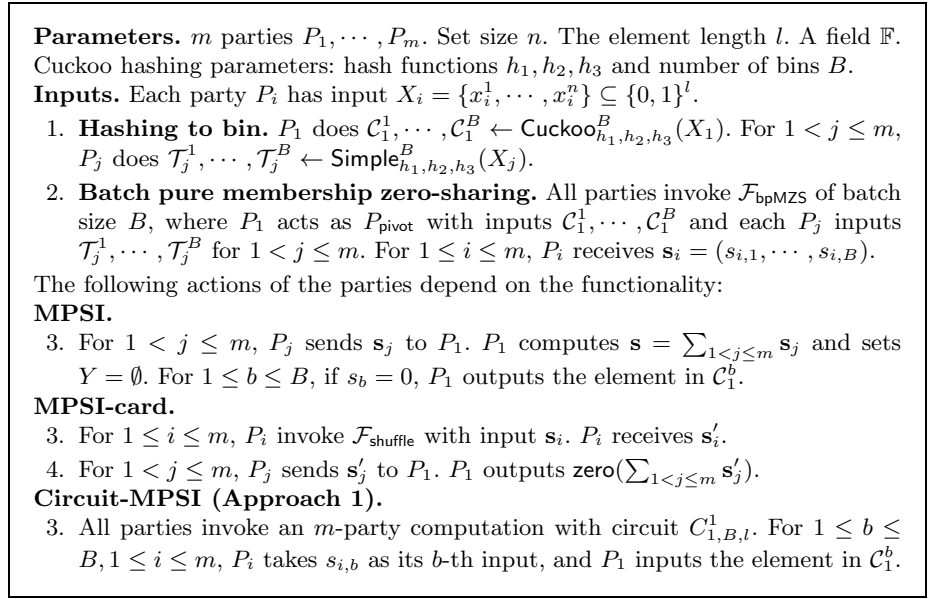

\begin{framed}
\begin{minipage}[center]{\textwidth}
\textbf{Parameters.} $m$ parties $P_1, \cdots, P_m$. Set size $n$. The element length $l$. A field $\mathbb{F}$. Cuckoo hashing parameters: hash functions $h_1, h_2, h_3$ and number of bins $B$.

\textbf{Inputs.} Each party $P_i$ has input $X_i = \{x_i^1,\cdots, x_i^n\} \subseteq \{0,1\}^l$.

\begin{enumerate}[itemsep=2pt,topsep=0pt,parsep=0pt]
\item \textbf{Hashing to bin.} $P_1$ does $\mathcal{C}_1^1, \cdots, \mathcal{C}_1^B \gets \Cuckoo_{h_1,h_2,h_3}^B(X_1)$. For $1 < j  \le m$, $P_j$ does $\mathcal{T}_j^1, \cdots, \mathcal{T}_j^B \gets \Simple_{h_1,h_2,h_3}^B(X_j)$.
\item \textbf{Batch pure membership zero-sharing.} All parties invoke $\FuncbpMZS$ of batch size $B$, where $P_1$ acts as $P_\mathsf{pivot}$ with inputs $\mathcal{C}_1^1, \cdots, \mathcal{C}_1^B$ and each $P_j$ inputs $\mathcal{T}_j^1, \cdots, \mathcal{T}_j^B$ for $1 < j  \le m$. For $1 \le i \le m$, $P_i$ receives $\textbf{s}_i=(s_{i,1},\cdots,s_{i,B})$.
\end{enumerate}

The following actions of the parties depend on the functionality:

    \textbf{MPSI.} 
    \begin{enumerate}[itemsep=2pt,topsep=0pt,parsep=0pt] \addtocounter{enumi}{2}
        \item For $1 < j  \le m$, $P_j$ sends $\textbf{s}_j$ to $P_1$. $P_1$ computes $\textbf{s} = \sum_{1 < j  \le m} \textbf{s}_j$ and sets $Y = \emptyset$. For $1 \le b \le B$, if $s_b = 0$, $P_1$ outputs the element in $\mathcal{C}_1^b$.
    \end{enumerate}
    \textbf{MPSI-card.} 
    \begin{enumerate}[itemsep=2pt,topsep=0pt,parsep=0pt] \addtocounter{enumi}{2}
        \item For $1 \le i \le m$, $P_i$ invoke $\FuncMS$ with input $\textbf{s}_i$. $P_i$ receives $\textbf{s}_{i}'$.
        \item For $1 < j  \le m$, $P_j$ sends $\textbf{s}_j'$ to $P_1$. $P_1$ outputs $\Zero(\sum_{1 < j  \le m} \textbf{s}_j')$.
    \end{enumerate}
    \textbf{Circuit-MPSI (Approach 1).}
    \begin{enumerate}[itemsep=2pt,topsep=0pt,parsep=0pt] \addtocounter{enumi}{2}
        \item All parties invoke an $m$-party computation with circuit $C_{1,B,l}^1$. For $1 \le b \le B, 1 \le i \le m$, $P_i$ takes $s_{i,b}$ as its $b$-th input, and $P_1$ inputs the element in $\mathcal{C}_1^b$.
    \end{enumerate}
\end{minipage}
\end{framed}
\caption{MPSI/MPSI-card/Circuit-MPSI}\label{fig:proto-mpsi}
\end{figure}

\begin{figure}[!hbtp]
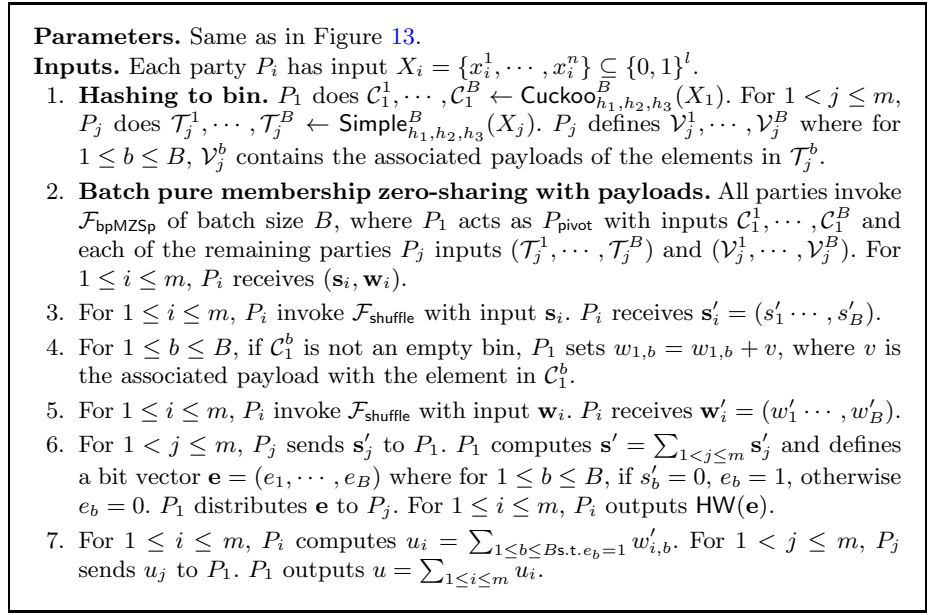

\begin{framed}
\begin{minipage}[center]{\textwidth}
\begin{trivlist}
\item \textbf{Parameters.} Same as in Figure~\ref{fig:proto-mpsi}.
\item \textbf{Inputs.} Each party $P_i$ has input $X_i = \{x_i^1,\cdots, x_i^n\} \subseteq \{0,1\}^l$.
\begin{enumerate}[itemsep=2pt,topsep=0pt,parsep=0pt]
    \item \textbf{Hashing to bin.} $P_1$ does $\mathcal{C}_1^1, \cdots, \mathcal{C}_1^B \gets \Cuckoo_{h_1,h_2,h_3}^B(X_1)$. For $1 < j \le m$, $P_j$ does $\mathcal{T}_j^1, \cdots, \mathcal{T}_j^B \gets \Simple_{h_1,h_2,h_3}^B(X_j)$. $P_j$ defines $\mathcal{V}_j^1, \cdots, \mathcal{V}_j^B$ where for $1 \le b \le B$, $\mathcal{V}_j^b$ contains the associated payloads of the elements in $\mathcal{T}_j^b$.
    \item \textbf{Batch pure membership zero-sharing with payloads.} All parties invoke $\FuncbpMZSp$ of batch size $B$, where $P_1$ acts as $P_\mathsf{pivot}$ with inputs $\mathcal{C}_1^1, \cdots, \mathcal{C}_1^B$ and each of the remaining parties $P_j$ inputs $(\mathcal{T}_j^1, \cdots, \mathcal{T}_j^B)$ and $(\mathcal{V}_j^1, \cdots, \mathcal{V}_j^B)$. For $1 \le i \le m$, $P_i$ receives $(\textbf{s}_i, \textbf{w}_i)$.
    \item For $1 \le i \le m$, $P_i$ invoke $\FuncMS$ with input $\textbf{s}_i$. $P_i$ receives $\textbf{s}_{i}' = (s_1' \cdots, s_B')$.
    \item For $1 \le b \le B$, if $\mathcal{C}_1^b$ is not an empty bin, $P_1$ sets $w_{1,b} = w_{1,b} + v$, where $v$ is the associated payload with the element in $\mathcal{C}_1^b$.
    \item For $1 \le i \le m$, $P_i$ invoke $\FuncMS$ with input $\textbf{w}_i$. $P_i$ receives $\textbf{w}_{i}' = (w_1' \cdots, w_B')$.
    \item For $1 < j  \le m$, $P_j$ sends $\textbf{s}_j'$ to $P_1$. $P_1$ computes $\textbf{s}' = \sum_{1 < j  \le m} \textbf{s}_j'$ and defines a bit vector $\textbf{e} = (e_1, \cdots, e_B)$ where for $1 \le b \le B$, if $s_b' = 0$, $e_b = 1$, otherwise $e_b = 0$. $P_1$ distributes $\textbf{e}$ to $P_j$. For $1 \le i  \le m$, $P_i$ outputs $\HW(\textbf{e})$.
    \item For $1 \le i  \le m$, $P_i$ computes $u_i = \sum_{1 \le b \le B \mathsf{ s.t. } e_b = 1} w_{i,b}'$. For $1 < j  \le m$, $P_j$ sends $u_j$ to $P_1$. $P_1$ outputs $u = \sum_{1 \le i  \le m} u_i$.
\end{enumerate}

\end{trivlist}
\end{minipage}
\end{framed}
\caption{MPSI-card-sum}\label{fig:proto-mpsi-sum}
\end{figure}

The most commonly used protocols in this case are MPSI, MPSI-card and circuit MPSI. Let $C_{s,B,l}^1$ be a circuit that has $sB(m\log_2\lvert \mathbb{F} \rvert+ l)$ input wires, divided to $s$ sections of $B(m\log_2\lvert \mathbb{F} \rvert + l)$ inputs wires each. In the $i$-th section ($1 \le i \le s$), the $k$-th group of $B$ inputs on $\mathbb{F}$ is associated with $P_k$ for $1 \le k \le m$, and we denote the $b$-th input in this group ($1 \le b \le B$) as $u_{i,k,b} \in \mathbb{F}$; The last $B$ $l$-length inputs are associated with $P_j$ for certain $1 \le j \le m$, where we denote the $b$-th input ($1 \le b \le B$) as $z_{i,b} \in \{0,1\}^l$. The circuit first consists a subcircuit producing a bit $w_{i,b} = 1$ if $u_{i,1,b} + \cdots + u_{i,m,b} = 0$ and 0 otherwise for $1 \le i \le s, 1 \le b \le B$. Then, the circuit computes and outputs $f(Z)$ where $Z = \{z_{i,b} \vert w_{i,b} = 1\}_{1 \le i \le s, 1 \le b \le B}$ and $f$ is the function to be computed on the constructible set $Y$.
The complete MPSI, MPSI-card and circuit MPSI protocols are described in Figure~\ref{fig:proto-mpsi}.
Additionally, The MPSI-card-sum protocol based on pure membership zero-sharing with payloads is outlined in Figure~\ref{fig:proto-mpsi-sum}.

\subsection{MPSU, MPSU-card and Circuit MPSU}\label{sec:mpsu}
Consider a constructible set $Y$, whose CPF representation $\psi(x,X_1,\cdots,X_m)$ is a disjunction of several subformulas, one is an atomic proposition $x \in X_i$ for some $1 \le i \le m$. For instance, $X_1 \cup \cdots \cup X_m$ can be represented as $(x \in X_1) \lor ((x \notin X_1) \land (x \in X_2)) \lor \cdots \lor ((x \notin X_1) \land \cdots \land (x \notin X_{m-1}) \land (x \in X_m))$. In this case, the subformula $x \in X_i$ only involves $P_i$, so $P_i$ simply shares its elements among the parties. Especially, if $i = 1$, then the subformula can be ignored, as long as $P_1$ finally appends its elements to the reconstructed elements to obtain $Y_1$.

The most commonly used protocols in this case are MPSU, MPSU-card and circuit MPSU. Let $C_{N,l'}^2$ be a circuit with $m$ groups of $N$ inputs on $\mathbb{F}$. The $k$-th group is associated with $P_k$ $(1 \le k \le m)$, where the $i$-th inputs is denoted by $z_{k,i}$ $(1 \le i \le N)$. The circuit computes and outputs $f(Z)$ where $Z = \{z_i \vert z_{1,i} + \cdots + z_{m,i} = z_i \Vert 0^{l'}\}_{1 \le i \le N}$ and $f$ is the function to be computed on $Y$. The complete MPSU, MPSU-card and circuit MPSU protocols are described in Figure~\ref{fig:proto-mpsu}.

\begin{figure}[!hbtp]
\begin{framed}
\begin{minipage}[center]{\textwidth}
\begin{trivlist}
\item \textbf{Parameters.} $m$ parties $P_1, \cdots, P_m$. Set size $n$ . The element length $l$. The all-zero string length $l'$. A field $\mathbb{F}$. An encoding function $\mathsf{code}:\mathbb{F} \to \{0,1\}^{l+l'}$. Cuckoo hashing parameters: hash functions $h_1, h_2, h_3$ and number of bins $B$.
\item \textbf{Inputs.} Each party $P_i$ has input $X_i = \{x_i^1,\cdots, x_i^n\} \subseteq \{0,1\}^l$.
\begin{enumerate}[itemsep=2pt,topsep=0pt,parsep=0pt]
\item \textbf{Hashing to bin.} $P_1$ does $\mathcal{T}_1^1, \cdots, \mathcal{T}_1^B \gets \Simple_{h_1,h_2,h_3}^B(X_1)$. For $1 < j \le m $, $P_j$ does $\mathcal{C}_j^1, \cdots, \mathcal{C}_j^B \gets \Cuckoo_{h_1,h_2,h_3}^B(X_j)$ and $\mathcal{T}_j^1, \cdots, \mathcal{T}_j^B \gets \Simple_{h_1,h_2,h_3}^B(X_j)$.
\item \textbf{Batch pure membership zero-sharing.} For $1 < j \le m$, $P_1, \cdots, P_j$ invoke $\FuncbpNMZS$ of batch size $B$, where $P_j$ acts as $P_\mathsf{pivot}$ with inputs $\mathcal{C}_j^1, \cdots, \mathcal{C}_j^B$ and each $P_{j'}$ inputs $\mathcal{T}_{j'}^1, \cdots, \mathcal{T}_{j'}^B$ for $j' \in \{{i_1},\cdots,{i_q}\} \setminus \{j\}$. For $1 \le i \le j$, $P_i$ receives $\textbf{s}_{j,i}=(s_{j,i,1},\cdots,s_{j,i,B})$.
\end{enumerate}

The following actions of the parties depend on the functionality:

    \item \textbf{MPSU.} 
    \begin{enumerate}[itemsep=2pt,topsep=0pt,parsep=0pt] \addtocounter{enumi}{2}
        \item For $1 < j \le m$, $1 \le b \le B$, if $\mathcal{C}_j^b$ is not an empty bin, $P_j$ computes $s_{j,j,b} = \mathsf{code}(s_{j,j,b}) \oplus (x \Vert 0^{l'})$, where $x$ is the element in $\mathcal{C}_j^b$, otherwise $P_j$ samples $s_{j,j,b} \gets \{0,1\}^{l+l'}$.
        \item For $1 \le i \le m$, $P_i$ computes $\textbf{u}_i \in (\{0,1\}^{l+l'})^{(m-1)B}$ as follows: For $max(2,i) \le j \le m, 1 \le b \le B$, $u_{i,(j-2)B+b} = s_{j,i,b}$. Set other positions to $0$.
        \item For $1 \le i \le m$, $P_i$ invoke $\FuncMS$ with input $\textbf{u}_i$. $P_i$ receives $\textbf{u}_{i}'$.
        \item For $1 < j \le m$, $P_j$ sends $\textbf{u}_j'$ to $P_1$. $P_1$ computes $\textbf{u}' = \sum_{1 < j \le m} \textbf{u}_j'$ and sets $Y = \emptyset$. For $1 \le b \le B$, if $u_b' = y \Vert 0^{l'}$ for some $y \in \{0,1\}^l$, $P_1$ outputs $y$.
    \end{enumerate}
    \item \textbf{MPSU-card.} 
    \begin{enumerate}[itemsep=2pt,topsep=0pt,parsep=0pt] \addtocounter{enumi}{2}
        \item For $1 < j \le m$, $1 \le b \le B$, $P_j$ chooses $s_{j,j,b}$ at random if $\mathcal{C}_j^b$ is an empty bin.
        \item For $1 \le i \le m$, $P_i$ computes $\textbf{u}_i \in \mathbb{F}^{(m-1)B}$ as follows: For $max(2,i) \le j \le m, 1 \le b \le B$, $u_{i,(j-2)B+b} = s_{j,i,b}$. Set other positions to $0$.
        \item For $1 \le i \le m$, $P_i$ invoke $\FuncMS$ with input $\textbf{u}_i$. $P_i$ receives $\textbf{u}_{i}'$.
        \item For $1 < j \le m$, $P_j$ sends $\textbf{u}_j'$ to $P_1$. $P_1$ outputs $\Zero(\sum_{1 < j \le m} \textbf{u}_j')$.
    \end{enumerate}
    \item \textbf{Circuit-MPSU (Approach 2).}
    \begin{enumerate}[itemsep=2pt,topsep=0pt,parsep=0pt] \addtocounter{enumi}{2}
        \item For $1 < j \le m$, $1 \le b \le B$, if $\mathcal{C}_j^b$ is not an empty bin, $P_j$ computes $s_{j,j,b} = \mathsf{code}(s_{j,j,b}) \oplus (x \Vert 0^{l'})$, where $x$ is the element in $\mathcal{C}_j^b$, otherwise $P_j$ samples $s_{j,j,b} \gets \{0,1\}^{l+l'}$.
        \item For $1 \le i \le m$, $P_i$ computes $\textbf{u}_i \in (\{0,1\}^{l+l'})^{(m-1)B}$ as follows: For $max(2,i) \le j \le m, 1 \le b \le B$, $u_{i,(j-2)B+b} = s_{j,i,b}$. Set other positions to $0$.
        \item All parties invoke an $m$-party computation with the circuit $C_{(m-1)B,l'}^2$. For $1 \le i \le m, 1 \le k \le (m-1)B$, $P_i$ inputs $u_{i,c}$ to the circuit.
    \end{enumerate}
\end{trivlist}
\end{minipage}
\end{framed}
\caption{MPSU/MPSU-card/Circuit-MPSU}\label{fig:proto-mpsu}
\end{figure}

\subsection{MPSO, MPSO-card and Circuit MPSO}\label{sec:MPSO} 

Following the overview of our framework, we formally present the MPSO, MPSO-card, and circuit-MPSO protocols for any constructible set in Figure~\ref{fig:proto-MPSO}.

\begin{figure}[!hbtp]
\begin{framed}
\begin{minipage}[center]{\textwidth}
\begin{trivlist}
\item \textbf{Parameters.} $m$ parties $P_1, \cdots, P_m$. Set size $n$ . The element length $l$. The all-zero string length $l'$. A field $\mathbb{F}$. An encoding function $\mathsf{code}:\mathbb{F} \to \{0,1\}^{l+l'}$. A constructible set $Y$ represented as a CPF representation $\psi(x,X_1,\cdots,X_m)$. Cuckoo hashing parameters: hash functions $h_1, h_2, h_3$ and number of bins $B$.

\item \textbf{Inputs.} Each party $P_i$ has input $X_i = \{x_i^1,\cdots, x_i^n\} \subseteq \{0,1\}^l$.
\begin{enumerate}[itemsep=2pt,topsep=0pt,parsep=0pt]
\item \textbf{Hashing to bin.} For $1 \le i \le m $, $P_i$ does $\mathcal{C}_i^1, \cdots, \mathcal{C}_i^B \gets \Cuckoo_{h_1,h_2,h_3}^B(X_i)$ and $\mathcal{T}_i^1, \cdots, \mathcal{T}_i^B \gets \Simple_{h_1,h_2,h_3}^B(X_i)$.
\item \textbf{Single  subformula evaluation.} \label{step2} Let $\psi = Q_1 \lor \cdots \lor Q_s$. For the $i$-th subformula $Q_i(x,X_{i_1},\cdots,X_{i_q})$ in $\psi$, where $1 \le i \le s, \{{i_1},\cdots,{i_q}\} \subseteq \{1,\cdots,m\}$, 
\begin{enumerate}
    \item If $q = 1$, suppose $i_1 = \cdots = i_q = j$, then $Q_i(x,X_j) = (x \in X_j)$. For $1 \le b \le B$, if $\mathcal{C}_j^b$ is not an empty bin, $P_j$ sets $s_{i,j,b} = 0$, otherwise $P_j$ chooses $s_{i,j,b}$ at random. For $j' \in \{1,\cdots,m\} \setminus \{j\}$, $P_{j'}$ sets $s_{i,j',b} = 0$.
    \item If $q > 1$, suppose $Q_i$ is set-separable with respect to $X_j$ for some $j \in \{{i_1},\cdots,{i_q}\}$ and $Q_i(x,X_{i_1},\cdots,X_{i_q}) = (x \in X_j) \land Q'_i(x,X_{i_1},\cdots,X_{j-1},X_{j+1},\cdots,X_{i_q})$. The parties invoke $\FuncbMZS^{Q'_i}$ where $P_j$ acts as $P_{\mathsf{pivot}}$ with inputs $\mathcal{C}_j^1, \cdots, \mathcal{C}_j^B$ and each $P_{j'}$ inputs $\mathcal{T}_{j'}^1, \cdots, \mathcal{T}_{j'}^B$ for $j' \in \{{i_1},\cdots,{i_q}\} \setminus \{j\}$. For $1 \le b \le B$, each $P_{i'}$ receives $\textbf{s}_{i,i'}=(s_{i,i',1},\cdots,s_{i,i',B})$ for $i' \in \{{i_1},\cdots,{i_q}\}$, and each $P_{k}$ sets $s_{i,k,b} = 0$ for $k \in \{1,\cdots,m\} \setminus \{{i_1},\cdots,{i_q}\}$.
\end{enumerate}
\end{enumerate}

The following actions of the parties depend on the functionality:

    \item \textbf{MPSO.} 
    \begin{enumerate}[itemsep=2pt,topsep=0pt,parsep=0pt] \addtocounter{enumi}{2}
        \item For $1 < i \le s$, $1 \le b \le B$, if $\mathcal{C}_j^b$ is not an empty bin, $P_j$ (the same $j$ as step~\ref{step2}) computes $s'_{i,j,b} = \mathsf{code}(s_{i,j,b}) \oplus (x \Vert 0^{l'})$, where $x$ is the element in $\mathcal{C}_j^b$, otherwise $P_j$ samples $s'_{i,j,b} \gets \{0,1\}^{l+l'}$.
        \item For $1 \le k \le m$, each $P_k$ computes $\textbf{u}_k \in (\{0,1\}^{l+l'})^{sB}$ as follows: For $1 \le i \le s, 1 \le b \le B$, $u_{k,(i-1)B+b} = s'_{i,k,b}$.
        \item For $1 \le k \le m$, $P_k$ invoke $\FuncMS$ with input $\textbf{u}_k$. $P_k$ receives $\textbf{u}_{k}'$.
        \item For $1 < j \le m$, $P_j$ sends $\textbf{u}_j'$ to $P_1$. $P_1$ computes $\textbf{u}' = \sum_{1 < j \le m} \textbf{u}_j'$ and sets $Y = \emptyset$. For $1 \le b \le B$, if $u_b' = y \Vert 0^{l'}$ for some $y \in \{0,1\}^l$, $P_1$ outputs $y$.
    \end{enumerate}
    \item \textbf{MPSO-card.} 
    \begin{enumerate}[itemsep=2pt,topsep=0pt,parsep=0pt] \addtocounter{enumi}{2}
        \item For $1 \le k \le m$, $P_k$ computes $\textbf{u}_k \in \mathbb{F}^{sB}$ as follows: For $1 \le i \le s, 1 \le b \le B$, $u_{k,(i-1)B+b} = s_{i,k,b}$.
        \item For $1 \le k \le m$, $P_k$ invoke $\FuncMS$ with input $\textbf{u}_k$. $P_k$ receives $\textbf{u}_{k}'$.
        \item For $1 < j \le m$, $P_j$ sends $\textbf{u}_j'$ to $P_1$. $P_1$ outputs $\Zero(\sum_{1 < j \le m} \textbf{u}_j')$.
    \end{enumerate}
    \item \textbf{Circuit-MPSO (Approach 1).}
    \begin{enumerate}[itemsep=2pt,topsep=0pt,parsep=0pt] \addtocounter{enumi}{2}
        \item All parties invoke an $m$-party computation with the circuit $C_{s,B,l}^1$. For $1 \le i \le s, 1 \le k \le m$, $P_k$ inputs $s_{i,k,1},\cdots,s_{i,k,B}$ to the $i$-th section, and $P_j$ (the same $j$ as step~\ref{step2}) inputs the elements in $\mathcal{C}_j^1,\cdots,\mathcal{C}_j^B$ in addition.
    \end{enumerate}
    \item \textbf{Circuit-MPSO (Approach 2).}
    \begin{enumerate}[itemsep=2pt,topsep=0pt,parsep=0pt] \addtocounter{enumi}{2}
        \item For $1 < i \le s$, $1 \le b \le B$, if $\mathcal{C}_j^b$ is not an empty bin, $P_j$ (the same $j$ as step~\ref{step2}) computes $s'_{i,j,b} = \mathsf{code}(s_{i,j,b}) \oplus (x \Vert 0^{l'})$, where $x$ is the element in $\mathcal{C}_j^b$, otherwise $P_j$ samples $s'_{i,j,b} \gets \{0,1\}^{l+l'}$.
        \item All parties invoke an $m$-party computation with the circuit $C_{sB,l'}^2$. For $1 \le i \le s, 1 \le b \le B, 1 \le k \le m$, $P_k$ takes $s'_{i,k,b}$ as its $((i-1)B+b)$-th input.
    \end{enumerate}
\end{trivlist}
\end{minipage}
\end{framed}
\caption{MPSO/MPSO-card/Circuit-MPSO}\label{fig:proto-MPSO}
\end{figure}

\begin{theorem}\label{theorem:mpso}
The MPSO, MPSO-card and circuit-MPSO protocols in Figure~\ref{fig:proto-MPSO} are secure against any semi-honest adversary corrupting $t < m$ parties in the $(\FuncbMZS^Q, \FuncMS)$-hybrid model.    
\end{theorem}

\begin{trivlist}
\item \textbf{Correctness.} The correctness of these protocols comes from the existence and qualities of CPF representations in Theorem~\ref{theorem:set2} and Definition~\ref{def:cpf}, and the correctness of batch membership zero-sharing. To ensure the correctness of all batch membership zero-sharing protocols, the field size must satisfy $\lvert \mathbb{F} \rvert \ge \lvert \mathsf{OR} \rvert \cdot B \cdot 2^\sigma$, where $\lvert \mathsf{OR} \rvert$ is the total number of $\mathsf{OR}$ operators in all $Q_i$ for $1 \le i \le s$.

Let $Y_i$ denote the set represented by $Q_i$. 
In the MPSO and circuit-MPSO (Approach 2) protocols, the parties hold $\lvert Y_i \rvert$ secret-sharings of the elements in $Y_i$, and $B-\lvert Y_i \rvert$ secret-sharings of random values after each batch membership zero-sharing for $Q_i$, for $1 \le i \le s$. Given that $\{Y_1, \cdots, Y_s\}$ is a partition of $Y$, the parties hold $\lvert Y \rvert$ secret-sharings of the elements in $Y$, and $sB-\lvert Y \rvert$ secret-sharings of random values in total. Finally, $P_1$ and the circuit identify all set elements by checking whether the last $l'$ bits are all zero. An error occurs when a random value collides with $0^{l'}$. Thereby, the overall false positive error probability is at most $sB \cdot 2^{-l'}$. To make this failure probability negligible, we set $l' \ge \sigma + \log s +\log B$; 
In the MPSO-card and circuit-MPSO (Approach 1) protocols, the parties hold $\lvert Y \rvert$ secret-sharings of 0, and $B-\lvert Y \rvert$ secret-sharings of random values. To bound the overall false positive error probability by $2^{-\sigma}$, we set $\lvert \mathbb{F} \rvert \ge sB \cdot 2^\sigma$.

\item \textbf{Security.}  This security proof is deferred to Appendix~\ref{appdix:proof-MPSO}.
\end{trivlist}

\begin{trivlist}
\item \textbf{Complexity Analysis.} The computation and communication complexity are both  dominated by the form of the CPF representation $\psi(x, X_1, \cdots, X_m)$ of the constructible set $Y$ being computed, where $\psi = Q_1 \lor \cdots \lor Q_s$. 

In the subformula evaluation stage, the computation complexity of each party $P_j$ includes two parts: 
(1) $O(\sum_{1 \le i \le s}(i_q \cdot \lvert \mathsf{AND}_i + \mathsf{OR}_i \rvert \cdot n))$, where $Q_i$ is set-separable with respect to $X_j$ (we use $Q'_i$ to denote the
separation formula of $Q_i$ with respect to $X_j$), while $i_q$ is the number of literals and $\lvert \mathsf{AND}_i + \mathsf{OR}_i \rvert$ is the total number of $\mathsf{AND}$ and $\mathsf{OR}$ operators in $Q'_i$;
(2) $O(\sum_{1 \le i \le s}(\lvert \mathsf{AND}_i + \mathsf{OR}_i \rvert \cdot n))$, where $Q_i$ is not set-separable with respect to $X_j$ while includes $X_j$, and $\lvert \mathsf{AND}_i + \mathsf{OR}_i \rvert$ is the total number of $\mathsf{AND}$ and $\mathsf{OR}$ operators in the separation formula of $Q_i$ with respect to some other set. 
The communication complexity of $P_j$ can also be computed as two parts: 
(1) $O(\sum_{1 \le i \le s}(i_q \cdot \lvert \mathsf{OR}_i \rvert \cdot n))$, where $Q_i$ is set-separable with respect to $X_j$ (we use $Q'_i$ to denote the
separation formula of $Q_i$ with respect to $X_j$), while $i_q$ is the number of literals and $\lvert \mathsf{OR}_i \rvert$ is the number of $\mathsf{OR}$ operators in $Q'_i$; 
(2) $O(\sum_{1 \le i \le s}(\lvert \mathsf{OR}_i \rvert \cdot n))$, where $Q_i$ is not set-separable with respect to $X_j$ while includes $X_j$, and $\lvert \mathsf{OR}_i \rvert$ is the number of $\mathsf{OR}$ operators in the separation formula of $Q_i$ with respect to some other set.

In the multi-party secret-shared shuffle and reconstruction steps, the leader's computation and communication complexity are both $O(smn)$ while each client's computation and communication complexity are both $O(sn)$, where $s$ is the number of subformulas in the CPF representation $\psi$.

A more detailed of complexity analysis for our MPSI (and its variants) and our MPSU (and its variants) protocols is provided in Appendix \ref{appdix:compare}.
\end{trivlist}

\section{Performance Evaluation}

We demonstrate the practicality of our framework with implementations for its typical instantiations, including MPSI, MPSI-card, MPSI-card-sum and MPSU protocols. We assume a commonly used setting where Beaver triples are pre-computed offline and stored locally. This follows real scenarios where Beaver triples are pre-generated by parties themselves or with the help of a Trusted-Third Party under the Trusted Dealer model~\cite{ZCLZL-USENIX-2023,LG-ASIACRYPT-2023}. We report on the online performance of our protocols in comparison with the respective state of the art:
\begin{itemize}
    \item The state-of-the-art MPSI~\cite{WuYC24}: This work proposes two MPSI protocols, O-Ring and K-Star, with publicly available implementations \cite{libOring}. We select K-Star for comparison since it is faster than O-Ring with the same total communication costs. The reported data in Table \ref{tab:mpsi} is obtained by running their full protocol, as there is no correlated randomness pre-generated offline in their implementation. 
    We set the corruption threshold $t = m-1$ in their code and report the leader's running time and the total communication costs of all parties for both their and our MPSI protocols.  

    \item The state-of-the-art MPSI-card~\cite{ChenDGB22}: This work does not provide open-source code, thus we take the experimental results of the online phase from their paper, whose experiments was run on Intel i7-12700H 2.30GHz CPU and 28GB RAM. We report the leader's running time and communication costs for both their and our MPSI-card protocols in Table \ref{tab:mpsi-card}.
    \item The state-of-the-art MPSU~\cite{DongCZB24}: This work proposes two MPSU protocols with publicly available implementation \cite{libMPSU}. We compare with the symmetric-key based one for its better online performance. In the benchmark of MPSU, we set the item length as 64 bits and report the leader's running time and communication costs in the online phase in Table \ref{tab:mpsu}.
\end{itemize}
To our knowledge, there is no available implementation or reported experimental data for the MPSI-card-sum protocol. We provide the first MPSI-card-sum implementation and report experimental data in the same setting as MPSI-card.

We conduct our experiments on a cloud virtual machine with Intel(R) Xeon(R) 2.70GHz CPU (32 physical cores) and 128 GB RAM.
In the LAN setting, the bandwidth is set to be 10 Gbps with 0.1 ms RTT latency.
In the WAN setting, the bandwidth is set to be 400 Mbps with 80 ms RTT latency. 
The implementation details and parameter settings are detailed in Appendix~\ref{appdix:implementation}.

The experimental results, presented in Table \ref{tab:mpsi}-\ref{tab:mpsu}, show that our protocols exhibit online performance that is either superior to or comparable with the state
of the art in all cases. In terms of computation costs, our MPSI protocol achieves a $2.5-8.3 \times$ speedup (LAN) and a $1.7-3.4 \times$ speedup (WAN) compared to \cite{WuYC24}; our MPSI-card protocol achieves an $18.0-63.4 \times$ speedup (LAN) compared to \cite{ChenDGB22}. In terms of communication costs, our MPSI achieves an improvement up to $14.4\times$ while our MPSI-card achieves an improvement up to $20.3\times$. The computation and communication costs of our MPSI-card-sum is only double our MPSI approximately, while realizing a richer functionality. The computation and communication costs of our MPSU are comparable with~\cite{DongCZB24}. 


\begin{table}[]
\centering
\resizebox{\textwidth}{!}{%
\begin{tabular}{|c|c|cccccccccc|ccccc|}
\hline
\multirow{3}{*}{$m$} & \multirow{3}{*}{protocol} & \multicolumn{10}{c|}{Time (second)} & \multicolumn{5}{c|}{\multirow{2}{*}{Comm. (MB)}} \\ \cline{3-12}
 &  & \multicolumn{5}{c|}{LAN} & \multicolumn{5}{c|}{WAN} & \multicolumn{5}{c|}{} \\ \cline{3-17} 
 &  & \multicolumn{1}{c|}{$2^{12}$} & \multicolumn{1}{c|}{$2^{14}$} & \multicolumn{1}{c|}{$2^{16}$} & \multicolumn{1}{c|}{$2^{18}$} & \multicolumn{1}{c|}{$2^{20}$} & \multicolumn{1}{c|}{$2^{12}$} & \multicolumn{1}{c|}{$2^{14}$} & \multicolumn{1}{c|}{$2^{16}$} & \multicolumn{1}{c|}{$2^{18}$} & $2^{20}$ & \multicolumn{1}{c|}{$2^{12}$} & \multicolumn{1}{c|}{$2^{14}$} & \multicolumn{1}{c|}{$2^{16}$} & \multicolumn{1}{c|}{$2^{18}$} & $2^{20}$ \\ \hline
 & \cite{WuYC24} & \multicolumn{1}{c|}{0.064} & \multicolumn{1}{c|}{0.188} & \multicolumn{1}{c|}{0.860} & \multicolumn{1}{c|}{4.389} & \multicolumn{1}{c|}{19.81} & \multicolumn{1}{c|}{5.283} & \multicolumn{1}{c|}{5.733} & \multicolumn{1}{c|}{6.680} & \multicolumn{1}{c|}{10.24} & 30.10 & \multicolumn{1}{c|}{2.109} & \multicolumn{1}{c|}{3.418} & \multicolumn{1}{c|}{\textbf{8.110}} & \multicolumn{1}{c|}{\textbf{28.54}} & \textbf{111.9} \\ \cline{2-17} 
\multirow{-2}{*}{3} & Ours & \multicolumn{1}{c|}{\textbf{0.013}} & \multicolumn{1}{c|}{\textbf{0.047}} & \multicolumn{1}{c|}{\textbf{0.227}} & \multicolumn{1}{c|}{\textbf{1.533}} & \multicolumn{1}{c|}{\textbf{7.945}} & \multicolumn{1}{c|}{\textbf{1.540}} & \multicolumn{1}{c|}{\textbf{2.070}} & \multicolumn{1}{c|}{\textbf{2.912}} & \multicolumn{1}{c|}{\textbf{5.136}} & \textbf{16.42} & \multicolumn{1}{c|}{\textbf{0.641}} & \multicolumn{1}{c|}{\textbf{2.551}} & \multicolumn{1}{c|}{10.20} & \multicolumn{1}{c|}{40.91} & 164.2 \\ \hline
 & \cite{WuYC24} & \multicolumn{1}{c|}{0.077} & \multicolumn{1}{c|}{0.211} & \multicolumn{1}{c|}{0.905} & \multicolumn{1}{c|}{4.482} & \multicolumn{1}{c|}{20.78} & \multicolumn{1}{c|}{5.719} & \multicolumn{1}{c|}{6.159} & \multicolumn{1}{c|}{7.174} & \multicolumn{1}{c|}{11.24} & 33.45 & \multicolumn{1}{c|}{4.689} & \multicolumn{1}{c|}{7.464} & \multicolumn{1}{c|}{17.39} & \multicolumn{1}{c|}{\textbf{60.20}} & \cellcolor[HTML]{FFFFFF}\textbf{235.4} \\ \cline{2-17} 
\multirow{-2}{*}{4} & Ours & \multicolumn{1}{c|}{\textbf{0.015}} & \multicolumn{1}{c|}{\textbf{0.051}} & \multicolumn{1}{c|}{\textbf{0.242}} & \multicolumn{1}{c|}{\textbf{1.632}} & \multicolumn{1}{c|}{\textbf{8.133}} & \multicolumn{1}{c|}{\textbf{1.944}} & \multicolumn{1}{c|}{\textbf{2.476}} & \multicolumn{1}{c|}{\textbf{3.339}} & \multicolumn{1}{c|}{\textbf{6.345}} & \textbf{19.23} & \multicolumn{1}{c|}{\textbf{0.961}} & \multicolumn{1}{c|}{\textbf{3.826}} & \multicolumn{1}{c|}{\textbf{15.31}} & \multicolumn{1}{c|}{61.36} & 246.2 \\ \hline
 & \cite{WuYC24} & \multicolumn{1}{c|}{0.104} & \multicolumn{1}{c|}{0.235} & \multicolumn{1}{c|}{1.019} & \multicolumn{1}{c|}{4.839} & \multicolumn{1}{c|}{22.47} & \multicolumn{1}{c|}{6.049} & \multicolumn{1}{c|}{6.511} & \multicolumn{1}{c|}{7.806} & \multicolumn{1}{c|}{12.02} & 37.27 & \multicolumn{1}{c|}{8.285} & \multicolumn{1}{c|}{13.07} & \multicolumn{1}{c|}{30.16} & \multicolumn{1}{c|}{103.5} & 403.9 \\ \cline{2-17} 
\multirow{-2}{*}{5} & Ours & \multicolumn{1}{c|}{\textbf{0.016}} & \multicolumn{1}{c|}{\textbf{0.053}} & \multicolumn{1}{c|}{\textbf{0.260}} & \multicolumn{1}{c|}{\textbf{1.724}} & \multicolumn{1}{c|}{\textbf{8.255}} & \multicolumn{1}{c|}{\textbf{2.348}} & \multicolumn{1}{c|}{\textbf{2.889}} & \multicolumn{1}{c|}{\textbf{3.770}} & \multicolumn{1}{c|}{\textbf{6.789}} & \textbf{22.04} & \multicolumn{1}{c|}{\textbf{1.282}} & \multicolumn{1}{c|}{\textbf{5.102}} & \multicolumn{1}{c|}{\textbf{20.41}} & \multicolumn{1}{c|}{\textbf{81.81}} & \textbf{328.3} \\ \hline
 & \cite{WuYC24} & \multicolumn{1}{c|}{0.215} & \multicolumn{1}{c|}{0.465} & \multicolumn{1}{c|}{1.691} & \multicolumn{1}{c|}{7.310} & \multicolumn{1}{c|}{32.91} & \multicolumn{1}{c|}{8.653} & \multicolumn{1}{c|}{9.511} & \multicolumn{1}{c|}{11.18} & \multicolumn{1}{c|}{21.12} & 70.74 & \multicolumn{1}{c|}{41.51} & \multicolumn{1}{c|}{64.46} & \multicolumn{1}{c|}{146.3} & \multicolumn{1}{c|}{493.7} & 1921 \\ \cline{2-17} 
\multirow{-2}{*}{10} & Ours & \multicolumn{1}{c|}{\textbf{0.026}} & \multicolumn{1}{c|}{\textbf{0.075}} & \multicolumn{1}{c|}{\textbf{0.354}} & \multicolumn{1}{c|}{\textbf{1.910}} & \multicolumn{1}{c|}{\textbf{8.898}} & \multicolumn{1}{c|}{\textbf{4.363}} & \multicolumn{1}{c|}{\textbf{4.915}} & \multicolumn{1}{c|}{\textbf{5.769}} & \multicolumn{1}{c|}{\textbf{9.423}} & \textbf{33.38} & \multicolumn{1}{c|}{\textbf{2.884}} & \multicolumn{1}{c|}{\textbf{11.48}} & \multicolumn{1}{c|}{\textbf{45.92}} & \multicolumn{1}{c|}{\textbf{184.1}} & \textbf{738.7} \\ \hline
\end{tabular}}
\caption{Running time and communication costs for MPSI protocols in LAN and WAN settings. $m$ is the number of parties. }
\label{tab:mpsi}
\end{table}

\begin{table}[]
\centering
\resizebox{\textwidth}{!}{%
\begin{tabular}{|c|c|cccccccccc|ccccc|}
\hline
\multirow{3}{*}{$m$} & \multirow{3}{*}{protocol} & \multicolumn{10}{c|}{Time (second)} & \multicolumn{5}{c|}{\multirow{2}{*}{Comm. (MB)}} \\ \cline{3-12}
 &  & \multicolumn{5}{c|}{LAN} & \multicolumn{5}{c|}{WAN} & \multicolumn{5}{c|}{} \\ \cline{3-17} 
 &  & \multicolumn{1}{c|}{$2^{12}$} & \multicolumn{1}{c|}{$2^{14}$} & \multicolumn{1}{c|}{$2^{16}$} & \multicolumn{1}{c|}{$2^{18}$} & \multicolumn{1}{c|}{$2^{20}$} & \multicolumn{1}{c|}{$2^{12}$} & \multicolumn{1}{c|}{$2^{14}$} & \multicolumn{1}{c|}{$2^{16}$} & \multicolumn{1}{c|}{$2^{18}$} & $2^{20}$ & \multicolumn{1}{c|}{$2^{12}$} & \multicolumn{1}{c|}{$2^{14}$} & \multicolumn{1}{c|}{$2^{16}$} & \multicolumn{1}{c|}{$2^{18}$} & $2^{20}$ \\ \hline
3 & Ours & \multicolumn{1}{c|}{\textbf{0.015}} & \multicolumn{1}{c|}{\textbf{0.052}} & \multicolumn{1}{c|}{\textbf{0.237}} & \multicolumn{1}{c|}{\textbf{1.581}} & \multicolumn{1}{c|}{\textbf{8.050}} & \multicolumn{1}{c|}{\textbf{1.784}} & \multicolumn{1}{c|}{\textbf{2.800}} & \multicolumn{1}{c|}{\textbf{3.681}} & \multicolumn{1}{c|}{\textbf{6.498}} & \textbf{19.50} & \multicolumn{1}{c|}{\textbf{0.761}} & \multicolumn{1}{c|}{\textbf{3.035}} & \multicolumn{1}{c|}{\textbf{12.15}} & \multicolumn{1}{c|}{\textbf{48.76}} & \textbf{195.8} \\ \hline
4 & Ours & \multicolumn{1}{c|}{\textbf{0.016}} & \multicolumn{1}{c|}{\textbf{0.054}} & \multicolumn{1}{c|}{\textbf{0.252}} & \multicolumn{1}{c|}{\textbf{1.684}} & \multicolumn{1}{c|}{\textbf{8.300}} & \multicolumn{1}{c|}{\textbf{2.267}} & \multicolumn{1}{c|}{\textbf{3.780}} & \multicolumn{1}{c|}{\textbf{5.005}} & \multicolumn{1}{c|}{\textbf{8.565}} & \textbf{24.88} & \multicolumn{1}{c|}{\textbf{1.122}} & \multicolumn{1}{c|}{\textbf{4.472}} & \multicolumn{1}{c|}{\textbf{17.91}} & \multicolumn{1}{c|}{\textbf{71.83}} & \textbf{288.4} \\ \hline
\multirow{2}{*}{5} & \cite{ChenDGB22} & \multicolumn{1}{c|}{0.670} & \multicolumn{1}{c|}{1.789} & \multicolumn{1}{c|}{6.289} & \multicolumn{1}{c|}{31.24} & \multicolumn{1}{c|}{$-$} & \multicolumn{1}{c|}{$-$} & \multicolumn{1}{c|}{$-$} & \multicolumn{1}{c|}{$-$} & \multicolumn{1}{c|}{$-$} & $-$ & \multicolumn{1}{c|}{20.70} & \multicolumn{1}{c|}{94.49} & \multicolumn{1}{c|}{425.6} & \multicolumn{1}{c|}{1894} & $-$ \\ \cline{2-17} 
 & Ours & \multicolumn{1}{c|}{\textbf{0.018}} & \multicolumn{1}{c|}{\textbf{0.056}} & \multicolumn{1}{c|}{\textbf{0.270}} & \multicolumn{1}{c|}{\textbf{1.735}} & \multicolumn{1}{c|}{\textbf{8.616}} & \multicolumn{1}{c|}{\textbf{2.753}} & \multicolumn{1}{c|}{\textbf{4.747}} & \multicolumn{1}{c|}{\textbf{6.076}} & \multicolumn{1}{c|}{\textbf{10.42}} & \textbf{28.87} & \multicolumn{1}{c|}{\textbf{1.482}} & \multicolumn{1}{c|}{\textbf{5.909}} & \multicolumn{1}{c|}{\textbf{23.66}} & \multicolumn{1}{c|}{\textbf{94.90}} & \textbf{381.0} \\ \hline
\multirow{2}{*}{10} & \cite{ChenDGB22} & \multicolumn{1}{c|}{1.477} & \multicolumn{1}{c|}{4.503} & \multicolumn{1}{c|}{12.81} & \multicolumn{1}{c|}{95.23} & \multicolumn{1}{c|}{$-$} & \multicolumn{1}{c|}{$-$} & \multicolumn{1}{c|}{$-$} & \multicolumn{1}{c|}{$-$} & \multicolumn{1}{c|}{$-$} & $-$ & \multicolumn{1}{c|}{46.58} & \multicolumn{1}{c|}{212.6} & \multicolumn{1}{c|}{957.7} & \multicolumn{1}{c|}{4262} & $-$ \\ \cline{2-17} 
 & Ours & \multicolumn{1}{c|}{\textbf{0.026}} & \multicolumn{1}{c|}{\textbf{0.071}} & \multicolumn{1}{c|}{\textbf{0.375}} & \multicolumn{1}{c|}{\textbf{2.001}} & \multicolumn{1}{c|}{\textbf{9.226}} & \multicolumn{1}{c|}{\textbf{5.174}} & \multicolumn{1}{c|}{\textbf{9.578}} & \multicolumn{1}{c|}{\textbf{11.93}} & \multicolumn{1}{c|}{\textbf{18.43}} & \textbf{50.33} & \multicolumn{1}{c|}{\textbf{3.285}} & \multicolumn{1}{c|}{\textbf{13.09}} & \multicolumn{1}{c|}{\textbf{52.42}} & \multicolumn{1}{c|}{\textbf{210.2}} & \textbf{844.0} \\ \hline
\end{tabular}}
\caption{Running time and communication costs for MPSI-card protocols in LAN and WAN settings. The data of~\cite{ChenDGB22} originates from their paper for lack of available implementation. Cells with $-$ denote missing data that is not reported.}
\label{tab:mpsi-card}
\end{table}

\begin{table}[]
\centering
\resizebox{\textwidth}{!}{%
\begin{tabular}{|>{\centering\arraybackslash}p{1cm}|cccccccccc|ccccc|}
\hline
\multicolumn{1}{|c|}{\multirow{2}{*}[-0.5ex]{\centering $m$}} & \multicolumn{10}{c|}{Time (second)} & \multicolumn{5}{c|}{\multirow{2}{*}{Comm.(MB)}} \\ \cline{2-11}
\multicolumn{1}{|l|}{} & \multicolumn{5}{c|}{LAN} & \multicolumn{5}{c|}{WAN} & \multicolumn{5}{c|}{} \\ \cline{2-16}
 & \multicolumn{1}{c|}{$2^{12}$} & \multicolumn{1}{c|}{$2^{14}$} & \multicolumn{1}{c|}{$2^{16}$} & \multicolumn{1}{c|}{$2^{18}$} & \multicolumn{1}{c|}{$2^{20}$} & \multicolumn{1}{c|}{$2^{12}$} & \multicolumn{1}{c|}{$2^{14}$} & \multicolumn{1}{c|}{$2^{16}$} & \multicolumn{1}{c|}{$2^{18}$} & $2^{20}$ & \multicolumn{1}{c|}{$2^{12}$} & \multicolumn{1}{c|}{$2^{14}$} & \multicolumn{1}{c|}{$2^{16}$} & \multicolumn{1}{c|}{$2^{18}$} & $2^{20}$ \\ \hline
3 & \multicolumn{1}{c|}{0.023} & \multicolumn{1}{c|}{0.088} & \multicolumn{1}{c|}{0.417} & \multicolumn{1}{c|}{2.810} & \multicolumn{1}{c|}{14.67} & \multicolumn{1}{c|}{2.519} & \multicolumn{1}{c|}{3.541} & \multicolumn{1}{c|}{4.686} & \multicolumn{1}{c|}{9.417} & 31.88 & \multicolumn{1}{c|}{1.283} & \multicolumn{1}{c|}{5.107} & \multicolumn{1}{c|}{20.43} & \multicolumn{1}{c|}{81.89} & 328.6 \\ \hline
4 & \multicolumn{1}{c|}{0.025} & \multicolumn{1}{c|}{0.091} & \multicolumn{1}{c|}{0.436} & \multicolumn{1}{c|}{3.044} & \multicolumn{1}{c|}{15.16} & \multicolumn{1}{c|}{3.084} & \multicolumn{1}{c|}{4.680} & \multicolumn{1}{c|}{5.948} & \multicolumn{1}{c|}{12.40} & 38.17 & \multicolumn{1}{c|}{1.885} & \multicolumn{1}{c|}{7.499} & \multicolumn{1}{c|}{29.99} & \multicolumn{1}{c|}{120.2} & 482.4 \\ \hline
5 & \multicolumn{1}{c|}{0.027} & \multicolumn{1}{c|}{0.094} & \multicolumn{1}{c|}{0.474} & \multicolumn{1}{c|}{3.150} & \multicolumn{1}{c|}{15.49} & \multicolumn{1}{c|}{3.650} & \multicolumn{1}{c|}{5.729} & \multicolumn{1}{c|}{7.288} & \multicolumn{1}{c|}{13.17} & 43.06 & \multicolumn{1}{c|}{2.486} & \multicolumn{1}{c|}{9.890} & \multicolumn{1}{c|}{39.56} & \multicolumn{1}{c|}{158.6} & 636.2 \\ \hline
10 & \multicolumn{1}{c|}{0.039} & \multicolumn{1}{c|}{0.120} & \multicolumn{1}{c|}{0.632} & \multicolumn{1}{c|}{3.610} & \multicolumn{1}{c|}{16.65} & \multicolumn{1}{c|}{6.474} & \multicolumn{1}{c|}{10.93} & \multicolumn{1}{c|}{13.79} & \multicolumn{1}{c|}{23.56} & 71.36 & \multicolumn{1}{c|}{5.493} & \multicolumn{1}{c|}{21.85} & \multicolumn{1}{c|}{87.37} & \multicolumn{1}{c|}{350.2} & 1405 \\ \hline
\end{tabular}}
\caption{Running time and communication costs for our MPSI-card-sum protocol.}
\label{tab:mpsi-card-sum}
\end{table}

\begin{table}[]
\centering
\resizebox{\textwidth}{!}{%
\begin{tabular}{|c|c|cccccccccc|ccccc|}
\hline
\multirow{3}{*}{$m$} & \multirow{3}{*}{protocol} & \multicolumn{10}{c|}{Time (second)} & \multicolumn{5}{c|}{\multirow{2}{*}{Comm. (MB)}} \\ \cline{3-12}
 &  & \multicolumn{5}{c|}{LAN} & \multicolumn{5}{c|}{WAN} & \multicolumn{5}{c|}{} \\ \cline{3-17} 
 &  &  \multicolumn{1}{c|}{$2^{12}$} & \multicolumn{1}{c|}{$2^{14}$} & \multicolumn{1}{c|}{$2^{16}$} & \multicolumn{1}{c|}{$2^{18}$} & \multicolumn{1}{c|}{$2^{20}$} & \multicolumn{1}{c|}{$2^{12}$} & \multicolumn{1}{c|}{$2^{14}$} & \multicolumn{1}{c|}{$2^{16}$} & \multicolumn{1}{c|}{$2^{18}$} & $2^{20}$ & \multicolumn{1}{c|}{$2^{12}$} & \multicolumn{1}{c|}{$2^{14}$} & \multicolumn{1}{c|}{$2^{16}$} & \multicolumn{1}{c|}{$2^{18}$} & $2^{20}$ \\ \hline
 & \cite{DongCZB24} & \multicolumn{1}{c|}{\textbf{0.017}} & \multicolumn{1}{c|}{\textbf{0.050}} & \multicolumn{1}{c|}{\textbf{0.215}} & \multicolumn{1}{c|}{\textbf{1.005}} & \multicolumn{1}{c|}{\textbf{4.352}} & \multicolumn{1}{c|}{\textbf{3.157}} & \multicolumn{1}{c|}{\textbf{3.734}} & \multicolumn{1}{c|}{\textbf{4.444}} & \multicolumn{1}{c|}{\textbf{9.705}} & \textbf{33.10} & \multicolumn{1}{c|}{\textbf{1.690}} & \multicolumn{1}{c|}{\textbf{6.788}} & \multicolumn{1}{c|}{\textbf{27.87}} & \multicolumn{1}{c|}{\textbf{112.7}} & \textbf{455.6} \\ \cline{2-17}
 \multirow{-2}{*}{3} & Ours & \multicolumn{1}{c|}{0.022} & \multicolumn{1}{c|}{0.068} & \multicolumn{1}{c|}{0.298} & \multicolumn{1}{c|}{1.892} & \multicolumn{1}{c|}{9.607} & \multicolumn{1}{c|}{3.327} & \multicolumn{1}{c|}{3.774} & \multicolumn{1}{c|}{4.878} & \multicolumn{1}{c|}{11.04} & 37.77 & \multicolumn{1}{c|}{1.930} & \multicolumn{1}{c|}{7.755} & \multicolumn{1}{c|}{31.76} & \multicolumn{1}{c|}{128.3} & 518.7 \\ \hline
 & \cite{DongCZB24} & \multicolumn{1}{c|}{\textbf{0.023}} & \multicolumn{1}{c|}{\textbf{0.071}} & \multicolumn{1}{c|}{\textbf{0.286}} & \multicolumn{1}{c|}{\textbf{1.393}} & \multicolumn{1}{c|}{\textbf{5.645}} & \multicolumn{1}{c|}{\textbf{3.976}} & \multicolumn{1}{c|}{\textbf{4.618}} & \multicolumn{1}{c|}{\textbf{6.507}} & \multicolumn{1}{c|}{\textbf{17.10}} & \multicolumn{1}{c|}{\textbf{59.21}} & \multicolumn{1}{c|}{\textbf{3.145}} & \multicolumn{1}{c|}{\textbf{12.81}} & \multicolumn{1}{c|}{\textbf{51.69}} & \multicolumn{1}{c|}{\textbf{208.8}} & \textbf{843.7}   \\ \cline{2-17} 
 \multirow{-2}{*}{4} & Ours & \multicolumn{1}{c|}{0.029} & \multicolumn{1}{c|}{0.089} & \multicolumn{1}{c|}{0.415} & \multicolumn{1}{c|}{2.345} & \multicolumn{1}{c|}{11.25} & \multicolumn{1}{c|}{3.996} & \multicolumn{1}{c|}{4.820} & \multicolumn{1}{c|}{6.756} & \multicolumn{1}{c|}{17.45} & 64.04 & \multicolumn{1}{c|}{3.624} & \multicolumn{1}{c|}{14.74} & \multicolumn{1}{c|}{59.46} & \multicolumn{1}{c|}{240.1} & 969.6 \\ \hline
 & \cite{DongCZB24} & \multicolumn{1}{c|}{\textbf{0.030}} & \multicolumn{1}{c|}{\textbf{0.087}} & \multicolumn{1}{c|}{\textbf{0.368}} & \multicolumn{1}{c|}{\textbf{1.714}} & \multicolumn{1}{c|}{\textbf{7.003}} & \multicolumn{1}{c|}{\textbf{4.800}} & \multicolumn{1}{c|}{\textbf{5.521}} & \multicolumn{1}{c|}{\textbf{8.938}} & \multicolumn{1}{c|}{\textbf{25.95}} & \textbf{95.40} & \multicolumn{1}{c|}{\textbf{5.007}} & \multicolumn{1}{c|}{\textbf{20.36}} & \multicolumn{1}{c|}{\textbf{82.11}} & \multicolumn{1}{c|}{\textbf{331.5}} & \textbf{1339} \\ \cline{2-17}
 \multirow{-2}{*}{5} & Ours & \multicolumn{1}{c|}{0.039} & \multicolumn{1}{c|}{0.114} & \multicolumn{1}{c|}{0.542} & \multicolumn{1}{c|}{2.796} & \multicolumn{1}{c|}{13.18} & \multicolumn{1}{c|}{4.872} & \multicolumn{1}{c|}{5.705} & \multicolumn{1}{c|}{8.992} & \multicolumn{1}{c|}{26.09} & 101.6 & \multicolumn{1}{c|}{5.805} & \multicolumn{1}{c|}{23.57} & \multicolumn{1}{c|}{95.05} & \multicolumn{1}{c|}{383.6} & 1548 \\ \hline
 & \cite{DongCZB24} & \multicolumn{1}{c|}{\textbf{0.088}} & \multicolumn{1}{c|}{\textbf{0.286}} & \multicolumn{1}{c|}{\textbf{1.183}} & \multicolumn{1}{c|}{$-$} & \multicolumn{1}{c|}{$-$} & \multicolumn{1}{c|}{9.203} & \multicolumn{1}{c|}{14.54} & \multicolumn{1}{c|}{38.31} & \multicolumn{1}{c|}{$-$} & \multicolumn{1}{c|}{$-$} & \multicolumn{1}{c|}{\textbf{20.48}} & \multicolumn{1}{c|}{\textbf{82.39}} & \multicolumn{1}{c|}{\textbf{332.0}} & \multicolumn{1}{c|}{$-$} & \multicolumn{1}{c|}{$-$} \\ \cline{2-17}
 \multirow{-2}{*}{10} & Ours & \multicolumn{1}{c|}{0.110} & \multicolumn{1}{c|}{0.337} & \multicolumn{1}{c|}{1.483} & \multicolumn{1}{c|}{7.167} & \multicolumn{1}{c|}{$-$} & \multicolumn{1}{c|}{\textbf{8.187}} & \multicolumn{1}{c|}{\textbf{12.06}} & \multicolumn{1}{c|}{\textbf{30.56}} & \multicolumn{1}{c|}{105.7} & $-$ & \multicolumn{1}{c|}{24.07} & \multicolumn{1}{c|}{96.82} & \multicolumn{1}{c|}{390.1} & \multicolumn{1}{c|}{1572} & $-$ \\ \hline
\end{tabular}}
\caption{Running time and communication costs for MPSU protocols in LAN and WAN settings. 
Cells with $-$ denote trials running out of memory.}
\label{tab:mpsu}
\end{table}

%
%
%
\bibliographystyle{splncs04}

\bibliography{MPSU}

\appendix

\section{Theoretical Analysis and Comparison}\label{appdix:compare}

\subsection{Complexity of Our MPSI and Its Variants}
In the following analyses of asymptotic complexity, we consider the only dependency in $n$ and $m$, omitting security parameters.

In Figure~\ref{fig:proto-mpsi}, the parties ($P_1$ as $P_\mathsf{pivot}$) invoke the batch pure membership zero-sharing protocol of size $B = O(n)$. In this stage, the computation and communication complexity of $P_1$ are $O(mn)$, while the computation and communication complexity of each $P_j$ ($1 < j \le m$) are $O(n)$.
In MPSI/circuit-MPSI, each $P_j$ directly sends its shares to $P_1$, thereby, the overall computation and communication complexity of $P_1$ are $O(mn)$, while the overall computation and communication complexity of each $P_j$ are $O(n)$;
In MPSI-card, the parties invoke the multi-party secret-shared shuffle protocol before the straightforward reconstruction. We use the multi-party secret-shared shuffle protocol in~\cite{EB22} and designate $P_1$ as the leader.
In this stage, the computation and communication complexity of $P_1$ are $O(mn)$, while the computation and communication complexity of $P_j$ are $O(n)$. In all, the computation and communication complexity of MPSI-card are identical to MPSI/circuit-MPSI.

In Figure~\ref{fig:proto-mpsi-sum}, the parties invoke the batch pure membership zero-sharing with payloads protocol of size $B = O(n)$. In this stage, the computation and communication complexity of $P_1$ are $O(mn)$, while the computation and communication complexity of each $P_j$ ($1 < j \le m$) are $O(n)$. Then, the parties invoke the multi-party secret-shared shuffle protocol twice, $P_j$ reconstructs the cardinality to $P_1$, and $P_1$ broadcasts the indicator vector for the shuffle payloads. In all, the computation and communication complexity of MPSI-card-sum remain the same as MPSI/MPSI-card/circuit-MPSI.

Notably, in the naive (insecure) solution, the clients directly sends their input sets to the leader and the leader computes the result locally, where the leader's computation and communication complexity are $O(m n)$ and each client's computation and communication complexity are $O(n)$. Therefore, our MPSI/MPSI-card/circuit-MPSI/MPSI-card-sum constructions achieve optimal complexity that matches the naive solution while ensuring security.

\subsection{Complexity of Our MPSU and Its Variants}
In the following analyses of asymptotic complexity, we consider the only dependency in $n$ and $m$, omitting security parameters.

In Figure~\ref{fig:proto-mpsu}, $1 < j \le m$, $P_1, \cdots, P_j$ ($P_j$ as $P_\mathsf{pivot}$) invoke the batch pure non-membership zero-sharing protocol of size $B = O(n)$. Each $P_j$ engages in $m-j+1$ invocations of batch pure non-membership zero-sharing protocols, acting as $P_\mathsf{pivot}$ in the first time. $P_1$ engages in $m-1$ invocations of batch pure non-membership zero-sharing protocols without acting as $P_\mathsf{pivot}$. In this stage, the computation and communication complexity of each party are $O(mn)$. After that, the parties hold $(m-1)B$ secret-sharings.
Then, they invoke the multi-party secret-shared shuffle protocol (with $P_1$ as the leader) with their $(m-1)B = O(mn)$ shares, and finally each $P_j$ sends its shuffled shares to $P_1$. Thereby, the computation and communication complexity of $P_1$ are $O(m^2n)$, while the computation and communication complexity of each $P_j$ are $O(mn)$. As a result, the overall computation and communication complexity of $P_1$ are $O(m^2n)$, while the computation and communication complexity of each $P_j$ are $O(mn)$.

Our MPSU protocol follows the secret-sharing based MPSU paradigm, where the leader's optimal computation and communication complexity are $O(m^2 n)$, while each client's optimal computation and communication complexity are $O(m n)$. This optimal complexity is determined by the core design of secret-sharing $O(mn)$ elements among $m$ parties, since the necessary reconstruction step requires the optimal complexity. Therefore, our MPSU construction achieves optimal computation and communication complexity of this MPSU paradigm.

\subsection{Comparison with Prior Works}

Table~\ref{tab:mpsi-comparisons} shows a theoretical comparison of the computation and communication required by various MPSI protocols.
Table~\ref{tab:mpcsi-comparisons} shows a theoretical comparison between the related MPSI-card/MPSI-card-sum protocols and ours.
Table~\ref{tab:mpsu-comparisons} shows a theoretical comparison between the related MPSU protocols and ours.
\begin{table}
\centering
\resizebox{\textwidth}{!}{%
\begin{tabular}{|c|cc|cc|c|c|}
\hline
\multirow{2}{*}{\textbf{Protocol}} & \multicolumn{2}{c|}{\textbf{Computation}} & \multicolumn{2}{c|}{\textbf{Communication}} & \multirow{2}{*}{\textbf{Security}} & \multirow{2}{*}{\textbf{Operation}} \\ \cline{2-5}
 & \multicolumn{1}{c|}{Leader} & Client & \multicolumn{1}{c|}{Leader} & Client &  &  \\ \hline
 \cite{FreedmanNP04} & \multicolumn{1}{c|}{$O(m^2 n^2)$} & $O(m^2 n^2)$ & \multicolumn{1}{c|}{$O(m^2 n^2 \lambda)$} & $O(m^2 n^2 \lambda)$ & standard & PK \\ \hline
\cite{KS-CRYPTO-2005} & \multicolumn{1}{c|}{$O(m t n^2)$} & $O(m t n^2)$ & \multicolumn{1}{c|}{$O(m t n \log \lvert U \rvert \lambda)$} & $O(m t n \log \lvert U \rvert \lambda)$ & standard semi-honest & PK \\ \hline
\cite{PKC-HazayV17} & \multicolumn{1}{c|}{$O(m n^2)$} & $O(n)$ & \multicolumn{1}{c|}{$O(m n \lambda)$} & $O(n \lambda)$ & standard semi-honest & PK \\ \hline
\multirow{2}{*}{\cite{KMPRT-CCS-2017}} & \multicolumn{1}{c|}{\multirow{2}{*}{$O(m n)$}} & $O(n)$ & \multicolumn{1}{c|}{\multirow{2}{*}{$O(m n (\lambda + \sigma + \log n))$}} & $O(n (\lambda + \sigma + \log n))$ & augmented semi-honest & SK \\ \cline{3-3} \cline{5-7} 
 & \multicolumn{1}{c|}{} & $O(t n)$ & \multicolumn{1}{c|}{} & $O(t n (\lambda + \sigma + \log n))$ & standard semi-honest & SK \\ \hline
\multirow{2}{*}{\cite{InbarOP18}} & \multicolumn{2}{c|}{\multirow{2}{*}{$O(m n)$}} & \multicolumn{1}{c|}{\multirow{2}{*}{$O(m n)$}} & $O(\log (m) n \sigma^2)$ & augmented semi-honest & SK \\ \cline{5-7}
 & \multicolumn{2}{c|}{} & \multicolumn{1}{c|}{} & $O(m n \sigma^2)$ & standard semi-honest & SK \\ \hline
\cite{GhoshN19} & \multicolumn{1}{c|}{$O(m n \log n)$} & $O(n \log n)$ & \multicolumn{2}{c|}{$O((m^2 + m n) \lambda)$} & malicious & SK \\ \hline
~\cite{BNOP-AsiaCCS-2022} & \multicolumn{2}{c|}{$O(m n)$} & \multicolumn{1}{c|}{$O(m n \lambda (\log (n \lambda)+\lambda))$} & $O(n \lambda (\log (n \lambda)+\lambda))$ & augmented semi-honest/malicious & SK \\ \hline
~\cite{NTY-CCS-2021} & \multicolumn{1}{c|}{$O(m n)$} & $O(t n)$ & \multicolumn{1}{c|}{$O(m n (\lambda + \sigma + \log n)$} & $O(n (\lambda + \sigma + \log n)$ & augmented semi-honest/malicious & SK \\ \hline
\cite{WuYC24} & \multicolumn{1}{c|}{$O(m n)$} & $O(t n)$ & \multicolumn{1}{c|}{$O(m n (\lambda + \sigma + \log n))$} & $O(t n (\lambda + \sigma + \log n))$ & standard semi-honest & SK \\ \hline
Ours & \multicolumn{1}{c|}{$O(m n)$} & $O(n)$ & \multicolumn{1}{c|}{$O(m n (\sigma + \log n))$} & $O(n (\sigma + \log n))$ & standard semi-honest & SK \\ \hline
\end{tabular}}
\caption{Asymptotic communication and computation costs of MPSI protocols in the semi-honest setting, where $n$ is the set size. $m$ is the number of parties. $t$ is the number of colluding parties. $U$ is the domain of elements. We use ``PK'' to denote the protocols based on public-key operations, and ``SK'' to denote the protocols based on OT and symmetric-key operations. We use $\lambda, \sigma$ as the computational and statistical security parameters respectively. We use ``augmented semi-honest/malicious'' to denote the malicious protocols that implies augmented semi-honest security while is insecure in standard semi-honest model (A detailed discussion of the relations between malicious model and augmented / standard semi-honest model can be found in \cite{HazayL10} Section 2.4.4). }
\label{tab:mpsi-comparisons}
\end{table}

\begin{table}
\centering
\resizebox{\textwidth}{!}{%
\begin{tabular}{|c|cc|cc|c|c|}
\hline
\multirow{2}{*}{\textbf{Protocol}} & \multicolumn{2}{c|}{\textbf{Computation}} & \multicolumn{2}{c|}{\textbf{Communication}} & \multirow{2}{*}{\textbf{Security}} & \multirow{2}{*}{\textbf{Operation}} \\ \cline{2-5}
 & \multicolumn{1}{c|}{Leader} & Client & \multicolumn{1}{c|}{Leader} & Client &  &  \\ \hline
\cite{ChenDGB22} & \multicolumn{1}{c|}{$O((m - t) n + t n \log n)$} & $O(t n)$ & \multicolumn{1}{c|}{$O(((m - t) n + t n \log n) (\sigma + \log n))$} & $O(t n (\sigma + \log n))$ & standard semi-honest & SK \\ \hline
Ours & \multicolumn{1}{c|}{$O(m n)$} & $O(n)$ & \multicolumn{1}{c|}{$O(m n (\sigma + \log n))$} & $O(n (\sigma + \log n))$ & standard semi-honest & SK \\ \hline
\end{tabular}}
\caption{Asymptotic communication and computation costs of MPSI-card/MPSI-card-sum protocols in the semi-honest setting, where $n$ is the set size. $m$ is the number of parties. We set the number of colluding parties $t$ as the maximum $m-1$. We use ``SK'' to denote the protocols based on OT and symmetric-key operations. We use $\lambda, \sigma$ as the computational and statistical security parameters respectively.}
\label{tab:mpcsi-comparisons}
\end{table}

\begin{table}
\centering
\resizebox{\textwidth}{!}{%
\begin{tabular}{|c|cc|cc|c|c|}
\hline
\multirow{2}{*}{\textbf{Protocol}} & \multicolumn{2}{c|}{\textbf{Computation}} & \multicolumn{2}{c|}{\textbf{Communication}} & \multirow{2}{*}{\textbf{Security}} & \multirow{2}{*}{\textbf{Operation}} \\ \cline{2-5}
 & \multicolumn{1}{c|}{Leader} & Client & \multicolumn{1}{c|}{Leader} & Client &  &  \\ \hline
 \cite{GaoNT24} & \multicolumn{2}{c|}{$O(m n (\log n / \log \log n))$} & \multicolumn{2}{c|}{$\lambda m n (\log n / \log \log n)$} & standard semi-honest & PK \\ \hline
\cite{DongCZB24} & \multicolumn{1}{c|}{$O(m^2 n)$} & $O(m^2 n)$ & \multicolumn{1}{c|}{$O(m^2 n) (l + \sigma + \log m+ \log n)$} & $O(m^2 n (l + \sigma + \log m+ \log n))$ & standard semi-honest & SK \\ \hline
Ours & \multicolumn{1}{c|}{$O(m^2 n)$} & $O(m n)$ & \multicolumn{1}{c|}{$O(m^2 n (l + \sigma + \log m+ \log n))$} & $O(m n (l + \sigma + \log m+ \log n))$ & standard semi-honest & SK \\ \hline
\end{tabular}}
\caption{Asymptotic communication and computation costs of MPSU protocols in the semi-honest setting, where $n$ is the set size. $m$ is the number of parties. $l$ is the length of elements. We use ``PK'' to denote the protocols based on public-key operations, and ``SK'' to denote the protocols based on OT and symmetric-key operations. We use $\lambda, \sigma$ as the computational and statistical security parameters.}
\label{tab:mpsu-comparisons}
\end{table}

\section{The Proof of Theorem~\ref{theorem:set1}}\label{proof:set1}

\begin{proof}
We prove this theorem by constructing the predicate formula $\varphi$ using mathematical induction.
\begin{itemize}
    \item \textbf{Base Case.} If $Y = X_i$ for some $i \in \{1,\cdots,m\}$, then $\varphi(x,X_1,\cdots,X_m) = M(x,X_i): x \in X_i$.
    \item \textbf{Induction Hypothesis.} Assume that for any sets $A$ and $B$ obtained from $X_1,\cdots,X_m$ through $k$ set operations, there exist set predicate formulas $\varphi_A$ and $\varphi_B$ such that
    $$x \in A \iff \varphi_A(x,X_1,\cdots,X_m) = 1, \quad x \in B \iff \varphi_B(x,X_1,\cdots,X_m) = 1.$$
    \item \textbf{Induction Step.} We proceed to construct $\varphi$ for a set $Y$ obtained from $A$ and $B$ through one additional set operation (intersection, union, difference), conducting $k+1$ set operations in total.
    \begin{enumerate}
        \item \textbf{Union.} If $Y = A \cup B$, then $\varphi(x,X_1,\cdots,X_m) = \varphi_A(x,X_1,\cdots,X_m) \lor \varphi_B(x,X_1,\cdots,X_m)$.
        \item \textbf{Intersection.} If $Y = A \cap B$, $\varphi(x,X_1,\cdots,X_m) = \varphi_A(x,X_1,\cdots,X_m) \land \varphi_B(x,X_1,\cdots,X_m)$.
        \item \textbf{Difference.} If $Y = A \setminus B$, $\varphi(x,X_1,\cdots,X_m) = \varphi_A(x,X_1,\cdots,X_m) \land \neg \varphi_B(x,X_1,\cdots,X_m)$.
    \end{enumerate}
\end{itemize}
By repeating the above steps, we construct the set predicate formula $\varphi$ for any constructible set $Y$. Thus, the theorem is proven.
\end{proof}

\section{The Proof of Theorem~\ref{theorem:set2}}\label{proof:set2}

\begin{proof}
Given that any set predicate formula can be transformed into disjunctive normal form (DNF), Theorem~\ref{theorem:set1} can be further extended to represent $Y$ as a DNF formula $$\varphi(x,X_1,\cdots,X_m) = C_1 \lor \cdots \lor C_n$$ where each $C_i$ ($1 \le i \le n$) is a conjunctive clause. We now show that $\varphi$ can be transform into another DNF formula $\psi$ with $s$ conjunctive clauses ($s > n$) such that each $C_i$ ($1 \le i \le s$) contains at least one atomic proposition of the form $x \in X_j$, i.e., $C_i$ can be written in the form $C_i = (x \in X_j) \land D_i$ for some $X_j$.

The proof is by contradiction. Suppose there is a clause $C_i$ containing no atomic propositions of the form $x \in X_j$, i.e. $C_i$ is the conjunction of atomic propositions of the form $x \notin X_j$. Consider two cases:
\begin{itemize}
    \item \textbf{Case 1:} $C_i = (x \notin X_{i_1}) \land \cdots \land (x \notin X_{i_{t}})$ where $t < m$. As $C_i$ is a conjunctive clause of $\varphi$, the corresponding set $Y'_i$ is a subset of the constructible set $Y$, and $Y'_i$ is also a constructible set. Hence, we can augment $C_i$ by $$C_i = C_i \land ((x \in X_1) \lor \cdots \lor (x \in X_m)) = (C_i \land (x \in X_1)) \lor \cdots \lor (C_i \land (x \in X_m)).$$ For any clause that contains both $x \in X_{i_d}$ and its negation $x \notin X_{i_d}$ ($1 \leq d \leq t$), it evaluates to 0 and can be discarded. The remaining formula is $$C_i = (C_i \land (x \in X_{j_1})) \lor \cdots \lor (C_i \land (x \in X_{j_{m-t}})) = C_{i,1} \lor \cdots \lor C_{i,m-t}.$$ This splits each $C_i$ into $m-t$ conjunctive clauses where each clause contains at least one literal of the form $x \in X_j$. We substitute each $C_i$ with the above equation and have the new DNF formula $$\psi(x,X_1,\cdots,X_m) = C_1 \lor \cdots \lor C_s,$$ where each $C_i$ ($1 \le i \le s$) represents an $X_j$-constructible set for some $X_j$ ($1 \le j \le m$). Note that $C_i$ is not set-separable with respect to $X_j$ yet, since $D_i$ might involve atomic propositions relevant to $X_j$.
    \item \textbf{Case 2:} $C_i = (x \notin X_1) \land \cdots \land (x \notin X_m)$. This contradicts that the set $Y_i$ represented by $C_i$ is constructible from $X_1,\cdots,X_m$, so this case is not valid.
\end{itemize}

Next we transform the new DNF formula $\psi$ into a disjunction of subformulas that represent disjoint sets $\{Y_1, \cdots, Y_s\}$. Since the disjunction form of $\psi$ implies $Y_1 \cup \cdots \cup Y_s = Y$, this will demonstrate that $\{Y_1, \cdots, Y_s\}$ form a partition of $Y$.

Let $\psi_1(x,X_1,\cdots,X_m) = C_2 \lor \cdots \lor C_s$, then we have $$\psi(x,X_1,\cdots,X_m) = C_1 \lor \psi_1(x,X_1,\cdots,X_m).$$ We augment $C_1 \lor \psi_1(x,X_1,\cdots,X_m)$ as $C_1 \lor ((C_1 \land \neg C_1) \lor \psi_1(x,X_1,\cdots,X_m))$, which can expand into $C_1 \lor (C_1 \land \psi_1(x,X_1,\cdots,X_m)) \lor (\neg C_1 \land \psi_1(x,X_1,\cdots,X_m))$. Given that $C_1 \land \psi_1(x,X_1,\cdots,X_m) = 1$ necessitate $C_1 = 1$, we have $C_1 \lor (C_1 \land \psi_1(x,X_1,\cdots,X_m)) = C_1$, hence $$\psi(x,X_1,\cdots,X_m) = C_1 \lor (\neg C_1 \land \psi_1(x,X_1,\cdots,X_m)).$$
By repeating this process for all $C_i$ ($1 \le i \le s$), we obtain $$\psi(x,X_1,\cdots,X_m) = C_1 \lor (\neg C_1 \land C_2) \lor \cdots \lor (\neg C_1 \land \neg C_2 \land \cdots \land \neg C_{s-1} \land C_s).$$
We denote as $\psi(x,X_1,\cdots,X_m) = C'_1 \lor C'_2 \cdots \lor C'_s$. For any two distinct $C'_i$ and $C'_k$, it is easy to see that $C'_i \land C'_k = 0$, so the sets $Y_i$ and $Y_k$ represented by $C'_i$ and $C'_k$ satisfy that $Y_i \cap Y_k = \emptyset$. Thus each $C'_i$ represents a disjoint set.


Finally, we prove that each $C'_i$ can be reduced to be set-separable with respect to $X_j$.
By definition, $C'_i$ is the conjunction of negations of previous clauses $C_k$ ($1 \le k < i$) and $C_i$: $$C'_i = \bigwedge_{k=1}^{i-1} \neg C_k \land C_i.$$
Since $C_i$ can be written as $(x \in X_j) \land D_i$, where $D_i$ is the remaining part of the conjunctive clause. Substituting this into $C'_i$, we get $$C'_i = \bigwedge_{j=1}^{i-1} \neg C_j \land (x \in X_j) \land D_i = (x \in X_j) \land D'_i.$$ 

At this point, $D'_i$ may also contain atomic propositions relevant to $X_j$. We now show how to eliminate these redundant atomic propositions:
A key observation is that for $C'_i = 1$ to hold, the condition $x \in X_j$ must be true. Therefore, we reduce $C'_i$ by assigning a truth value of 1 to all terms of the form $x \in X_j$ and a truth value of 0 to all terms of the form $x \notin X_j$. After this reduction, we obtain a reduced formula $Q_i = (x \in X_j) \land Q'_i$, which is equivalent to $C'_i$ but $Q'_i$ contains no atomic propositions relevant to $X_j$. Thus, $Q_i$ is set-separable with respect to $X_j$. The proof is complete.
\end{proof}

\section{The Proof of Theorem~\ref{security}}\label{appdix:security}

\begin{proof} We prove the theorem by induction on the number of parties corrupted by the adversary $\mathcal{A}$. 
  
\begin{trivlist}
\item[$-$] \textbf{Base Case: $\mathbf{t = m-1}$}

Assume $\mathcal{A}$ corrupts $t = m-1$ parties. Denote the set of corrupted parties as $\textbf{P}_{\mathcal{A}} = \{P_{i_1}, \cdots, P_{i_{m-1}}\}$, leaving only one honest party $P_h$. According to the privacy requirement, there exists a simulator $\mathsf{Sim}$ such that
\begin{gather*}
    \{\mathsf{Sim}(\textbf{P}_{\mathcal{A}},\textbf{x}_{\mathcal{A}},\textbf{s}_{\mathcal{A}})\}_{\textbf{x}} \overset{c}{\approx} \{\mathsf{View}_{\mathcal{A}}^\Pi(\textbf{x})\}_{\textbf{x}}
\end{gather*}

We use $\textbf{r}$ (resp. $\textbf{r}^\Pi$) to denote the randomness in $f(\textbf{x})$ in the ideal (resp. real) execution. It is easy to see that in the ideal execution, $\textbf{r}$ is independent of $\mathsf{Sim}(\textbf{P}_{\mathcal{A}},\textbf{x}_{\mathcal{A}},\textbf{s}_{\mathcal{A}})$. Meanwhile, by the independence requirement, $\textbf{r}^\Pi$ is independent of $\mathsf{View}_{\mathcal{A}}^\Pi(\textbf{x})$ in the real execution, so we can obtain 
\begin{gather*}
    \{\mathsf{Sim}(\textbf{P}_{\mathcal{A}},\textbf{x}_{\mathcal{A}},\textbf{s}_{\mathcal{A}}),\textbf{r}\}_{\textbf{x}} \overset{c}{\approx} \{\mathsf{View}_{\mathcal{A}}^\Pi(\textbf{x}),\textbf{r}^\Pi\}_{\textbf{x}}.
\end{gather*}
We further extend the indistinguishability into
\begin{gather*}
    \{\mathsf{Sim}(\textbf{P}_{\mathcal{A}},\textbf{x}_{\mathcal{A}},\textbf{s}_{\mathcal{A}}),\textbf{s}_{\mathcal{A}}, \textbf{r}\}_{\textbf{x}} \overset{c}{\approx} \{\mathsf{View}_{\mathcal{A}}^\Pi(\textbf{x}),\textbf{s}^\Pi_{\mathcal{A}}, \textbf{r}^\Pi\}_{\textbf{x}}
\end{gather*}
where $\textbf{s}^\Pi_{\mathcal{A}} = (s^\Pi_{i_1}, \cdots, s^\Pi_{i_{m-1}})$, since each corrupted party $P_i$'s output $s_{i}$ (resp. $s^\Pi_{i}$) can be computed from its own view in the ideal (resp. real) execution ($i \in \{i_1,\cdots,i_{m-1}\}$).

By the functionality, the output of $P_h$ satisfies $s_h = -(\sum_{s_i \in \textbf{s}_{\mathcal{A}}} s_i) + f(\textbf{x})$. By the correctness requirement, $s^\Pi_h = -(\sum_{s^\Pi_i \in \textbf{s}^\Pi_{\mathcal{A}}} s^\Pi_i) + f(\textbf{x})$. Thus, we extend the previous distributions by including $s_h$ and $s^\Pi_h$ respectively and obtain
\begin{gather*}
    \{\mathsf{Sim}(\textbf{P}_{\mathcal{A}},\textbf{x}_{\mathcal{A}},\textbf{s}_{\mathcal{A}}),\textbf{s}_{\mathcal{A}},s_h, \textbf{r}\}_{\textbf{x}} \overset{c}{\approx} \{\mathsf{View}_{\mathcal{A}}^\Pi(\textbf{x}),\textbf{s}^\Pi_{\mathcal{A}},s^\Pi_h, \textbf{r}^\Pi\}_{\textbf{x}}.
\end{gather*}
The indistinguishability holds because $s_h$ (resp. $s^\Pi_h$) is determined by $\textbf{s}_{\mathcal{A}}$ (resp. $\textbf{s}^\Pi_{\mathcal{A}}$), the randomness $\textbf{r}$ (resp. $\textbf{r}^\Pi$), and the parties' inputs $\textbf{x}$. This implies 
\begin{gather*}
    \{\mathsf{Sim}(\textbf{P}_{\mathcal{A}},\textbf{x}_{\mathcal{A}},\textbf{s}_{\mathcal{A}}),\textbf{s}_{\mathcal{A}},s_h\}_{\textbf{x}} \overset{c}{\approx} \{\mathsf{View}_{\mathcal{A}}^\Pi(\textbf{x}),\textbf{s}^\Pi_{\mathcal{A}},s^\Pi_h\}_{\textbf{x}}.
\end{gather*}
Namely, $\Pi$ securely computes $f$ when $\mathcal{A}$ corrupting $t = m-1$ parties. Note that the independence requirement in this case implies that $\textbf{r}^\Pi$ is independent of the joint view of any $m-1$ parties in the real execution, which will be used in the subsequent proof.

\item[$-$] \textbf{Inductive Hypothesis: $\mathbf{t = m-k}$}

Assume that for any adversary $\mathcal{A}$ corrupting $t = m-k$ parties ($1 \le k < m-1$), $\Pi$ securely computes $f$. Namely, there exists a simulator $\mathsf{Sim}$ such that
\begin{gather*}
    \{\mathsf{Sim}(\{P_1, \cdots, P_m\} \setminus \textbf{P}_\mathcal{H},\{x_{1},\cdots,x_{m}\} \setminus \textbf{x}_\mathcal{H},\{s_{1},\cdots,s_{m}\} \setminus \textbf{s}_\mathcal{H}),\{s_{1},\cdots,s_{m}\} \setminus \textbf{s}_\mathcal{H}, \textbf{s}_\mathcal{H}\}_{\textbf{x}} \\ \overset{c}{\approx} \{\{\mathsf{View}_{1}^\Pi(\textbf{x}),\cdots,\mathsf{View}_{m}^\Pi(\textbf{x})\} \setminus \{\mathsf{View}_{\mathcal{H}}^\Pi(\textbf{x})\},\{s^\Pi_{1},\cdots,s^\Pi_{m}\} \setminus \textbf{s}^\Pi_{\mathcal{H}}, \textbf{s}^\Pi_{\mathcal{H}}\}_{\textbf{x}},
\end{gather*}
where $\textbf{P}_\mathcal{H}$ is the set of honest parties, $\mathsf{View}_{\mathcal{H}}^\Pi(\textbf{x})$ denotes the joint view of $\textbf{P}_\mathcal{H}$, while $\textbf{s}_{\mathcal{H}}$ and $\textbf{s}^\Pi_{\mathcal{H}}$ are the respective outputs in the ideal and real executions.

It is easy to see that in the ideal execution, $\textbf{s}_{\mathcal{H}}$ is independent of the joint distribution $\{\mathsf{Sim}(\{P_1, \cdots, P_m\} \setminus \textbf{P}_\mathcal{H},\{x_{1},\cdots,x_{m}\} \setminus \textbf{x}_\mathcal{H},\{s_{1},\cdots,s_{m}\} \setminus \textbf{s}_\mathcal{H}),\{s_{1},\cdots,s_{m}\} \setminus \textbf{s}_\mathcal{H}\}$. Thus, we can conclude that for any subset of parties $\textbf{P}_\mathcal{H} \subset \{P_1, \cdots, P_m\}$ of size $k$, $\textbf{s}^\Pi_{\mathcal{H}}$ is independent of the joint distribution $\{\{\mathsf{View}_{1}^\Pi(\textbf{x}),\cdots,\mathsf{View}_{m}^\Pi(\textbf{x})\} \setminus \mathsf{View}_{\mathcal{H}}^\Pi(\textbf{x}),\{s^\Pi_{1},\cdots,s^\Pi_{m}\} \setminus \textbf{s}^\Pi_{\mathcal{H}}\}$ in the real execution.

\item[$-$] \textbf{Inductive Step: $\mathbf{t = m-k-1}$}

We proceed to prove the case where $\mathcal{A}$ corrupts $t = m-k-1$ parties:

Let $\textbf{P}_{\mathcal{H}'}$ represent a subset of $\{P_1, \cdots, P_m\}$ with size $k+1$. We decompose it into $\textbf{P}_{\mathcal{H}'} = \{\textbf{P}_{\mathcal{H}}, P_h\}$, where $\textbf{P}_\mathcal{H} \subset \{P_1, \cdots, P_m\}$ contains exact $k$ parties while $P_h$ is the remaining one party. By the privacy, we have $$\{\mathsf{Sim}(\textbf{P}_{\mathcal{A}},\textbf{x}_{\mathcal{A}},\textbf{s}_{\mathcal{A}}),\textbf{s}_{\mathcal{A}}\}_{\textbf{x}} \overset{c}{\approx} \{\mathsf{View}_{\mathcal{A}}^\Pi(\textbf{x}),\textbf{s}^\Pi_{\mathcal{A}}\}_{\textbf{x}}.$$ By the correctness, we also have $$\{\textbf{s}_{\mathcal{H}}\}_{\textbf{x}} \overset{c}{\approx} \{\textbf{s}^\Pi_{\mathcal{H}}\}_{\textbf{x}}.$$

From the inductive hypothesis, $\textbf{s}_\mathcal{H}$ is independent of the joint distribution $\{\mathsf{View}_{\mathcal{A}}^\Pi(\textbf{x}),\textbf{s}^\Pi_{\mathcal{A}}\}$, given that it is a subdistribution of $\{\{\mathsf{View}_{1}^\Pi(\textbf{x}),\cdots,\mathsf{View}_{m}^\Pi(\textbf{x})\} \setminus \mathsf{View}_{\mathcal{H}}^\Pi(\textbf{x}),\{s^\Pi_{1},\cdots,s^\Pi_{m}\} \setminus \textbf{s}^\Pi_{\mathcal{H}}\}$. 
It is easy to see that $\textbf{s}_{\mathcal{H}}$ is independent of the joint distribution $\{\mathsf{Sim}(\textbf{P}_{\mathcal{A}},\textbf{x}_{\mathcal{A}},\textbf{s}_{\mathcal{A}}),\textbf{s}_{\mathcal{A}}\}_{\textbf{x}}$. 
Combining the above, 
$$\{\mathsf{Sim}(\textbf{P}_{\mathcal{A}},\textbf{x}_{\mathcal{A}},\textbf{s}_{\mathcal{A}}),\textbf{s}_{\mathcal{A}},\textbf{s}_{\mathcal{H}}\}_{\textbf{x}} \overset{c}{\approx} \{\mathsf{View}_{\mathcal{A}}^\Pi(\textbf{x}),\textbf{s}^\Pi_{\mathcal{A}},\textbf{s}^\Pi_{\mathcal{H}}\}_{\textbf{x}}.$$

Recall that in the base case we derived that $\textbf{r}^\Pi$ is independent of the joint view of any $m-1$ parties in the real execution, thereby, $\textbf{r}^\Pi$ is independent of $\{\mathsf{View}_{\mathcal{A}}^\Pi(\textbf{x}),\mathsf{View}_{\mathcal{H}}^\Pi(\textbf{x})\}$. As $\textbf{s}^\Pi_{\mathcal{A}}$ and $\textbf{s}^\Pi_{\mathcal{H}}$ can be determined by $\mathsf{View}_{\mathcal{A}}^\Pi(\textbf{x})$ and $\mathsf{View}_{\mathcal{H}}^\Pi(\textbf{x})$ respectively, $\textbf{r}^\Pi$ is independent of $\{\mathsf{View}_{\mathcal{A}}^\Pi(\textbf{x}),\textbf{s}^\Pi_{\mathcal{A}},\textbf{s}^\Pi_{\mathcal{H}}\}$. Furthermore, in the ideal execution, $\textbf{r}$ is independent of $\{\mathsf{Sim}(\textbf{P}_{\mathcal{A}},\textbf{x}_{\mathcal{A}},\textbf{s}_{\mathcal{A}}),\textbf{s}_{\mathcal{A}},\textbf{s}_{\mathcal{H}}\}$, which can extend the previous indistinguishability into
$$\{\mathsf{Sim}(\textbf{P}_{\mathcal{A}},\textbf{x}_{\mathcal{A}},\textbf{s}_{\mathcal{A}}),\textbf{s}_{\mathcal{A}},\textbf{s}_{\mathcal{H}}, \textbf{r}\}_{\textbf{x}} \overset{c}{\approx} \{\mathsf{View}_{\mathcal{A}}^\Pi(\textbf{x}),\textbf{s}^\Pi_{\mathcal{A}},\textbf{s}^\Pi_{\mathcal{H}}, \textbf{r}^\Pi\}_{\textbf{x}}.$$

By the functionality, the output of $P_h$ satisfies $s_{h} = -(\sum_{s_i \in \textbf{s}_{\mathcal{A}}} s_i + \sum_{s_i \in \textbf{s}_{\mathcal{H}}} s_i) + f(\textbf{x})$. 
By the correctness, $s^\Pi_{h} = -(\sum_{s^\Pi_i \in \textbf{s}^\Pi_{\mathcal{A}}} s^\Pi_i + \sum_{s^\Pi_i \in \textbf{s}^\Pi_{\mathcal{H}}} s^\Pi_i) + f(\textbf{x})$. 
Thus, we extend the previous distributions by including $s_h$ and $s^\Pi_h$ respectively and obtain
$$\{\mathsf{Sim}(\textbf{P}_{\mathcal{A}},\textbf{x}_{\mathcal{A}},\textbf{s}_{\mathcal{A}}),\textbf{s}_{\mathcal{A}},\textbf{s}_{\mathcal{H}},s_{h},\textbf{r}\}_{\textbf{x}} \overset{c}{\approx} \{\mathsf{View}_{\mathcal{A}}^\Pi(\textbf{x}),\textbf{s}^\Pi_{\mathcal{A}},\textbf{s}^\Pi_{\mathcal{H}},s^\Pi_{h},\textbf{r}^\Pi\}_{\textbf{x}}$$
The indistinguishability holds because $s_h$ (resp. $s^\Pi_h$) is uniquely determined by $\textbf{s}_{\mathcal{A}}$ and $\textbf{s}_{\mathcal{H}}$ (resp. $\textbf{s}^\Pi_{\mathcal{A}}$ and $\textbf{s}^\Pi_{\mathcal{H}}$), the randomness $\textbf{r}$ (resp. $\textbf{r}^\Pi$), and the parties' inputs $\textbf{x}$. This implies 
\begin{gather*}
    \{\mathsf{Sim}(\textbf{P}_{\mathcal{A}},\textbf{x}_{\mathcal{A}},\textbf{s}_{\mathcal{A}}),\textbf{s}_{\mathcal{A}},s_h\}_{\textbf{x}} \overset{c}{\approx} \{\mathsf{View}_{\mathcal{A}}^\Pi(\textbf{x}),\textbf{s}^\Pi_{\mathcal{A}},s^\Pi_h\}_{\textbf{x}}.
\end{gather*}
Namely, $\Pi$ securely computes $f$ in the presence of $\mathcal{A}$ corrupting $t = m-k-1$ parties. This completes the inductive step.

\end{trivlist}
\end{proof}


\section{The Optimization in Section~\ref{transform-pzs}}\label{beaver triples}

We first recall the technique of Beaver triples. A Beaver triple consists of three secret-sharings $([a], [b], [c])$, where $[a], [b]$ are random secret-sharings and $c = a \cdot b$. Typically, a Beaver triple is used to reduce one multiplication to two reconstructions in the online phase, while here since the multiplier $b$ is random, a Beaver triple can be used to reduce one multiplication to one reconstruction in the online phase. Concretely,
\begin{gather*}
s = r \cdot b = (r+a-a) \cdot b = (r+a) \cdot b + a \cdot b
\end{gather*}
hence we can compute
\begin{gather*}
[s] = (r+a) \cdot [b] + [a \cdot b] = (r+a) \cdot [b] + [c]
\end{gather*}
The above equation suggests that we can locally compute $[s]$ once $u = r + a$ is publicly known. Therefore, the task of generating $[s]$ boils down to reconstructing $[u] = [r] + [a]$. Let each party $P_i$ locally compute $u_i = r_i + a_i$ and send $u_i$ to the leader $P_1$, then $P_1$ computes $u = u_1 + \cdots + u_m$ and opens it to all parties. As we can see, the transformation only consumes one Beaver triple generated in the offline phrase and requires one opening with $2 (m-1) \log_2 \log_2\lvert \mathbb{F} \rvert$ communication overhead for the leader and $2 \log_2 \log_2\lvert \mathbb{F} \rvert$ communication overhead for each client in the online phrase. 

\section{Membership Zero-Sharing Appendix}\label{MZS-appendix}
\subsection{Pure Membership Zero-Sharing}\label{bpMZS}

The (batch) pure membership zero-sharing functionality is a special case of (batch) membership zero-sharing when $Q$ is a conjunction of $m-1$ set membership predicates (i.e., $\bigwedge_{j \in \{1,\cdots,m\} \setminus \{\mathsf{pivot}\}} x \in X_j$). 
The ideal functionality $\FuncbpMZS$ is formally described in Figure~\ref{fig:func-bpmzs}.
The complete protocol is given in Figure~\ref{fig:proto-bpmzs}.

\begin{figure}[!hbtp]
\begin{framed}
\begin{minipage}[center]{\textwidth}
\begin{trivlist}
\item \textbf{Parameters:} $m$ parties $P_1, \cdots P_m$, where $P_\mathsf{pivot}$ is the only party holding $n$ single elements as inputs instead of $n$ sets. Batch size $n$. A field $\mathbb{F}$. 
\item \textbf{Functionality:} On input $\textbf{x} = (x_1, \cdots, x_n)$ from $P_\mathsf{pivot}$ and $\textbf{X}_j = (X_{j,1}, \cdots, X_{j,n})$ from each $P_j$ ($j \in \{1,\cdots m\} \setminus \{\mathsf{pivot}\}$), sample $\textbf{s}_i = (s_{i,1}, \cdots, s_{i,n}) \gets \mathbb{F}^n$ for $1 \le i \le m$, s.t. for $1 \le d \le n$, if $\bigwedge_{j \in \{1,\cdots,m\} \setminus \{\mathsf{pivot}\}} (x_d \in X_{j,d}) = 1$, $\sum_{1 \le i \le m} s_{i,d} = 0$. Give $\textbf{s}_i$ to $P_i$.
\end{trivlist}
\end{minipage}
\end{framed}
\caption{Batch Pure Membership Zero-Sharing Functionality $\FuncbpMZS$}\label{fig:func-bpmzs}
\end{figure}

\begin{figure}[!hbtp]
\begin{framed}
\begin{minipage}[center]{\textwidth}
\begin{trivlist}
\item \textbf{Parameters:} $m$ parties $P_1, \cdots P_m$. Batch size $n$. A field $\mathbb{F}$. $n$ Beaver triples $([\textbf{a}], [\textbf{b}], [\textbf{c}])$ generated in the offline phrase, where $[\textbf{a}] = ([a_1], \cdots, [a_n])$, $[\textbf{b}] = ([b_1], \cdots, [b_n])$, $[\textbf{c}] = ([c_1], \cdots, [c_n])$ and $c_i = a_i \cdot b_i$ for $1 \le i \le n$.
\item \textbf{Inputs:} $P_\mathsf{pivot}$ inputs a vector $\textbf{x} = (x_1, \cdots, x_n)$. $P_j$ inputs $\textbf{X}_j = (X_{j,1}, \cdots, X_{j,n})$ for $j \in \{1,\cdots m\} \setminus \{\mathsf{pivot}\}$.
\item \textbf{Protocol:}
\begin{enumerate}[itemsep=2pt,topsep=0pt,parsep=0pt]
\item For the $i$-th instance ($1 \le i \le n$), $P_j$ samples $r_{j,i}$ and sets $K_{j,i} = X_{j,i}$ and $V_{j,i} = \{-r_{j,i}, \cdots, -r_{j,i}\}$, where $\lvert K_{j,i} \rvert = \lvert V_{j,i} \rvert$. 
\item $P_\mathsf{pivot}$ and $P_j$ invoke $\FuncbOPPRF$ where $P_j$ acts as $\Sd$ inputting $(K_{j,1},\cdots,K_{j,n})$ and $(V_{j,1},\cdots,V_{j,n})$, and $P_\mathsf{pivot}$ acts as $\Rcv$ with input $\textbf{x}$ and receives $\textbf{u}_{j}$.
\item $P_\mathsf{pivot}$ sets its shares $\textbf{r}_{\mathsf{pivot}} = \sum_{j\in \{1,\cdots m\} \setminus \{\mathsf{pivot}\}} \textbf{u}_{j}$. $P_j$ sets its shares $\textbf{r}_{j} = (r_{j,1}, \cdots, r_{j,n})$. All parties hold a vector of $n$ secret-sharings $[\textbf{r}] = ([r_1], \cdots, [r_n])$.
\item All parties compute $[\textbf{s}]$ by performing $n$ secure multiplications $[s_i] = [r_i] \cdot [b_i]$ ($1 \le i \le n$), using $n$ Beaver triples $([\textbf{a}], [\textbf{b}], [\textbf{c}])$.
\end{enumerate}
\end{trivlist}
\end{minipage}
\end{framed}
\caption{Batch Pure Membership Zero-Sharing $\ProtobpMZS$}\label{fig:proto-bpmzs}
\end{figure}

\begin{trivlist}
\item \textbf{Complexity Analysis.} In the batch pure membership zero-sharing protocol (Figure~\ref{fig:func-bpmzs}), the costs of each stage are calculated as follows.
\begin{itemize}
  \item $P_\mathsf{pivot}$ executes batch OPPRF of size $n$ with each $P_j$ for $j \in \{1,\cdots m\} \setminus \{\mathsf{pivot}\}$. Suppose that in the subsequent invocations of batch OPPRF, each $\lvert X_{j,i} \rvert = N_{j,i} = O(1)$ for $1 \le i \le n, j \in \{1,\cdots m\} \setminus \{\mathsf{pivot}\}$, which is consistent with the use of batch membership zero-sharing protocols in our MPSO protocols (combined with hashing to bins technique). We follow the paradigm in~\cite{PSTY-EUROCRYPT-2019} to construct batch OPPRF from batch OPRF and OKVS. By leveraging the technique to amortize communication, the total communication of computing $n$ instances of OPPRF is equal to the total number of items $3n$. Furthermore, we utilize vector oblivious linear evaluation (VOLE)~\cite{BCGIKRS19,BCGIKS19,RRT23} to instantiate batch OPRF and the construction in~\cite{RR-CCS-2022} to instantiate OKVS. This ensures the computation complexity of batch OPPRF of size $n$ to scale linearly with $n$. Therefore, in this stage, the computation and communication complexity of $P_\mathsf{pivot}$ are $O(mn)$, while the computation and communication complexity of each $P_j$ are $O(n)$.
  \item The parties perform $n$ secure multiplications using the optimization outlined in the previous section and designate $P_\mathsf{pivot}$ as the leader. This requires $n$ opening with $O(mn)$ computation/communication complexity for $P_\mathsf{pivot}$ and $O(n)$ computation/communication complexity for each $P_j$. 
\end{itemize}

To sum up, in the online phase of the batch pure membership zero-sharing protocol, the computation and communication complexity of $P_\mathsf{pivot}$ are $O(mn)$, while the computation and communication complexity of each $P_j$ are $O(n)$.
\end{trivlist}

\subsection{Pure Non-Membership Zero-Sharing}\label{bpMZS}

The (batch) pure non-membership zero-sharing functionality is a special case of (batch) membership zero-sharing when $Q$ is a conjunction of $m-1$ set non-membership predicates (i.e., $\bigwedge_{j \in \{1,\cdots,m\} \setminus \{\mathsf{pivot}\}} x \notin X_j$). 
The ideal functionality $\FuncbpNMZS$ is formally described in Figure~\ref{fig:func-bpnmzs}.
The complete protocols is given in Figure~\ref{fig:proto-bpnmzs}.

\begin{figure}[!hbtp]
\begin{framed}
\begin{minipage}[center]{\textwidth}
\begin{trivlist}
\item \textbf{Parameters:} $m$ parties $P_1, \cdots P_m$, where $P_\mathsf{pivot}$ is the only party holding $n$ single elements as inputs instead of $n$ sets. Batch size $n$. A field $\mathbb{F}$. 
\item \textbf{Functionality:} On input $\textbf{x} = (x_1, \cdots, x_n)$ from $P_\mathsf{pivot}$ and $\textbf{X}_j = (X_{j,1}, \cdots, X_{j,n})$ from each $P_j$ ($j \in \{1,\cdots m\} \setminus \{\mathsf{pivot}\}$), sample $\textbf{s}_i = (s_{i,1},\cdots,s_{i,n}) \gets \mathbb{F}^n$ for $1 \le i \le m$, s.t. for $1 \le d \le n$, if $\bigwedge_{j \in \{1,\cdots,m\} \setminus \{\mathsf{pivot}\}} x_d \notin X_{j,d} = 1$, $\sum_{1 \le i \le m} s_{i,d} = 0$. Give $\textbf{s}_i$ to $P_i$.
\end{trivlist}
\end{minipage}
\end{framed}
\caption{Batch Pure Non-Membership Zero-Sharing Functionality $\FuncbpNMZS$}\label{fig:func-bpnmzs}
\end{figure}

\begin{figure}[!hbtp]
\begin{framed}
\begin{minipage}[center]{\textwidth}
\begin{trivlist}
\item \textbf{Parameters:} $m$ parties $P_1, \cdots P_m$. Batch size $n$. A field $\mathbb{F}$. $n$ Beaver triples $([\textbf{a}], [\textbf{b}], [\textbf{c}])$ generated in the offline phrase, where $[\textbf{a}] = ([a_1], \cdots, [a_n])$, $[\textbf{b}] = ([b_1], \cdots, [b_n])$, $[\textbf{c}] = ([c_1], \cdots, [c_n])$ and $c_i = a_i \cdot b_i$ for $1 \le i \le n$.
\item \textbf{Inputs:} $P_\mathsf{pivot}$ inputs a vector $\textbf{x} = (x_1, \cdots, x_n)$. $P_j$ inputs $\textbf{X}_j = (X_{j,1}, \cdots, X_{j,n})$ for $j \in \{1,\cdots m\} \setminus \{\mathsf{pivot}\}$.
\item \textbf{Protocol:}
\begin{enumerate}[itemsep=2pt,topsep=0pt,parsep=0pt]
\item $P_\mathsf{pivot}$ and $P_j$ invoke $\FuncbssPMT$ where in the $i$-th instance ($1 \le i \le n$), $P_j$ inputs $X_{j,i}$ and receives $e_{j,i}^0$, while $P_\mathsf{pivot}$ inputs $x_i$ and receives $e_{j,i}^1$. 
\item $P_\mathsf{pivot}$ and $P_j$ invoke $n$ instances of ROT where in the $i$-th instance ($1 \le i \le n$), $P_j$ acts as $\Sd$ and receives $r_{j,i}^0,r_{j,i}^1$, while $P_\mathsf{pivot}$ acts as $\Rcv$ inputting $e_{j,i}^1$ and receives $r_{j,i}^{e_{j,i}^1}$. $P_\mathsf{pivot}$ sets $\textbf{r}'_{j} = (r_{j,1}^{e_{j,1}^1},\cdots,r_{j,n}^{e_{j,n}^1})$. 
\item $P_\mathsf{pivot}$ sets its shares $\textbf{r}_{\mathsf{pivot}} = \sum_{j\in \{1,\cdots m\} \setminus \{\mathsf{pivot}\}} \textbf{r}'_{j}$. $P_j$ sets its shares $\textbf{r}_{j} = (-r_{j,1}^{e_{j,1}^0}, \cdots, -r_{j,n}^{e_{j,n}^0})$. All parties hold a vector of $n$ secret-sharings $[\textbf{r}] = ([r_1], \cdots, [r_n])$.
\item All parties compute $[\textbf{s}]$ by performing $n$ secure multiplications $[s_i] = [r_i] \cdot [b_i]$ ($1 \le i \le n$), using $n$ Beaver triples $([\textbf{a}], [\textbf{b}], [\textbf{c}])$.
\end{enumerate}
\end{trivlist}
\end{minipage}
\end{framed}
\caption{Batch Pure Non-Membership Zero-Sharing $\ProtobpNMZS$}\label{fig:proto-bpnmzs}
\end{figure}

\begin{trivlist}
\item \textbf{Complexity Analysis.} In the batch pure non-membership zero-sharing protocol (Figure~\ref{fig:func-bpnmzs}), the costs of each stage are calculated as follows.
\begin{itemize}
  \item $P_\mathsf{pivot}$ executes batch ssPMT of size $n$ with each $P_j$ for $j \in \{1,\cdots m\} \setminus \{\mathsf{pivot}\}$. We utilize the construction in~\cite{DongCZB24} based on batch OPPRF and secret-shared private equality test (ssPEQT)~\cite{PSTY-EUROCRYPT-2019,CGS22}, which achieves linear computation and communication complexity. Therefore, in this stage, the computation and communication complexity of $P_\mathsf{pivot}$ are $O(mn)$, while the computation and communication complexity of each $P_j$ are $O(n)$.
  \item $P_\mathsf{pivot}$ acts as $\Rcv$ and executes $n$ instances of ROT with each $P_j$ for $j \in \{1,\cdots m\} \setminus \{\mathsf{pivot}\}$. In the offline phases, $P_\mathsf{pivot}$ and each $P_j$ generate $n$ instances of random-choice-bit ROT, then in the online phase, $P_\mathsf{pivot}$ only needs to send $n$ choice bits masked by the random choice bits to each $P_j$. Therefore, the computation and communication complexity of $P_\mathsf{pivot}$ are $O(mn)$, while the computation and communication complexity of each $P_j$ are $O(n)$.
  \item The parties perform $n$ secure multiplications using the optimization outlined in the previous section and designate $P_\mathsf{pivot}$ as the leader. This requires $n$ opening with $O(mn)$ computation/communication complexity for $P_\mathsf{pivot}$ and $O(n)$ computation/communication complexity for each $P_j$. 
\end{itemize}

To sum up, in the online phase of the batch pure non-membership zero-sharing protocol, the computation and communication complexity of $P_\mathsf{pivot}$ are $O(mn)$, while the computation and communication complexity of each $P_j$ are $O(n)$.
\end{trivlist}

\subsection{Pure Membership Zero-Sharing with Payloads}\label{bpMZSpayloads}

Pure membership zero-sharing with payloads is an extension of the pure membership zero-sharing functionality, combined with a variant of relaxed pure membership payload-sharing, which we call relaxed pure membership payload-sharing. In this variant, $P_\mathsf{pivot}$ holds an element $x$ while each of the others holds a set of elements and a set of associated payloads. If the conjunction of set membership predicates holds true (i.e., $x$ belongs to all element sets), the parties receive secret shares of the sum of all payloads associated with $x$; otherwise they receive secret shares of a random value. The formal definition of batch pure membership zero-sharing with payloads functionality is in Figure~\ref{fig:func-mzsp}. Note that the payload-sharing only needs to satisfy the relaxed security definition in Section~\ref{relaxed-pzs}.

The construction of batch pure membership zero-sharing with payloads protocol resembles the batch pure membership zero-sharing protocol in Figure~\ref{fig:proto-bpmzs}. The core idea is to somehow encode the payload set into the senders' inputs of OPPRF, in each two-party protocol of the relaxed pure membership payload-sharing. 
Specifically, we start by implementing the relaxed pure membership zero-sharing with payloads in the two-party setting. Next, we show how to extend this primitive into multi-party setting. 

In the two-party relaxed membership zero-sharing with payloads protocol, there are two parties, the sender $\Sd$ with an element set $Y$ and a payload set $V$ and the receiver $\Rcv$ with an element $x$. $\Sd$ samples two secret shares $r, w$, and sets $Y$ as the key set and a set containing the pair $(-r, v_i - w)$ for $1 \le i \le n$ as the value set, where $v_i \in V$ is the associated payload with $y_i \in Y$. $\Sd$ outputs $(r, w)$ as its two secret shares. $\Sd$ and $\Rcv$ invoke OPPRF, where $\Rcv$ inputs $x$ and receives $(r', w')$ as its secret share. By the definition of OPPRF, if $x \in Y$, $(r', w') = (-r, v - w)$, where $v$ is the associated payload with $x$ in $Y$. Namely, if $x \in Y$, the parties hold one secret sharing of 0 and one secret sharing of the associated payload with $x$, otherwise they hold two secret sharings of pseudorandom values. 

In the multi-party membership zero-sharing with payloads protocol, there are $m$ ($m>2$) parties, where $P_\mathsf{pivot}$ holds an element $x$ and each $P_j$ ($j \in \{1,\cdots m\} \setminus \{\mathsf{pivot}\}$) holds an element set $X_j$ and a payload set $V_j$. $P_\mathsf{pivot}$ engages in the two-party version with each $P_j$, where $P_\mathsf{pivot}$ receives $r'_j$ and $w'_j$ while $P_j$ receives $r_j$ and $w_j$. By definition, we have that if $x \in X_j$, $r_j + r'_j = 0$ and $w_j + w'_j = v_j$, where $v_j$ is the associated payload with $x$ in $V_j$; otherwise $r_j + r'_j$ and $w_j + w'_j$ are both random values. $P_\mathsf{pivot}$ sets $r_\mathsf{pivot} = \sum_{j \in \{1,\cdots m\} \setminus \{\mathsf{pivot}\}} t_j$ as its first secret share and sets $w_\mathsf{pivot} = \sum_{j \in \{1,\cdots m\} \setminus \{\mathsf{pivot}\}} u_j$ as its second secret share. Meanwhile, $P_j$ sets $r_j$ as its first secret share and $w_j$ as its second secret share. Note that if and only if $x \in X_j$ for all $j$, $\sum_{1 \le i \le m} r_i = 0$ and $\sum_{1 \le i \le m} w_i = \sum_{j \in \{1,\cdots m\} \setminus \{\mathsf{pivot}\}} v_j$, otherwise $\sum_{1 \le i \le m} r_i$ and $\sum_{1 \le i \le m} w_i$ are random values. At this point, the first secret-sharing is a relaxed pure membership zero-sharing while the second secret-sharing is relaxed pure membership payload-sharing. In order to realize the membership zero-sharing with payloads functionality, the last step is to transform the first relaxed pure membership zero-sharing into the standard. The complete batch version is provided in Figure~\ref{fig:proto-bpmzsp}.

\begin{trivlist}
\item \textbf{Complexity Analysis.} In the batch pure membership zero-sharing with payloads protocol (Figure~\ref{fig:proto-bpmzsp}), the costs of each stage are calculated as follows.
\begin{itemize}
  \item $P_\mathsf{pivot}$ executes batch OPPRF of size $n$ with each $P_j$ for $j \in \{1,\cdots m\} \setminus \{\mathsf{pivot}\}$. In this stage, the computation and communication complexity of $P_\mathsf{pivot}$ are $O(mn)$, while the computation and communication complexity of each $P_j$ are $O(n)$.
  \item The parties perform $n$ secure multiplications using the optimization outlined in the previous section and designate $P_\mathsf{pivot}$ as the leader. This requires $n$ opening with $O(mn)$ computation/communication complexity for $P_\mathsf{pivot}$ and $O(n)$ computation/ communication complexity for each $P_j$. 
\end{itemize}

To sum up, in the online phase of the batch pure membership zero-sharing with payloads protocol, the computation and communication complexity of $P_\mathsf{pivot}$ are $O(mn)$, while the computation and communication complexity of each $P_j$ are $O(n)$.
\end{trivlist}

\begin{figure}[!hbtp]
\begin{framed}
\begin{minipage}[center]{\textwidth}
\begin{trivlist}
\item \textbf{Parameters:} $m$ parties $P_1, \cdots P_m$. Batch size $n$. A field $\mathbb{F}$ and payload field $\mathbb{F'}$. The mapping function $\mathsf{payload}_j()$ from element sets $\textbf{X}_j$ to the associated payload sets $\textbf{V}_j$. $n$ Beaver triples $([\textbf{a}], [\textbf{b}], [\textbf{c}])$ generated in offline phrase, where $[\textbf{a}] = ([a_1], \cdots, [a_n])$, $[\textbf{b}] = ([b_1], \cdots, [b_n])$, $[\textbf{c}] = ([c_1], \cdots, [c_n])$ and $c_i = a_i \cdot b_i$ for $1 \le i \le n$.
\item \textbf{Inputs:} $P_\mathsf{pivot}$ inputs a vector $\textbf{x} = (x_1, \cdots, x_n)$. $P_j$ inputs $\textbf{X}_j = (X_{j,1}, \cdots, X_{j,n})$ and $\textbf{V}_j = (V_{j,1}, \cdots, V_{j,n})$ for $j \in \{1,\cdots m\} \setminus \{\mathsf{pivot}\}$.
\item \textbf{Protocol:}
\begin{enumerate}[itemsep=2pt,topsep=0pt,parsep=0pt]
\item For the $i$-th instance ($1 \le i \le n$), $P_j$ samples $(r_{j,i}, w_{j,i})$. Suppose $\lvert X_{j,i} \rvert = N_{j,i}$, $P_j$ sets $K_{j,i} = X_{j,i} = (x_{j,i,1}, \cdots, x_{j,i,N_{j,i}})$ and $V_{j,i}' = \{(-r_{j,i}, v_{j,i,1} - w_{j,i}), \cdots, (-r_{j,i}, v_{j,i, N_{j,i}} - w_{j,i})\}$, where $\lvert V_{j,i}' \rvert = N_{j,i}$ and $v_{j,i,k} = \mathsf{payload}_j (x_{j,i,k})$ for $1 \le k \le N_{j,i}$. 
\item $P_\mathsf{pivot}$ and $P_j$ invoke $\FuncbOPPRF$ where $P_j$ acts as $\Sd$ inputting $(K_{j,1},\cdots,K_{j,n})$ and $(V_{j,1}',\cdots,V_{j,n}')$, and $P_\mathsf{pivot}$ acts as $\Rcv$ with input $\textbf{x}$ and receives $(\textbf{r}_{j}',\textbf{w}_{j}')$.
\item $P_\mathsf{pivot}$ sets its first shares $\textbf{r}_{\mathsf{pivot}} = \sum_{j\in \{1,\cdots m\} \setminus \{\mathsf{pivot}\}} \textbf{r}_{j}'$, and its second shares $\textbf{w}_{\mathsf{pivot}} = \sum_{j\in \{1,\cdots m\} \setminus \{\mathsf{pivot}\}} \textbf{w}_{j}'$. $P_j$ sets its first shares $\textbf{r}_{j} = (r_{j,1}, \cdots, r_{j,n})$, and its second shares $\textbf{w}_{j} = (w_{j,1}, \cdots, w_{j,n})$. All parties hold two vectors of $n$ secret-sharings $[\textbf{r}] = ([r_1], \cdots, [r_n])$ and $[\textbf{w}] = ([w_1], \cdots, [w_n])$.
\item All parties compute $[\textbf{s}]$ by performing $n$ secure multiplications $[s_i] = [r_i] \cdot [b_i]$ ($1 \le i \le n$), using $n$ Beaver triples $([\textbf{a}], [\textbf{b}], [\textbf{c}])$.
\end{enumerate}
\end{trivlist}
\end{minipage}
\end{framed}
\caption{Batch Pure Membership Zero-Sharing with Payloads $\ProtobpMZSp$}\label{fig:proto-bpmzsp}
\end{figure}

\section{Security Proof of Theorem~\ref{theorem:mpso}}\label{appdix:proof-MPSO}

Let $\textbf{P}_{\mathcal{A}}$ denote the set of corrupted parties controlled by $\mathcal{A}$. In the MPSO protocol, the simulator receives each corrupted party's input $X_c$ from $P_c \in \textbf{P}_{\mathcal{A}}$ and if $P_1 \in \textbf{P}_{\mathcal{A}}$, it receives the resulting set $Y$. For each $P_c$, its view consists of its input $X_c$, $B$ secret shares $\textbf{s}_{i, c}$ from each $\FuncbMZS^{Q'_i}$ for $1 \le i \le s$ (if $P_c$ belongs to the set of $Q_i$'s involving parties $\{P_{i_1},\cdots,P_{i_q}\}$), shuffled secret shares $\textbf{u}_{c}'$ from $\FuncMS$, and if $P_1 \in \textbf{P}_{\mathcal{A}}$, reconstruction messages $\textbf{u}_j'$ from $P_j$ for $1 < j \le m$. 

Suppose there are $s'$ subformulas $\textbf{Q}_{\mathcal{A}} = \{Q_{j_1},\cdots,Q_{j_{s'}}\} \subseteq \{Q_1,\cdots,Q_s\}$ without involving honest parties, and $\textbf{Q}_{\mathcal{H}} = \{Q_1,\cdots,Q_s\} \setminus \textbf{Q}_{\mathcal{A}}$, containing the subformulas that involve at least one honest party. The simulator emulates each $P_c$'s view by running the protocol honestly with these changes:
\begin{itemize}[itemsep=2pt,topsep=0pt,parsep=0pt]
\item It simulates uniform secret shares from each $\FuncbMZS^{Q'_h}$ for each $Q_h \in \textbf{Q}_{\mathcal{H}}$.
\item \textbf{Case $P_1 \notin \textbf{P}_{\mathcal{A}}$.} It samples uniform secret shares $\textbf{u}_{c}'$ from $\FuncMS$.
\item \textbf{Case $P_1 \in \textbf{P}_{\mathcal{A}}$.} After the corrupted parties honestly invoke batch membership zero-sharing protocols for all subformulas in $\textbf{Q}_{\mathcal{A}}$, the parties hold $s'B$ secret-sharings, where we denote all secrets of elements (appended with all-zero strings) as a set $Y_{\mathcal{A}} \in Y$ and $s'B - \lvert Y_{\mathcal{A}} \rvert$ random secrets as a set $R_{\mathcal{A}}$. Let $Y_{\mathcal{H}} = Y\setminus Y_{\mathcal{A}}$.
The simulator samples $(s-s') B - \lvert Y_{\mathcal{H}} \rvert$ random values as a set $R_{\mathcal{H}}$, shuffles the union $Y \cup R_{\mathcal{H}} \cup R_{\mathcal{A}}$ with a random permutation $\pi$ and secret-shares the shuffled union as $\textbf{u}_{1}', \cdots,\textbf{u}_{m}'$, where $\textbf{u}_{c}'$ is outputted to $P_c$ as secret shares from $\FuncMS$ for each $P_c \in \{P_{i_1},\cdots,P_{i_q}\}$. 
\end{itemize}

In the case $P_1 \notin \textbf{P}_{\mathcal{A}}$, it is easy to see that $P_c$'s secret shares from each $\FuncbMZS^{Q'_h}$ and $\FuncMS$ are uniformly distributed and independent of any other distributions in the real execution (as there exists at least an honest party holding one share), which is identical to the simulation.

In the case $P_1 \in \textbf{P}_{\mathcal{A}}$, $P_c$'s secret shares from each $\FuncbMZS^{Q'_h}$ ($Q_h \in \textbf{Q}_{\mathcal{H}}$) are also uniformly distributed and independent of any other distributions in the real execution, so 
\begin{gather*}
\{\mathsf{Sim}^{Q_h}(\textbf{P}^h_{\mathcal{A}},\textbf{X}^h_{\mathcal{A}},\{\textbf{s}_{h, c}\}_{P_c \in \textbf{P}_{\mathcal{A}}^h})_{Q_h \in \textbf{Q}_{\mathcal{H}}}, \{\textbf{s}_{h, c}\}_{Q_h \in \textbf{Q}_{\mathcal{H}}, P_c \in \textbf{P}_{\mathcal{A}}^h}\}_\textbf{X} \\ \overset{c}{\approx} \{\mathsf{View}_{\mathcal{A}}^{Q_h}(\textbf{X}^h)_{Q_h \in \textbf{Q}_{\mathcal{H}}},\{\textbf{s}^{\Pi}_{h, c}\}_{Q_h \in \textbf{Q}_{\mathcal{H}},P_c \in \textbf{P}_{\mathcal{A}}^h}\}_\textbf{X},
\end{gather*}
where $\textbf{X} = \{X_1,\cdots,X_m\}$. $\textbf{P}_{\mathcal{A}}^h$ denotes the corrupted parties involving in $Q_h$, while $\textbf{X}_{\mathcal{A}}^h$ denotes the set of their inputs sets. $\textbf{X}^h$ denotes the set of all involved parties’ inputs sets in $Q_h$. $\mathsf{Sim}^{Q_h}$ is the view emulated by the simulator of $\FuncbMZS^{Q'_h}$, while $\mathsf{View}_{\mathcal{A}}^{Q_h}$ is the real view of adversary in the batch membership zero-sharing protocol for $Q_h$. The distinctions with a superscript $\Pi$ are in the real execution, otherwise in simulation. As the corrupted parties honestly invoke batch membership zero-sharing protocols for all subformulas in $\textbf{Q}_{\mathcal{A}}$, we obtain
\begin{gather*}
\{\mathsf{Sim}^{Q_i}(\textbf{P}^i_{\mathcal{A}},\textbf{X}^i_{\mathcal{A}},\{\textbf{s}_{i, c}\}_{P_c \in \textbf{P}^i_{\mathcal{A}}})_{1 \le i \le s}, \{\textbf{s}_{i, c}\}_{1 \le i \le s, P_c \in \textbf{P}_{\mathcal{A}}}\}_\textbf{X} \\ \overset{c}{\approx} \{\mathsf{View}_{\mathcal{A}}^{Q_i}(X_{i_1},\cdots,X_{i_q})_{1 \le i \le s},\{\textbf{s}^{\Pi}_{i, c}\}_{1 \le i \le s, P_c \in \textbf{P}_{\mathcal{A}}}\}_\textbf{X}.
\end{gather*}

By correctness, after invoking all $\FuncbMZS^{Q'_h}$ for each $Q_h \in \textbf{Q}_{\mathcal{H}}$, the parties hold $\lvert Y_{\mathcal{H}} \rvert$ secret-sharings of the elements in $Y_{\mathcal{H}}$, and $(s-s')B-\lvert Y_{\mathcal{H}} \rvert$ secret-sharings of random values ( the set of these random secrets is denoted by $R^\Pi_{\mathcal{H}}$). By the independence requirement of $\FuncbMZS^{Q'_h}$, all random values in $R^\Pi_{\mathcal{H}}$ are independent of the joint view of any $m-1$ parties, i.e. the view of adversary, in the real execution of batch membership zero-sharing protocols. In simulation, the random values in $R_{\mathcal{H}}$ are sampled using independent randomness so they are also independent of the emulated view for $\FuncbMZS^{Q'_h}$. After the execution of multi-party secret-shared shuffle, the order of elements in $Y \cup R^\Pi_{\mathcal{H}} \cup R^\Pi_{\mathcal{A}}$ is shuffled. By the functionality of $\FuncMS$, the random permutation $\pi^\Pi$ is sampled independently. Thereby,
\begin{gather*}
\{\mathsf{Sim}^{Q_i}(\textbf{P}^i_{\mathcal{A}},\textbf{X}^i_{\mathcal{A}},\{\textbf{s}_{i, c}\}_{P_c \in \textbf{P}^i_{\mathcal{A}}})_{1 \le i \le s}, \{\textbf{s}_{i, c}\}_{1 \le i \le s, P_c \in \textbf{P}_{\mathcal{A}}}, \pi(Y \cup R_{\mathcal{H}} \cup R_{\mathcal{A}})\}_\textbf{X} \\ \overset{c}{\approx} \{\mathsf{View}_{\mathcal{A}}^{Q_i}(X_{i_1},\cdots,X_{i_q})_{1 \le i \le s},\{\textbf{s}^{\Pi}_{i, c}\}_{1 \le i \le s, P_c \in \textbf{P}_{\mathcal{A}}}, \pi^\Pi(Y \cup R^\Pi_{\mathcal{H}} \cup R^\Pi_{\mathcal{A}})\}_\textbf{X}.
\end{gather*}

Given that $\textbf{u}_{1}', \cdots,\textbf{u}_{m}'$ and $\textbf{u}_{1}'^{\Pi}, \cdots,\textbf{u}_{m}'^{\Pi}$ are secret shares of $\pi(Y \cup R_{\mathcal{H}} \cup R^\Pi_{\mathcal{A}})$ and $\pi^\Pi(Y \cup R^\Pi_{\mathcal{H}} \cup R^\Pi_{\mathcal{A}})$ respectively, we derive that
\begin{gather*}
\{\mathsf{Sim}^{Q_i}(\textbf{P}^i_{\mathcal{A}},\textbf{X}^i_{\mathcal{A}},\{\textbf{s}_{i, c}\}_{P_c \in \textbf{P}^i_{\mathcal{A}}})_{1 \le i \le s}, \{\textbf{s}_{i, c}\}_{1 \le i \le s, P_c \in \textbf{P}_{\mathcal{A}}}, \textbf{u}_{1}', \cdots,\textbf{u}_{m}'\}_\textbf{X} \\ \overset{c}{\approx} \{\mathsf{View}_{\mathcal{A}}^{Q_i}(X_{i_1},\cdots,X_{i_q})_{1 \le i \le s},\{\textbf{s}^{\Pi}_{i, c}\}_{1 \le i \le s, P_c \in \textbf{P}_{\mathcal{A}}}, \textbf{u}_{1}'^{\Pi}, \cdots,\textbf{u}_{m}'^{\Pi}\}_\textbf{X}.
\end{gather*}

By invoking the simulator for multi-party secret-shared shuffle $\mathsf{Sim}^{sh}$, 
\begin{gather*}
\{\mathsf{Sim}^{Q_i}(\textbf{P}^i_{\mathcal{A}},\textbf{X}^i_{\mathcal{A}},\{\textbf{s}_{i, c}\}_{P_c \in \textbf{P}^i_{\mathcal{A}}})_{1 \le i \le s}, \{\textbf{s}_{i, c}\}_{1 \le i \le s, P_c \in \textbf{P}_{\mathcal{A}}}, \\ \mathsf{Sim}^{sh}(\textbf{P}_{\mathcal{A}},\{\textbf{u}_c,\textbf{u}_c'\}_{P_c \in \textbf{P}_{\mathcal{A}}}), \textbf{u}_{1}', \cdots,\textbf{u}_{m}'\}_\textbf{X} \\ \overset{c}{\approx} \\ \{\mathsf{View}_{\mathcal{A}}^{Q_i}(X_{i_1},\cdots,X_{i_q})_{1 \le i \le s},\{\textbf{s}^{\Pi}_{i, c}\}_{1 \le i \le s, P_c \in \textbf{P}_{\mathcal{A}}}, \\ \mathsf{View}_{\mathcal{A}}^{sh}(\textbf{u}_{1}^{\Pi},\cdots,\textbf{u}_{m}^{\Pi}),\textbf{u}_{1}'^{\Pi},\cdots,\textbf{u}_{m}'^{\Pi}\}_\textbf{X},
\end{gather*}
where $\textbf{u}_k$ is computed by $\{\textbf{s}_{i, k}\}_{1 \le i \le s}$ We conclude that the adversary's view in real execution is indistinguishable to its view in the simulation. 

The security proof for the circuit-MPSO (Approach 2) protocol is the same. The security proof for the MPSO-card and circuit-MPSO (Approach 1) protocols are similar, except that the simulator replaces all elements in the simulation with 0s, since it only obtains the cardinality rather than the set itself if $P_1 \in \textbf{P}_{\mathcal{A}}$.

\section{Implementation Details and Parameter Settings}\label{appdix:implementation}

\subsection{Implementation Details}
Our protocols are written in C++, where each party uses $m - 1$ threads to interact simultaneously with all other parties. We instantiate batch OPPRF with VOLE and OKVS~\cite{PRTY20,GPRTY-CRYPTO-2021,RR-CCS-2022,BPSY-USENIX-2023}, following~\cite{BuiC23}; We instantiate batch ssPMT with batch OPPRF and ssPEQT, following~\cite{DongCZB24}. We use the following libraries in our implementation.

\begin{itemize}
\item VOLE: We use VOLE implemented in \textsf{libOTe} \cite{libOTe}, instantiating the code family with Expand-Convolute codes~\cite{RRT23}.
\item OKVS and GMW: We use the optimized OKVS construction in \cite{RR-CCS-2022}\footnote{Since the existence of suitable parameters for the new OKVS construction of the recent work \cite{BPSY-USENIX-2023} is unclear when the set size is less than $2^{10}$, we choose to use the OKVS construction of \cite{RR-CCS-2022}.} and re-use the OKVS implementation in \cite{lib-volepsi}. We also re-use the GMW implementation in \cite{lib-volepsi} to construct ssPEQT.
\item ROT: We use SoftSpokenOT \cite{Roy22} implemented in \textsf{libOTe}. 
\item Additionally, we use the \textsf{cryptoTools} \cite{lib-cryptoTools} library to compute hash functions and PRNG calls, and we adopt \textsf{Coproto} \cite{lib-coproto} to realize network communication.
\end{itemize}


\subsection{Choosing Suitable Parameters}
We set the computational security parameter $\lambda = 128$ and the statistical security parameter $\sigma = 40$. The other parameters are:

\begin{trivlist}
\item \textbf{Cuckoo hashing parameters.} To achieve linear communication of batch ssPMT, we use stash-less Cuckoo hashing \cite{PSTY-EUROCRYPT-2019}. To render the failure probability (failure is defined as the event where an item cannot be stored in the table and must be stored in the stash) less than $2^{-40}$, we set $B = 1.27n$ for 3-hash Cuckoo hashing.

\item \textbf{OKVS parameters.} We employ $w = 3$ scheme with a cluster size of $2^{14}$ in~\cite{RR-CCS-2022}, and the expansion rate (which is the size of OKVS divided by the number of encoding items) in this setting is $1.28$.

\item \textbf{ROT parameters.} We set field bits to 5 in SoftSpokenOT to balance computation and communication costs.

\item \textbf{Length of OPPRF outputs.} According to~\cite{DongCZB24}, to ensure the correctness of batch ssPMT, the output length of OPPRF in batch ssPMT is at least $\sigma + \log_2 T + \log_2 B$, where $T$ is the total number of the batch ssPMT invocations, which is $(m^2 - m) / 2$ in our MPSU protocol. Thereby, the lower bound of output length of OPPRF in our MPSU protocol is $\sigma + \log_2((m^2 - m) / 2) + \log_2 (1.27n)$.

\item \textbf{Field size and all-zero string length.} The field size and all-zero string control the probability of a spurious collision in our protocols. According to the correctness analysis in Section~\ref{sec:MPSO}, for MPSI, MPSI-card and MPSI-card-sum protocols, field size of $B \cdot 2^\sigma = 1.27n \cdot 2^\sigma$ is sufficient to bound the probability of any spurious collision to $2^{-\sigma}$. For MPSU protocol, the field size should meet two requirements: $\lvert \mathbb{F} \rvert \ge B \cdot 2^\sigma$ and the length of elements in $\mathbb{F}$ equals $l+l'$. Given that the all-zero string length $l' \ge \sigma + \log (m-1) +\log B$, we have $\lvert \mathbb{F} \rvert \ge 2^l + (m-1)B \cdot 2^\sigma$ in our MPSU. Concretely, we use GF(64) for our MPSI, MPSI-card and MPSI-card-sum protocols, and GF(128) for our MPSU protocol (where $l'$ is set as 64 bits) in our experiments.

\end{trivlist}

\end{document}